\RequirePackage{lineno}
\documentclass[a4paper,aps,12pt,amsmath,amssymb]{revtex4-1}
\usepackage[dvips]{graphicx}
\usepackage{threeparttable}
\usepackage[yyyymmdd,hhmmss]{datetime}
\usepackage{subfigure}
\usepackage{hyperref}
\usepackage{indentfirst}
\usepackage{color}
\usepackage{rotating}
\usepackage{footmisc}
\DeclareGraphicsExtensions{.eps,.mps,.pdf,.jpg,.png}
\begin{document}
\setpagewiselinenumbers
\modulolinenumbers[100]
\linenumbers

\title{The ENUF Method --- Ewald Summation based on Non-Uniform Fast Fourier Transform: Implementation, Parallelization, and Application}

\author{Sheng-Chun Yang~$^{1,}$\footnote{These authors contributed equally to this work.~\label{equal}}, Bin Li~$^{2,}\footref{equal}$, You-Liang Zhu~$^{3,}\footref{equal}$, Aatto Laaksonen~$^{4,5,6,7}$, and Yong-Lei Wang~$^{4,}$\footnote{Author to whom correspondence should be addressed: wangyonl@gmail.com; yonglei.wang@mmk.su.se.}}

\affiliation{$^{1}$ School of Computer Science, Northeast Electric Power University, Jilin 132012, P. R. China\\
$^{2}$ School of Chemical Engineering and Technology, Sun Yat-sen University, Zhuhai 519082, P. R. China\\
$^{3}$ State Key Laboratory of Polymer Physics and Chemistry, Changchun Institute of Applied Chemistry, Chinese Academy of Sciences, Changchun 130022, P. R. China\\
$^{4}$ Department of Materials and Environmental Chemistry, Arrhenius Laboratory, Stockholm University, SE-10691 Stockholm, Sweden\\
$^{5}$ State Key Laboratory of Materials-Oriented and Chemical Engineering, Nanjing Tech University, Nanjing 210009, P. R. China\\
$^{6}$ Centre of Advanced Research in Bionanoconjugates and Biopolymers, Petru Poni Institute of Macromolecular Chemistry Aleea Grigore Ghica-Voda, 41A, 700487 Iasi, Romania\\
$^{7}$ Department of Engineering Sciences and Mathematics, Division of Energy Science, Lule\aa~University of Technology, SE-97187 Lule\aa, Sweden}
\date{\today}

\begin{abstract}
Computer simulations of model systems are widely used to explore striking phenomena in promising applications spanning from physics, chemistry, biology, to materials science and engineering.
The long range electrostatic interactions between charged particles constitute a prominent factor in determining structures and states of model systems.
How to efficiently calculate electrostatic interactions in model systems subjected to partial or full periodic boundary conditions has been a grand challenging task.
In the past decades, a large variety of computational schemes have been proposed, among which the Ewald summation method is the most reliable route to accurately deal with electrostatic interactions in model systems.
In addition, extensive effort has been done to improve computational efficiency of the Ewald summation based methods.
Representative examples are approaches based on cutoffs, reaction fields, multi-poles, multi-grids, and particle-mesh schemes.
We sketched an ENUF method, an abbreviation for the Ewald summation method based on Non-Uniform fast Fourier transform technique, and have implemented this method in particle-based simulation packages to calculate electrostatic energies and forces at micro- and mesoscopic levels.
Extensive computational studies of conformational properties of polyelectrolytes, dendrimer-membrane complexes, and ionic fluids demonstrated that the ENUF method and its derivatives conserve both energy and momentum to floating point accuracy, and exhibit a computational complexity of $\mathcal{O}(N\log N)$ with optimal physical parameters. 
These ENUF based methods are attractive alternatives in molecular simulations where high accuracy and efficiency of simulation methods are needed to accelerate calculations of electrostatic interactions at extended spatiotemporal scales.
\end{abstract}
\maketitle

\tableofcontents

\clearpage
\section{Introduction}

\par
Computer simulations of molecular systems have rapidly expanded over the past decades, and have widely used to study phase behaviors of materials and biological systems~\cite{allen1989computer, frenkel1996understanding}.
Electrostatic interactions between charged particles play a prominent role in determining structures, dynamics, and states of these physical system, leading to many important applications in academia and industrial communities~\cite{sharp1990electrostatic, holm2001electrostatic, naji2005electrostatic, sainis2008electrostatic, kobrak2010electrostatic, wang2018electrostatic}.
An accurate description of electrostatic interactions in model systems is a non-trivial task in computer simulations.
The slow decay feature of electrostatic interactions with respect to particle distance poses a significant challenge to model charged simulation systems as their computations are extremely time consuming~\cite{allen1989computer, frenkel1996understanding, holm2001electrostatic, naji2005electrostatic}.
The spherical cutoff treatment used for short-ranged interactions (such as Lennard-Jones) ignoring particle interactions beyond a certain range is inadequate for electrostatic interactions because any arbitrary truncation leads to nonphysical artifacts.
Therefore, one has to take electrostatic interactions between all pairs of ion species into consideration, which is termed as direct summation method, leading to an unfavorable computational complexity of $\mathcal{O}(N^2)$ (where N is the number of charged particles in simulation systems).
As simulation system size expands, the calculation of electrostatic interactions becomes the major computational bottleneck for a thorough understanding of phase behaviors of charged physical systems at extended spatiotemporal scales~\cite{allen1989computer, frenkel1996understanding, holm2001electrostatic, wang2018electrostatic}.

\par
A traditional way to sum electrostatic interactions between charged particles and all their infinite periodic images is the Ewald summation method~\cite{ewald1921berechnung}.
By introducing a differentiable, localized function, the Ewald summation method recast the total electrostatic interaction, a single slowly and conditionally convergent series, into a short range particle-particle interaction part that can be calculated using spherical cutoff treatment in real space and a long range interaction part for smeared charges that can be computed by solving the Poisson's equation in reciprocal space~\cite{allen1989computer, frenkel1996understanding}.
Although this representation is exact, it contains infinite summation terms and therefore calls for error-controlled approximations for their applicable utilizations in molecular simulations of charged model systems.
The parameters entering the Ewald summation method, $i.e.$, the interaction range of short range part, the number of Fourier modes in long range part, and the splitting parameter controlling the relative weight of short range and long range terms, can be optimized in such a way that the overall performance of the Ewald summation method is reduced to $\mathcal{O}(N^{3/2})$~\cite{allen1989computer, frenkel1996understanding, holm2001electrostatic}.

\par
Although the Ewald summation method represents a substantial improvement respect to the direct summation method and removes the quadratic complexity, the numerical effort is still too large for simulation systems extending to millions of charged particles and long time simulations.
Alternative splitting methods having the same underlying idea as the Ewald summation method were developed to accelerate solvation of the Poisson's equation in reciprocal space by taking advantages of fast Fourier transform (FFT) technique, leading to an $\mathcal{O}(N\log N)$ scaling of computation time.
Examples include the particle-particle particle-mesh (PPPM) Ewald summation method~\cite{eastwood1981computer, hockney1988particle, deserno1998mesh2, brown2012implementing}, the particle-mesh Ewald summation method (PME)~\cite{darden1993particle, batcho2001optimized}, and a variety of derivatives from these splitting methods~\cite{essmann1995smooth, duan2000ewald, shan2005gaussian, harvey2009implementation, wang2010optimizing}.
These methods have been successfully employed in the past decades at varied levels of molecular simulations, but the related physical parameters should be carefully optimized for speedy and accuracy.
Although these splitting methods are substantially faster than the standard Ewald summation method, their accuracy is inferior.
Errors are inevitably introduced in the particle-mesh scheme, which first interpolates particle charges onto a uniform mesh, and thereafter extrapolates the solution of the Poisson's equation represented on mesh back to charged particles~\cite{darden1993particle, batcho2001optimized, shan2005gaussian, harvey2009implementation, wang2010optimizing}.
Therefore, the quality of interpolation and extrapolation plays a critical role in determining computational accuracy of these splitting methods~\cite{deserno1998mesh2, shan2005gaussian, harvey2009implementation, wang2010optimizing, brown2012implementing}.

\par
In previous works, we sketched an ENUF method~\cite{hedman2006ewald}, an abbreviation for the Ewald summation method based on Non-Uniform fast Fourier transform (NFFT) technique~\cite{dutt1993fast, dutt1995fast}, to calculate electrostatic energies and forces between charged particles in molecular simulation systems.
The ENUF method is easy-to-implement and efficient for calculating long range electrostatic interactions, and additionally, both energy and momentum are conserved to floating point accuracy.
Indeed, the ENUF method is the starting point for the subsequent development of particle-particle NFFT with periodic boundary conditions~\cite{weeber2019accelerating}.
By choosing a set of optimal physical parameters, the ENUF method gives a good precision as desired and bears a computational complexity of $\mathcal{O}(N\log N)$~\cite{hedman2006ewald, wang2014non, wang2018electrostatic}.
Later, the ENUF method was implemented in the dissipative particle dynamics (DPD) framework to calculate electrostatic interactions between charge density distributions at mesoscopic level~\cite{wang2013electrostatic, wang2013implementation, wang2014non}.
The ENUF and ENUF-DPD methods were adopted to explore the dependence of conformational properties of polyelectrolytes on charge fraction, ion concentration and counterion valency of added salts~\cite{wang2013implementation}, to investigate specific binding structures of dendrimers on bilayer membranes and the corresponding permeation mechanisms~\cite{wang2012specific}, and to study heterogeneous structures and dynamics in ionic liquids (ILs) and how electrostatic interactions between charged particles affect these properties at extended spatiotemporal scales~\cite{wang2018electrostatic}.

\par
In addition, great endeavors have been made in recent years to accelerate scientific computation using, for example, dedicated and specialized hardware and high-performance accelerator processors~\cite{shaw2007anton}.
The commodity graphics processing unit (GPU) and the compute unified device architecture (CUDA) represent a disruptive technology advance in simulation hardware, provide a mature programming environment, and have been the majority of investigations in computational materials science~\cite{kirk2007nvidia, liu2007molecular, vetter2011keeneland, blumers2017gpu}.
It has been recognized that high-performance GPU accelerated molecular simulations would have a significant impact on all aspects of modelling physical systems~\cite{liu2007molecular, anderson2008general, harvey2009implementation, friedrichs2009accelerating}.
At present, most molecular simulation packages support GPU acceleration~\cite{brown2010porting, stone2010gpu, gotz2012routine, salomon2013routine, abraham2015gromacs}.
We have implemented the ENUF and EUNF-DPD methods in an open source GALAMOST (GPU‐accelerated large‐scale molecular simulation toolkit) package~\cite{zhu2013galamost, wang2018electrostatic, zhu2018employing}.
In addition, several (hybrid) parallelization strategies based on $gridding$~\cite{yang2018new} and $NearDistance$ algorithms~\cite{yang2016accelerating} were developed to accelerate the evaluation of electrostatic energies and forces using GPU and CUDA technology~\cite{yang2017hybrid, yang2020hybrid}.
These derivatives of the ENUF and ENUF-DPD methods exhibit distinct computational efficiencies in handling long range electrostatic interactions between charged particles at extended spatiotemporal scales.

\par
In current contribution, we present a comprehensive review on the detailed implementation of the ENUF method in molecular simulation and DPD frameworks based on CPU nodes, the hybrid parallelization of the ENUF and ENUF-DPD methods using GPU and CUDA toolkit, the determination of effective interactions parameters to achieve an optimal computational complexity, and representation applications of the ENUF and ENUF-DPD methods in treating long range electrostatic interactions between charges particles and charge density distributions in inorganic crystals, polyelectrolytes, dendrimer-membrane complexes, and IL systems.

\clearpage
\section{The Implementation and Parallelization of ENUF and ENUF-DPD Methods in CPU Framework}

\subsection{The Ewald summation method}

\par
We consider a simple cubic simulation system with box length $L$ consisting of $N$ charged particles, each one carrying partial charge $q_i$ at position $\textbf{r}_i$ and interacting with each other according to the Coulomb's law.
An overall charge neutrality is assumed in such a simulation system and the boundary condition without cutoff is represented by replicating the simulation box in three dimensional (3D) space.
The total charge-charge electrostatic interaction energy is given as
\begin{eqnarray}\label{eq:01}
\mathbf{U}^E(\textbf{r}^N)&=&\frac{1}{4\pi\epsilon_0}\sum_{\textbf{n}}^{\dag}\sum_{i}\sum_{j>i} \frac{q_iq_j}{|\textbf{r}_{ij}+\textbf{n}L|}\,,
\end{eqnarray}
where $\textbf{n}=(n_x,n_y,n_z)$, and $n_x$, $n_y$, and $n_z$ are arbitrary integers representing a replication of charged particles in the whole 3D space.
The summation over $\textbf{n}$ takes into account all periodic images, and the \dag{} symbol indicates that the self-interaction terms for all charged particles are omitted when $\textbf{n}=0$.
The variable $\epsilon_0$ is the permittivity (dielectric constant) of the vacuum space.

\par
The direct summation of Eq.~\ref{eq:01}, although simple to implement, suffers a major drawback as a direct numerical evaluation of Eq.~\ref{eq:01} is excessively computational demanding~\cite{allen1989computer, frenkel1996understanding, holm2001electrostatic}.
In the triply periodic case, the direct and infinite summations in Eq.~\ref{eq:01} are conditionally convergent for charge neutral systems, and the computational results (electrostatic energies and forces between charged particles) depend on the order of summation.
In fact it was discovered that any conditionally convergent series can be rearranged to yield a series which converge to any prescribed summation~\cite{allen1989computer, frenkel1996understanding}.
Such a situation is very similar to the case when a linear equation has an infinite number of solutions because it is under-determined; by adding a set of conditions a unique solution will be defined.
In other words, the summation result is not well defined unless one specifies a detailed procedure to sum up these terms.
For electrostatic energies and forces between charged particles, a physically relevant summation order has to be prescribed, and the boundary condition (spherical, cubic, cylindrical, $etc.$) of surrounding medium has to be specified~\cite{allen1989computer, frenkel1996understanding, holm2001electrostatic}.

\par
The Ewald summation formula for the triply periodic case was derived by Ewald in 1921~\cite{ewald1921berechnung}.
The resulting formula imposes two choices: a spherical summation order and an assumption that the dielectric constant of the surrounding medium is infinite, $i.e.$, it is a conductor, which is often referred to “tin~foil” boundary condition.
To gain more physical insights, we consider the electric field generated by a charged particle with partial charge $q_i$ located at $\textbf{r}_i$
\begin{eqnarray}\label{eq:02}
\phi_i(\textbf{r})&=&\frac{1}{4\pi\epsilon_0}\frac{q_i}{|\textbf{r}-\textbf{r}_{i}|}\,.
\end{eqnarray}
The electric field generated by $N$ charged particles and their periodic images at $\textbf{r}_i$ is
\begin{eqnarray}\label{eq:03}
\phi(\textbf{r}_i)&=&\frac{1}{4\pi\epsilon_0}\sum_{\textbf{n}}\sum_{j=1}^N\frac{q_j}{|\textbf{r}_j-\textbf{r}_{i}+\textbf{n}L|}\,.
\end{eqnarray}
Herein, we define $\phi_{[i]}(\textbf{r}_i)$ as the electric field generated by all other charged particles and their periodic images at $\textbf{r}_i$, excluding ion $i$ itself,
\begin{eqnarray}\label{eq:04}
\phi_{[i]}(\textbf{r}_i) \equiv \phi(\textbf{r}_i)-\phi_i(\textbf{r}_i) = \frac{1}{4\pi\epsilon_0}\sum_{\textbf{n}}\sum_{j\ne i}^{N}\frac{q_j}{|\textbf{r}_j-\textbf{r}_{i}+\textbf{n}L|}\,.
\end{eqnarray}
Therefore, the charge-charge electrostatic interaction energy in Eq.~\ref{eq:01} can be rewritten as 
\begin{eqnarray}\label{eq:05}
\mathbf{U}^E(\textbf{r}^N)&=&\frac{1}{2}\sum_{i}^N q_i\phi_{[i]}(\textbf{r}_i)\,.
\end{eqnarray}

\par
The Ewald summation method splits the slowly convergent in Eq.~\ref{eq:01} into two terms that exhibit exponentially fast and absolute convergence at fixed level of accuracy.
The partial charges, described by a collection of delta functions $\rho_i(\textbf{r})=q_i\delta(\textbf{r}-\textbf{r}_i)$, are decomposed into two parts by adding and subtracting a set of Gaussian charge density distributions~\footnote{The functional form can be chosen arbitrarily as long as this function leads to two fast decaying terms. Herein we choose the Gaussian charge density distribution function as an example to extract electrostatic energies and forces between charged particles.}
\begin{eqnarray}\label{eq:06}
\rho_i(\textbf{r})&=&\rho_i^R(\textbf{r})+\rho_i^K(\textbf{r})\,, \nonumber\\
\rho_i^R(\textbf{r})&=&q_i\delta(\textbf{r}-\textbf{r}_i) - q_i G_{\sigma}(\textbf{r}-\textbf{r}_i)\,,\\
\rho_i^K(\textbf{r})&=&q_i G_{\sigma}(\textbf{r}-\textbf{r}_i)\,.\nonumber
\end{eqnarray}
where
\begin{eqnarray}\label{eq:07}
G_{\sigma}(\textbf{r}) = \frac{1}{(2\pi\sigma^2)^{3/2}}e^{-\frac{|\textbf{r}|^2}{2\sigma^2}}\,.
\end{eqnarray}
$\sigma$ is the standard deviation parameter of Gaussian charge density distribution.
In literature, $\alpha \equiv 1/(\sqrt{2}\sigma)$ denoted as the Ewald convergence parameter is always used in molecular simulations.
This splitting scheme leads to the electric field $\phi_{i}(\textbf{r})$ being described by two terms,
\begin{eqnarray}\label{eq:08}
\phi_{i}(\textbf{r})&=&\phi_{i}^R(\textbf{r})+\phi_{i}^K(\textbf{r})\,, \nonumber\\
\phi_{i}^R(\textbf{r})&=&\frac{q_i}{4\pi\epsilon_0}\int\frac{\delta(\textbf{r}-\textbf{r}_i)-G_{\sigma}(\textbf{r}-\textbf{r}_i)}{|\textbf{r}-\textbf{r}_i|}d^3\textbf{r}\,,\\
\phi_{i}^K(\textbf{r})&=&\frac{q_i}{4\pi\epsilon_0}\int\frac{G_{\sigma}(\textbf{r}-\textbf{r}_i)}{|\textbf{r}-\textbf{r}_i|}d^3\textbf{r}\,.\nonumber
\end{eqnarray}
The electric field $\phi_{[i]}(\textbf{r}_i)$ generated by all other charged particles and their periodic images at $\textbf{r}_i$ excluding ion $i$ itself can be decomposed in a similar way,
\begin{eqnarray}\label{eq:09}
\phi_{[i]}(\textbf{r}_i)&=&\phi_{[i]}^R(\textbf{r}_i)+\phi_{[i]}^K(\textbf{r}_i)\,.
\end{eqnarray}
Correspondingly, the charge-charge electrostatic energy in Eq.~\ref{eq:05} is given as
\begin{eqnarray}\label{eq:10}
\mathbf{U}^E(\textbf{r}^N)&=&\frac{1}{2}\sum_{i}^N q_i\phi_{[i]}^R(\textbf{r}_i) + \frac{1}{2}\sum_{i}^N q_i\phi_{[i]}^K(\textbf{r}_i)\,.
\end{eqnarray}

\par
The electric field $\phi_G(\textbf{r})$ generated by the Gaussian charge density distribution $G_{\sigma}(\textbf{r})$ can be obtained by solving the Poisson’s equation,
\begin{eqnarray}\label{eq:11}
\nabla^2\phi_G(\textbf{r}) &=& -\frac{q_iG_{\sigma}(\textbf{r})}{\epsilon_0}\,.
\end{eqnarray}
In spherical coordinate system, $\phi_G(\textbf{r})$ is symmetric and only depends on the magnitude of $r=\textbf{r}$, and therefore it is described as 
\begin{eqnarray}\label{eq:12}
\phi_G(r)&=&\frac{q_i}{4\pi\epsilon_0 r}erf(\alpha r)\,,
\end{eqnarray}
where $erf(x)=\frac{2}{\sqrt{\pi}}\int_0^x e^{-t^2}dt$ is the well-known error function.
Therefore, the electric fields $\phi_{i}^R(\textbf{r})$ and $\phi_{i}^K(\textbf{r})$ in Eq.~\ref{eq:08} can be expressed as 
\begin{eqnarray}\label{eq:1314}
\phi_{i}^R(\textbf{r})&=&\frac{q_i}{4\pi\epsilon_0|\textbf{r}-\textbf{r}_i|}erfc(\alpha |\textbf{r}-\textbf{r}_i|)\,,\\
\phi_{i}^K(\textbf{r})&=&\frac{q_i}{4\pi\epsilon_0|\textbf{r}-\textbf{r}_i|} erf(\alpha |\textbf{r}-\textbf{r}_i|)\,,
\end{eqnarray}
in which $erfc(x)\equiv 1-erf(x)$ is the complementary error function.
As $\lim_{x\to\infty}erf(x)=1$, the $\phi_{i}^K(\textbf{r})$ is a long-ranged nonsingular potential and the $\phi_{i}^R(\textbf{r})$ is a short-ranged singular potential.
The short-ranged singular potential $\phi_{[i]}^R(\textbf{r})$ generated by all other charged particles and their periodic images at $\textbf{r}_i$ excluding ion $i$ itself is given by 
\begin{eqnarray}\label{eq:15}
\phi_{[i]}^R(\textbf{r}_i)&=&\frac{1}{4\pi\epsilon_0}\sum_{\textbf{n}}\sum_{j\ne i}^{N}\frac{q_j}{|\textbf{r}_j-\textbf{r}_i+\textbf{n}L|} erfc(\alpha|\textbf{r}_j-\textbf{r}_i+\textbf{n}L|)\,.
\end{eqnarray}

\par
The short range part of electrostatic energy ($\mathbf{U}^{E,R}(\textbf{r}^N)$ in Eq.~\ref{eq:10}) can be rewritten as 
\begin{eqnarray}\label{eq:16}
\mathbf{U}^{E,R}(\textbf{r}^N) = \frac{1}{2}\sum_{i}^N q_i\phi_{[i]}^R(\textbf{r}_i) = \frac{1}{2}\frac{1}{4\pi\epsilon_0}\sum_{\textbf{n}}\sum_{i}^N\sum_{j\ne i}^{N}\frac{q_iq_j}{|\textbf{r}_{ij}+\textbf{n}L|} erfc(\alpha|\textbf{r}_{ij}+\textbf{n}L|)\,.
\end{eqnarray}
Choosing a suitable value of the Ewald convergence parameter $\alpha$, the short range part of electrostatic energy extends no longer than a cutoff distance and can be expressed as
\begin{eqnarray}\label{eq:17}
\mathbf{U}^{E,R}(\textbf{r}^N)&=&\frac{1}{4\pi\epsilon_0}\sum_{i}^N\sum_{j>i}^{N}\frac{q_iq_j}{r_{ij}}erfc(\alpha r_{ij})\,,
\end{eqnarray}
and therefore, this part can be obtained from a direct summation in real space calculations.

\par
As $\phi^K_i(\textbf{r})$ is a long-ranged nonsingular potential, the long range part of electrostatic energy $\mathbf{U}^{E,K}(\textbf{r}^N)$ in Eq.~\ref{eq:10} is not feasible via a direct computation in real space summations.
It is noteworthy that all Gaussian charge density distributions constitute a periodic function and can be described as 
\begin{eqnarray}\label{eq:18}
\rho^K (\textbf{r})&=&\sum_{\textbf{n}}\sum_{i=1}^N \rho_i^K (\textbf{r}+\textbf{n}L)\,.
\end{eqnarray}
Considering electrostatic contributions from all charged particles, $\phi^K(\textbf{r})$ is the electric field generated by a periodic array of charged particles.
This indicates that $\phi^K(\textbf{r})$ and $\mathbf{U}^{E,K}(\textbf{r}^N)$ are smooth periodic functions and hence their Fourier transformations exhibit fast decay in reciprocal space.

\par
By solving the Poisson's equation in reciprocal space
\begin{eqnarray}\label{eq:19}
\nabla^2\phi^K(\textbf{r}) &=& -\frac{\rho^K(\textbf{r})}{\epsilon_0}\,,
\end{eqnarray}
we can get the electric field generated by all Gaussian charge density distributions as 
\begin{eqnarray}\label{eq:20}
\phi^K (\textbf{r}) &=& \frac{1}{V\epsilon_0}\sum_{\textbf{k}\ne 0}\sum_j^N \frac{q_j}{k^2} e^{i\textbf{k}\cdot (\textbf{r}-\textbf{r}_j)} e^{-k^2/4\alpha^2}\,,
\end{eqnarray}
where $V=L^3$ is the volume of the central simulation system.
The long range part of electrostatic energy can be expressed as
\begin{eqnarray}\label{eq:21}
\mathbf{U}^{E,K}(\textbf{r}^N) = \frac{1}{2}\sum_{i}^Nq_i \phi^K(\textbf{r}_i) = \frac{1}{2V\epsilon_0}\sum_{\textbf{k}\ne 0}\sum_{i}^N\sum_j^N \frac{q_iq_j}{k^2} e^{i\textbf{k}\cdot (\textbf{r}_i-\textbf{r}_j)} e^{-k^2/4\alpha^2}\,.
\end{eqnarray}
By defining a lattice structure factor $S(\textbf{k})=\sum_{i=1}^Nq_i e^{i\textbf{k}\cdot\textbf{r}_i}$, the long range part of electrostatic energy can be rewritten as 
\begin{eqnarray}\label{eq:22}
\mathbf{U}^{E,K}(\textbf{r}^N)&=&\frac{1}{2V\epsilon_0}\sum_{\textbf{k}\ne 0}\frac{e^{-k^2/4\alpha^2}}{k^2} |S(\textbf{k})|^2\,.
\end{eqnarray}

\par
The summation over all Gaussian charge density distributions in Eq.~\ref{eq:18} indicates that a self-interaction term is included for the calculation of long range electrostatic interactions between all Gaussian charge density distributions.
Therefore, this self-interaction energy should be subtracted from the total electrostatic energy.
Taking the electric field $\phi_G(\textbf{r})$ generated by the Gaussian charge density distribution $G_{\sigma}(\textbf{r})$ at $r=0$, we have 
\begin{eqnarray}\label{eq:23}
\phi_{self}=\phi_G(0)=\frac{q_i}{4\pi\epsilon_0 r}\frac{2\alpha}{\sqrt{\pi}}\,.
\end{eqnarray}
The total self-interaction energy becomes
\begin{eqnarray}\label{eq:24}
\mathbf{U}^{E,Self}(\textbf{r}^N)=\frac{1}{2}\sum_{i}^N q_i\phi_{self}(\textbf{r}_i) = \frac{1}{4\pi\epsilon_0}\frac{\alpha}{\sqrt{\pi}}\sum_{i}^N q_i^2\,.
\end{eqnarray}

\par
Therefore, in standard Ewald summation method, the electrostatic energy in Eq.~\ref{eq:01} is decomposed into three contribution terms 
\begin{eqnarray}\label{eq:25}
\mathbf{U}^E(\textbf{r}^N)&=&\mathbf{U}^{E,R}(\textbf{r}^N) + \mathbf{U}^{E,K}(\textbf{r}^N)+\mathbf{U}^{E,Self}(\textbf{r}^N) \nonumber\\
&=&\frac{1}{4\pi\epsilon_0}\bigg\{\sum_i^N\sum_{j>i}^N\frac{q_iq_j}{r_{ij}}\textrm{erfc}\big(\alpha r_{ij}\big)  +\frac{2\pi}{V}\sum_{\textbf{k}\neq 0}\frac{e^{-k^{2}/4\alpha^2}}{k^2}S(\textbf{k})S(-\textbf{k}) -\frac{\alpha}{\sqrt{\pi}}\sum_i^N q_i^2\bigg\}\,,
\end{eqnarray}
with
\begin{eqnarray}\label{eq:26}
S(\textbf{k})=\sum_{i=1}^{N}q_ie^{-\imath \textbf{k}\cdot\textbf{r}_i} \quad \textrm{and} \quad \textbf{k}=\frac{2\pi}{L}\textbf{n}\,.
\end{eqnarray}

\par
From electrostatic energy we can easily obtain the electrostatic force $\mathbf{F}^E_i$ acting on charged particle $i$ by taking a partial derivative of the electrostatic energy $\mathbf{U}_i^E$ with respect to its position $\textbf{r}_i$,
\begin{eqnarray}\label{eq:27}
\mathbf{F}^E_i(\textbf{r}^N)&=&-\nabla_i\mathbf{U}_i^E(\textbf{r}^N)\,.
\end{eqnarray}
Splitting electrostatic force using a similar procedure as recasting electrostatic energy, we can get the total electrostatic force acting on charged particle $i$ from real and reciprocal summations,
\begin{eqnarray}\label{eq:28}
\mathbf{F}^E_i(\textbf{r}^N)&=&-\nabla_i\Big[\mathbf{U}^{E,R}(\textbf{r}^N)+\mathbf{U}^{E,K}(\textbf{r}^N)+\mathbf{U}^{E,Self}(\textbf{r}^N)\Big]\nonumber\\
&=&\mathbf{F}_i^{E,R}(\textbf{r}^N)+\mathbf{F}_i^{E,K}(\textbf{r}^N)\,.
\end{eqnarray}
The short range part of electrostatic force acting on charged particle $i$ can be obtained as
\begin{eqnarray}\label{eq:29}
\mathbf{F}_i^{E,R}(\textbf{r}^N)&=&-\nabla_i\Big[\frac{1}{4\pi\epsilon_0}\sum_{i}^N\sum_{j>i}^{N}\frac{q_iq_j}{r_{ij}}erfc(\alpha r_{ij})\Big]\nonumber\\
&=&\frac{1}{4\pi\epsilon_0}\sum_{j=1}^{N}q_j \Big[\frac{erfc(\alpha\textbf{r}_{ij})}{r_{ij}^2}+\frac{2\alpha}{\sqrt{\pi}}\textrm{exp}(-\alpha^2 r_{ij}^2)\Big]\,.
\end{eqnarray}
Similarly, the long range part of the electrostatic force on charged particle $i$ can be obtained as
\begin{eqnarray}\label{eq:30}
\mathbf{F}_{i}^{E,K}&=&-\nabla_i\Big[\frac{1}{2V\epsilon_0}\sum_{\textbf{k}\ne 0}\sum_{i}^N\sum_j^N \frac{q_iq_j}{k^2} e^{i\textbf{k}\cdot (\textbf{r}_i-\textbf{r}_j)} e^{-k^2/4\alpha^2} \Big] \nonumber\\
&=&\frac{1}{2V\epsilon_0}\sum_{\textbf{k}\ne 0}\frac{1}{k^2}e^{-k^2/4\alpha^2}\bigg\{  \Big[\sum_i^N q_i \cos( \textbf{k} \cdot \textbf{r}_i)\Big]^2 +\Big[\sum_i^N q_i \sin(\textbf{k} \cdot \textbf{r}_i)\Big]^2  \bigg\}   \nonumber\\
&=&\frac{1}{2V\epsilon_0}\sum_{\textbf{k}\ne 0}\frac{1}{k^2}e^{-k^2/4\alpha^2}\bigg\{ \sin(\textbf{k} \cdot \textbf{r}_i)Re(S(\textbf{k}))+\cos(\textbf{k} \cdot \textbf{r}_i)Im(S(\textbf{k})) \bigg\}
\end{eqnarray}

\par
The implementation of Ewald summation method requires the Ewald convergence parameter $\alpha$, a presettled energy accuracy parameter $\delta (\ll 1)$, and two cutoffs ($r_c$ for real space and $n_c$ for reciprocal space).
These parameters are inter-correlated with the following conditions:
\begin{align}
	&e^{-\pi^2|\textbf{n}|^2/(\alpha L)^2}\leq\delta \Longrightarrow n_c\geq \frac{\alpha L}{\pi}\sqrt{-\log(\delta)}\Longrightarrow n_c\propto L\propto N^{1/3}\,,\label{eq:31}\\
	&\textrm{erfc}(\alpha r_c)\approx e^{-\alpha^2r_c^2}\leq\delta\Longrightarrow r_c\approx \frac{\pi n_c}{\alpha^2L}\,.\label{eq:32}
\end{align}
In practical applications with presettled $\delta$, it is straightforward to choose a suitable value for $n_c$, and thereafter one can determine $\alpha$ and $r_c$ directly from Eq.~\ref{eq:32}.

\subsection{Discrete Fourier transforms for non-equispaced data}

\par
The Fourier transform for non-equispaced data-points is a generalization of FFT~\cite{cooley1965algorithm, dutt1993fast}.
The basic idea of NFFT is to combine standard FFT algorithm with various window functions, which are well localized both in space and in frequency domains.
Representative window functions include Gaussian, B-spline, Sinc-power, and Kaiser-Bessel types.
A controlled approximation using a cutoff scheme in frequency domain and a limited number of terms in space domain results in an aliasing error and a truncation error, respectively.
The aliasing error is controlled by an oversampling factor $\sigma_s$, and the truncation error is determined by the number of terms, $m$, in spatial approximation~\cite{dutt1993fast}.
For representative window functions mentioned above, it was found that for a fixed oversampling factor, $\sigma_s > 1$, the truncation error decays exponentially with $m$~\cite{benedetto2001modern, hedman2006ewald, nestler2016parameter, hofmann2017nfft, weeber2019accelerating}.

\par
For a finite number of Fourier coefficients $\hat{\boldsymbol{f_k}}\in\boldsymbol{C}$ with $\boldsymbol{k}\in I_M$, we wish to evaluate the trigonometric polynomial $f(\boldsymbol{x})=\sum_{\boldsymbol{k}\in I_M}\hat{\boldsymbol{f_k}}e^{-2\pi \imath \boldsymbol{k}\cdot\boldsymbol{x}}$ at $N$ non-equispaced points ($\boldsymbol{x}_j\in\boldsymbol{D}^d:j=0,1,\ldots,N-1$).
The space of the $d$-variable function $f\in\boldsymbol{D}^d$ is restricted to the space of $d$-variable trigonometric polynomial $\left(e^{-2\pi \imath \boldsymbol{k}}:\boldsymbol{k}\in I_M \right)$ with a degree of $M_t$($t=0,1,\ldots,d-1$) in the $t$-th dimension.
The possible frequencies $\boldsymbol{k}$ are collected in the multi index set $I_M$ with
\begin{eqnarray}\label{eq:33}
I_M&=&\big\{\boldsymbol{k}=(k_t)_{t=0,1,\ldots,d-1}\in Z^d:-\frac{M_t}{2}\leq k_t\leq \frac{M_t}{2}\big\}\,.
\end{eqnarray}
The dimension of the $d$-variable function or the total number of data-points in the index set is $M_{\Pi}=\Pi_{t=0}^{d-1}M_t$.
As such, the resulting FFT algorithm has a computational complexity of $\mathcal{O}(M_\Pi\log M_\Pi + \log(N/\delta))$, where $\delta$ is the desired computational accuracy~\cite{benedetto2001modern}.

\par
With these preliminary definitions we can perform discrete Fourier transform for non-equispaced data.
The trigonometric polynomial for $N$ given data-points can be described by
\begin{eqnarray}\label{eq:34}
f_j=f(\boldsymbol{x}_j)=\sum_{\boldsymbol{k}\in I_M}\hat{\boldsymbol{f_k}}e^{-2\pi \imath \boldsymbol{k}\cdot\boldsymbol{x}_j}\quad (j=0,1,\ldots,N-1)\,.
\end{eqnarray}
Using a matrix-vector notation, all trigonometric polynomials can be rewritten as $\boldsymbol{f = A}\hat{\boldsymbol{f}}$, where $\boldsymbol{f}=(f_j)_{j=0,1,\ldots,N-1}$, $\boldsymbol{A}=(e^{-2\pi\imath\boldsymbol{k}\cdot\boldsymbol{x}_j})_{j=0,1,\ldots,N-1;\,\boldsymbol{k}\in I_M}$, and $\hat{\boldsymbol{f}}=(\hat{f}_{\boldsymbol{k}})_{\boldsymbol{k}\in I_M}$.

\par
In the following implementations, the related matrix-vector products are the conjugated form
\begin{eqnarray} \label{eq:35}
\boldsymbol{f}=\bar{\boldsymbol{A}}\hat{\boldsymbol{f}}\,, \qquad
\boldsymbol{f}_j=\sum_{\boldsymbol{k} \in I_M} \hat{\boldsymbol{f_k}} e^{2\pi \imath \boldsymbol{k} \cdot \boldsymbol{x}_j}\,,
\end{eqnarray}
and the transposed form
\begin{eqnarray} \label{eq:36}
\hat{\boldsymbol{f}}=\boldsymbol{A}^T\boldsymbol{f}\,, \qquad
\hat{\boldsymbol{f_k}}=\sum_{j=0}^{N-1}\boldsymbol{f}_je^{-2\pi \imath \boldsymbol{k}\cdot\boldsymbol{x}_j}\,,
\end{eqnarray}
in which $\bar{\boldsymbol{A}}$ and $\boldsymbol{A}^T$ are the conjugated and transposed complex of matrix $\boldsymbol{A}$, respectively.
With given Fourier coefficients $\hat{\boldsymbol{f}}$, the Fourier samples $\boldsymbol{f}$ can be transformed with suitable FFT algorithms in both directions.
More in-depth discussion and technical details can be found in publications~\cite{benedetto2001modern, hedman2006ewald, wang2013implementation, wang2014non, nestler2016parameter, hofmann2017nfft, weeber2019accelerating} and references therein.

\subsection{The ENUF method}

\subsubsection{Implementation of the ENUF method}

\par
The ENUF method combines standard Ewald summation method with Non-Uniform FFT technique to handle electrostatic interactions between charged particles.
It is noteworthy that in the ENUF method NFFT only makes approximations for computing long range part of electrostatic energies and forces between charged particles in reciprocal space summations~\cite{hedman2006ewald, wang2013implementation, wang2014non}.
Therefore in the following subsections, we present the detailed procedures to calculate reciprocal space summations of electrostatic energies and forces using NFFT technique.

\par
For electrostatic energies obtained from reciprocal space summations (Eq.~\ref{eq:22}), replacing parameter $\textbf{k}$ with $\textbf{n}$ in the structure factor $S(\textbf{k})$, and then normalizing particle positions $\textbf{x}_i=\textbf{r}_i/L$, the lattice structure factor in Eq.~\ref{eq:26} is rewritten to another form $S(\textbf{n})$ as
\begin{eqnarray}\label{eq:37}
S(\textbf{k})=\sum_{i=1}^{N}q_ie^{-\imath \textbf{k}\cdot\textbf{r}_i}=\sum_{i=1}^{N}q_i e^{-\frac{2\pi\imath}{L}\textbf{n}\cdot\textbf{r}_i}= \sum_{i=1}^{N}q_ie^{-2\pi \imath \textbf{n}\cdot\textbf{x}_i}=S(\textbf{n})\,.
\end{eqnarray}
For a fixed vector $\textbf{n}$, the lattice structure factor $S(\textbf{n})$ is just a complex number.

\par
It is clear that the structure factor $S(\textbf{n})$ in Eq.~\ref{eq:37} and the transposed FFT form in Eq.~\ref{eq:36} have similar structures.
Substituting $q_i$ with $\boldsymbol{f}_j$, the structure factor $S(\textbf{n})$ is then a 3D sample of the transposed FFT form.
By viewing the structure factor $S(\textbf{n})$ as a trigonometric polynomial $\hat{\boldsymbol{f}}_{\textbf{n}}$, the long range part of electrostatic energy determined from reciprocal space summations can be rewritten as
\begin{eqnarray}\label{eq:38}
\mathbf{U}^{E,K}(\textbf{r}^N) &=& \frac{1}{4\pi\epsilon_0}\frac{1}{2\pi L}\sum_{\textbf{n}\neq 0}\frac{e^{-(\pi n)^2/(\alpha L)^2}}{n^2} |\hat{\boldsymbol{f}}_{\textbf{n}}|^2 \,.
\end{eqnarray}
The reciprocal space summations are approximated by a linear combination of window functions sampled at non-equidistant $M_{\Pi}$ grids.
These grids are used as input to transposed FFT, with which we can calculate each component of the structure factor $S(\textbf{n})$, and thereafter the reciprocal space summations of electrostatic energy.
The reciprocal space part of electrostatic force on charged particle $i$ can be obtained in a similar procedure
\begin{eqnarray}\label{eq:39}
\mathbf{F}_{i}^{E,K}&=&-\nabla_i\mathbf{U}^{E,K}\nonumber\\
&=&-\frac{1}{4\pi\epsilon_0}\frac{1}{2\pi L} \sum_{\textbf{n}\neq0} \frac{e^{-(\pi n)^2/(\alpha L)^2}}{n^2} \left(\frac{4\pi q_j}{L}\textbf{n}\right) \nonumber\\
&&{}\bigg\{-\sin(\frac{2\pi}{L}\textbf{n}\cdot\textbf{r}_i)\sum_j q_j\cos(\frac{2\pi}{L}\textbf{n}\cdot\textbf{r}_j)+\cos(\frac{2\pi}{L}\textbf{n}\cdot\textbf{r}_i)\sum_j q_j\sin(\frac{2\pi}{L}\textbf{n}\cdot\textbf{r}_j)\bigg\} \nonumber \\
&=&\frac{1}{4\pi\epsilon_0}\frac{2q_j}{L^2}\sum_{\textbf{n}\neq0} \textbf{n}\frac{e^{-(\pi n)^2/(\alpha L)^2}}{n^2}\bigg\{\sin(\frac{2\pi}{L}\textbf{n} \cdot\textbf{r}_i)Re\big(S(\textbf{n})\big) +\cos(\frac{2\pi}{L}\textbf{n}\cdot\textbf{r}_i)Im\big(S(\textbf{n})\big)\bigg\}\,.
\end{eqnarray}

\par
Since the structure factor $S(\textbf{n})$ is a complex number, the expression in the bracket of Eq.~\ref{eq:39} can be written as the imaginary part of a product
\begin{eqnarray}\label{eq:40}
\sin(\frac{2\pi}{L}\textbf{n}\cdot\textbf{r}_i)Re\big(S(\textbf{n})\big) + \cos(\frac{2\pi}{L}\textbf{n}\cdot\textbf{r}_i)Im\big(S(\textbf{n})\big) &=& Im\bigg\{e^{\frac{2\pi}{L}\imath\textbf{n}\cdot\textbf{r}_i}S(\textbf{n})\bigg\}\,.
\end{eqnarray}
Following this expression, Eq.~\ref{eq:39} can be expressed as
\begin{eqnarray}\label{eq:41}
\mathbf{F}_{i}^{E,K}&=&\frac{1}{4\pi\epsilon_0\epsilon_r}\frac{2q_i}{L^2} \sum_{\textbf{n}\neq0}\textbf{n}\frac{e^{-(\pi n)^2/(\alpha L)^2}}{n^2} Im\bigg\{e^{\frac{2\pi}{L}\imath\textbf{n}\cdot\textbf{r}_i}S(\textbf{n})\bigg\} \nonumber\\
&=&\frac{1}{4\pi\epsilon_0\epsilon_r}\frac{2q_i}{L^2} Im\bigg\{\sum_{\textbf{n}\neq0}\textbf{n} \frac{e^{-(\pi n)^2/(\alpha L)^2}}{n^2}S(\textbf{n})e^{\frac{2\pi}{L}\imath\textbf{n}\cdot\textbf{r}_i}\bigg\}\nonumber\\
&=&\frac{1}{4\pi\epsilon_0\epsilon_r}\frac{2q_i}{L^2}Im\bigg\{\sum_{\textbf{n}\neq0} \hat{\textbf{g}}_{\textbf{n}}
e^{2\pi \imath\textbf{n}\cdot\textbf{x}_i}
\bigg\}\,,
\end{eqnarray}
where $\hat{\textbf{g}}_{\textbf{n}}=\textbf{n}\frac{e^{-(\pi n)^2/(\alpha L)^2}}{n^2}S(\textbf{n})$ with $\textbf{n}\neq0$.
Again, Eq.~\ref{eq:41} is a 3D sample of the conjugated FFT form (Eq.~\ref{eq:35}).
Assuming $\textbf{n}\in I_M$ and $\hat{\textbf{g}}_0=0$, we can reformulate Eq.~\ref{eq:41} into a series of Fourier terms
\begin{eqnarray}\label{eq:42}
\mathbf{F}_{i}^{E,K} =\frac{1}{4\pi\epsilon_0}\frac{2q_i}{L^2} Im\bigg\{\sum_{\textbf{n}\in I_M}\hat{\textbf{g}}_{\textbf{n}}e^{2\pi \imath \textbf{n} \cdot\textbf{x}_i}\bigg\} =\frac{1}{4\pi\epsilon_0r}\frac{2q_i}{L^2}Im(\textbf{g}_i)\,.
\end{eqnarray}
Therefore, we can calculate the reciprocal space summations of electrostatic force on charged particle $i$ using conjugated FFT algorithm based on the structure factor $S(\textbf{n})$ obtained from the transposed FFT algorithm in the calculation of electrostatic energy.

\subsubsection{Determination of optimal parameters for the ENUF method}

\par
Since the ENUF method makes an approximation of the reciprocal space summation of electrostatic energy and force, it is reasonable to expect that the ENUF method behaves in a consistent manner with the standard Ewald summation method.
For the ENUF method, besides the parameters used in the standard Ewald summation method (Eq.~\ref{eq:31} and Eq.~\ref{eq:32}), there are additional two parameters from NFFT controlling the approximation errors: the oversampling factor $\sigma_s$ and the number of terms $m$ in spatial domain approximation~\cite{dutt1993fast, hedman2006ewald, weeber2019accelerating}.
These two parameters are regarded as~\lq\lq knobs\rq\rq~controlling how accurately the structure factor is approximated in NFFT.

\par
\begin{figure}[t]
\centering\includegraphics[width=0.95\textwidth]{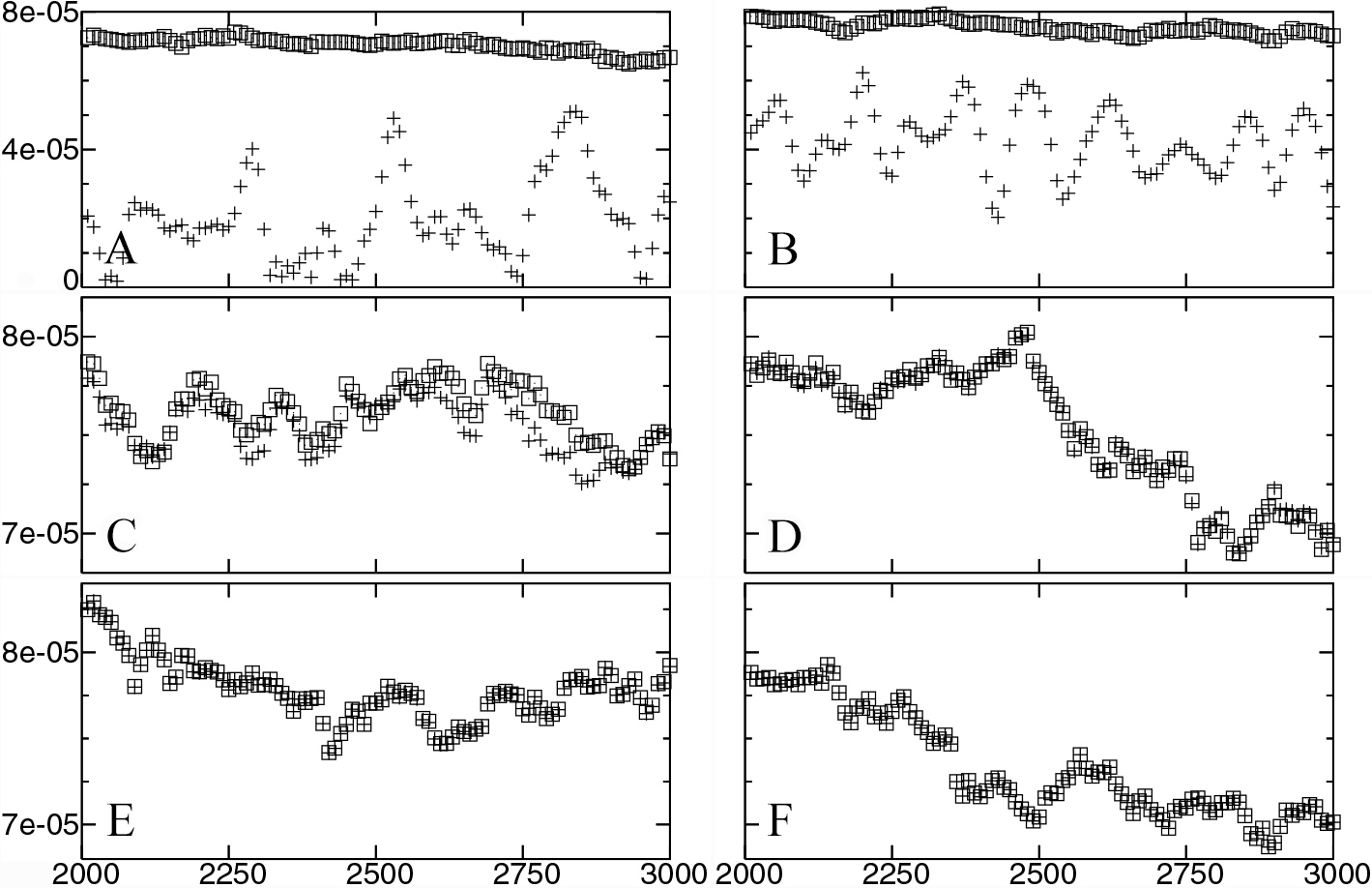}
\caption{Relative error ($\sqrt{(E-\widetilde{E})^2/E^2}$) of electrostatic energy for the standard Ewald summation method and the ENUF method with a given Ewald accuracy parameter $\delta \approx 10^{-5}$. $E$ is the electrostatic reference energy, and $\widetilde{E}$ is the electrostatic energy calculated via either the standard Ewald summation method ($\Box$) or the ENUF method (+) using different approximation parameters with (A) $m=1$ and $\sigma_s=1.5$, (B) $m=1$ and $\sigma_s=2.0$, (C) $m=2$ and $\sigma_s=1.5$, (D) $m=2$ and $\sigma_s=2.0$, (E) $m=3$ and $\sigma_s=1.5$, and (F) $m=3$ and $\sigma_s=2.0$.}\label{fig:enuf_energy}
\end{figure}

Through the calculation of electrostatic energies between charged particles using the standard Ewald summation method and the ENUF method, it is shown in Fig.~\ref{fig:enuf_energy} that for a fixed over-sampling factor $\sigma_s>1$, the relative error for electrostatic energy exhibits a significant decrease with $m$~\cite{hedman2006ewald, dutt1993fast}.
For a cutoff of $m \ge 2$, the approximation error is negligible in practice, indicating a strong similarity of the ENUF method with the standard Ewald summation method in handling electrostatic interactions between charged particles.
Similar computational results are also observed in the relative error and the maximum relative error of electrostatic force acting on charged particle $i$~\cite{hedman2006ewald}.

\par
The scaling characteristic of the ENUF method was estimated by varying the number of charged particles in simulation systems.
For the physical parameters related to the standard Ewald summation method shown in Eq.~\ref{eq:31} and Eq.~\ref{eq:32}, we fix $\delta$ and $n_{cut}$, and thereafter determine $\alpha$ and $r_{cut}$ using constraints as listed in these two equations.
The optimization procedure iterates from the smallest $n_{cut}$ permitted by the simulation system size, and stops when the resulting real space cutoff $r_{cut}$ is compatible with that for short-ranged interactions (such as Lennard-Jones).
With optimal parameters, the ENUF method exhibits a $\mathcal{O}(N\log N)$ scaling behavior.

\subsection{The ENUF-DPD method}

\subsubsection{The DPD method}

\par
The DPD method is a particle-based approach, originally introduced by Hoogerbrugge and Koelman in $1992$ as a novel scheme to simulate hydrodynamic phenomena of complex fluids at mesoscopic level~\cite{hoogerbrugge1992simulating, koelman1993dynamic}.
One important conceptual difference between DPD and atomistic molecular dynamics (MD) approach is the use of coarse-graining procedure allowing a mapping of several atoms or molecules in atomistic simulation systems onto large dissipative particles.
The time evolution of dissipative particles is governed by the Newton's equation of motion
\begin{eqnarray}\label{eq:43}
\frac{\partial \mathbf{r}_i}{\partial t} = \mathbf{v}_i\,,\qquad m_i\frac{\partial \mathbf{v}_i}{\partial t} = \mathbf{f}_i\,,
\end{eqnarray}
where $\mathbf{r}_i$, $\mathbf{v}_i$, and $m_i$ denote the coordinate, velocity, and mass of dissipative particle $i$, respectively.
The total force $\mathbf{f}_i$ acting on the dissipative particle $i$ is normally composed of three different pairwise additive forces: the conservative force $\mathbf{F}_{ij}^{C}$, the dissipative force $\mathbf{F}_{ij}^D$, and the random force $\mathbf{F}_{ij}^R$,
\begin{eqnarray}\label{eq:44}
\mathbf{f}_{i}&=&\sum_{i\neq{j}}(\mathbf{F}_{ij}^{C}+\mathbf{F}_{ij}^{D}+\mathbf{F}_{ij}^{R})\,,
\end{eqnarray}
with
\begin{eqnarray}
\mathbf{F}_{ij}^C&=&\alpha_{ij}\omega^C(r_{ij})\mathbf{\hat{r}}_{ij}\,,\label{eq:45}\\
\mathbf{F}_{ij}^D&=&-\gamma\omega^D(r_{ij})(\mathbf{v}_{ij}\cdot\mathbf{\hat{r}}_{ij})\mathbf{\hat{r}}_{ij}\,,\label{eq:46}\\
\mathbf{F}_{ij}^R&=&\sigma\omega^R(r_{ij})\theta_{ij}\mathbf{\hat{r}}_{ij}\,,\label{eq:47}
\end{eqnarray}
where $\mathbf{r}_{ij}=\mathbf{r}_i-\mathbf{r}_j$, $r_{ij}=|\,\mathbf{r}_{ij}|$, $\mathbf{\hat{r}}_{ij}=\mathbf{r}_{ij}/r_{ij}$, and $\mathbf{v}_{ij}=\mathbf{v}_i-\mathbf{v}_j$.
The parameters $\alpha_{ij}$, $\gamma$, and $\sigma$ determine the strength of conservative, dissipative, and random forces, respectively.
$\theta_{ij}$ is a random fluctuating variable with zero mean and unit variance.

\par
The pairwise conservative force is usually written as a weight function $\omega^C(r_{ij})$ with the form of 
\begin{eqnarray}\label{eq:48}
\omega^C(r_{ij})&=&
\left\{\begin{array}{cl}
(1-r_{ij}/r_c) &\left(r_{ij}\leq r_c\right) \\
0&\left(r_{ij}>r_c\right)
\end{array}\right.\,.
\end{eqnarray}
Compared with the Lennard-Jones 12-6 potential, the conservative force adopted in the DPD method is a soft repulsive force, and hence it allows a large time step in the integration of the equation of motion of all dissipative particles.
The unit of length $r_c$ is related to the volume of dissipative particles and can be determined from specific coarse-graining schemes.

\par
Two weight functions $\omega^D(r_{ij})$ and $\omega^R(r_{ij})$ for dissipative and random forces are coupled together via the fluctuation-dissipation theorem
\begin{eqnarray}\label{eq:49}
\omega^D\left(r\right)=\left[\omega^R\left(r\right)\right]^2 \quad \textrm{and}\quad
\sigma^2=2\gamma k_BT
\end{eqnarray}
to form a thermostat and generate natural canonical distribution~\cite{espanol1995statistical}.
In most applications, the weight function $\omega^D(r)$ adopts a simple form as~\cite{groot1997dissipative}
\begin{eqnarray}\label{eq:50}
\omega^D\left(r\right)=\left[\omega^R\left(r\right)\right]^2=\left\{\begin{array}{ll}
\left(1-r/r_c\right)^2&\left(r\leq r_c\right)\\
0&\left(r>r_c\right)
\end{array}\right.\,.
\end{eqnarray}

\par
One important consequence of the DPD formulation is that all interactions are pairwise additive and satisfy the Newton's third law, leading to both linear and angular momentum being conserved~\cite{hoogerbrugge1992simulating, koelman1993dynamic, groot1997dissipative}.
In addition, all three pairwise forces depend only on the relative positions and velocities between interacting dissipative particles, leading to the DPD model Galilean-invariant.
The satisfaction of these conditions makes the DPD method a consistent coarse-grained (CG) approach particularly appealing for studying soft matter systems at mesoscopic level~\cite{pagonabarraga2001dissipative, lu2013introduction, espanol2017perspective}.
Examples of these investigations are microphase separation of multiblock polymers~\cite{groot1998dynamic, qian2005computer}, polymeric surfactants in solution~\cite{rekvig2004chain}, colloidal suspensions~\cite{whittle2010dynamic, mai2015investigation}, structural and rheological behavior of biological membranes~\cite{kranenburg2005phase, shillcock2005tension, de2009effect} and red blood cells~\cite{li2012blood, blumers2017gpu, li2017computational}.

\subsubsection{Implementation of the ENUF-DPD method}

\par
In the DPD method, one critical advantage is the soft repulsive nature of the conservative potential, which enables to integrate the equation of motion of dissipative particles using a large time step.
However, such an advantage restricts the direct incorporation of electrostatic interactions in DPD model because dissipative particles carrying opposite point charges tend to collapse onto each other, forming artificial ion clusters due to stronger electrostatic interactions than soft repulsive conservative interactions~\cite{groot2003electrostatic, gonzalez2006electrostatic, ibergay2009electrostatic, warren2013screening, terron2016electrostatics}.
In order to avoid such non-physical phenomena, point charges at the center of dissipative particles are replaced by charge density distributions meshed around particles to remove the divergence of electrostatic interactions between point charges at $r=0$~\cite{groot2003electrostatic, gonzalez2006electrostatic, vaiwala2017electrostatic, eslami2019gaussian}.

\par
In the ENUF-DPD framework~\cite{wang2013electrostatic, wang2013implementation}, we used a Slater-type charge density distribution with the form of
\begin{eqnarray}\label{eq:51}
\rho_e(\textbf{r})=\frac{q}{\pi\lambda_e^3}e^{\frac{-2\textbf{r}}{\lambda_e}}\,,
\end{eqnarray}
in which $\lambda_e$ is the decay length of charge $q$.
The integration of Eq.~\eqref{eq:51} over the whole space gives the total charge $q$~\cite{groot2003electrostatic, gonzalez2006electrostatic}.
The electric field $\phi(r)$ generated by the Slater-type charge density distribution $\rho_e(r)$ can be obtained by solving the Poisson's equation
\begin{eqnarray}\label{eq:52}
\phi(\textbf{r}) &=& \frac{1}{4\pi\epsilon_0}\frac{q}{\textbf{r}} \Big(1-(1+\frac{\textbf{r}}{\lambda_e}) e^{\frac{-2\textbf{r}}{\lambda_e}}\Big)\,.
\end{eqnarray}
The electrostatic energy between two interacting Slater-type charge density distributions $i$ and $j$ is the product of the total charge density distribution $i$ and the electric field generated by the Slater-type charge density distribution $j$ at position $\textbf{r}_i$
\begin{eqnarray}\label{eq:53}
U_{ij}^{E,DPD}(\textbf{r}_{ij})=q_i\phi_j(\textbf{r}_i)
=\frac{1}{4\pi\epsilon_0}\frac{q_iq_j}{\textbf{r}_{ij}}
\Big(1-(1+\frac{\textbf{r}_{ij}}{\lambda_e})e^{\frac{-2\textbf{r}_{ij}}{\lambda_e}}\Big)\,.
\end{eqnarray}
The electrostatic force acting on the Slater-type charge density distribution $i$ is obtained by taking the negative of the derivative of electrostatic energy $U_{ij}^{E,DPD}$ respect to its position $\textbf{r}_i$
\begin{eqnarray}\label{eq:54}
F_{ij}^{E,DPD}(\textbf{r}_{ij})=-\nabla_i U_{ij}^{E,DPD}(\textbf{r}_{ij})
=\frac{1}{4\pi\epsilon_0}\frac{q_iq_j}{(\textbf{r}_{ij})^2} \bigg\{1-\Big(1+\frac{2\textbf{r}_{ij}}{\lambda_e}\big(1+\frac{\textbf{r}_{ij}}{\lambda_e} \big)\Big)e^{\frac{-2\textbf{r}_{ij}}{\lambda_e}}\bigg\}\,.
\end{eqnarray}
By defining a dimensionless parameter $\textbf{r}^*=\textbf{r}/r_c$ as the reduced center-to-center distance between two Slater-type charge density distributions and $\beta=r_c/\lambda_e$, the reduced electrostatic energy and force are given by
\begin{eqnarray}
U_{ij}^{E,DPD}(\textbf{r}_{ij}^*) &=&\frac{1}{4\pi\epsilon_0} \frac{q_iq_j}{r_c\textbf{r}_{ij}^*}\bigg\{1-\Big(1+\beta \textbf{r}_{ij}^*\Big) e^{-2\beta \textbf{r}_{ij}^*}\bigg\}\,,\label{eq:55}\\
F_{ij}^{E,DPD}(\textbf{r}_{ij}^*)&=&\frac{1}{4\pi\epsilon_0} \frac{q_iq_j}{(r_c\textbf{r}_{ij}^*)^2}\bigg\{1-\Big(1+2\beta \textbf{r}_{ij}^*(1+\beta \textbf{r}_{ij}^*)\Big)e^{-2\beta \textbf{r}_{ij}^*}\bigg\}\,.\label{eq:56}
\end{eqnarray}

\par
Now it is clear that the electrostatic energy and force between Slater-type charge density distributions in DPD simulations are those between point charges in MD simulations scaled with correction factors of
\begin{eqnarray}\label{eq:5758}
B_U &=& 1-\Big(1+\beta \textbf{r}^*\Big)e^{-2\beta \textbf{r}^*}\,,\\
B_F &=& 1-\Big(1+2\beta \textbf{r}^*(1+\beta r^*)\Big)e^{-2\beta \textbf{r}^*}\,.
\end{eqnarray}
Such similarities between electrostatic energies and forces in MD and DPD simulations imply that once we get electrostatic energies and forces between point charges in MD simulations, from which the electrostatic energies and forces between Slater-type charge density distributions in DPD simulations can be directly scaled with the corresponding correction factors.
In the limit of $r_{ij}^*\to 0$, the reduced electrostatic energy and force between the Slater-type charge density distributions are described by 
\begin{eqnarray}
\lim\limits_{\textbf{r}_{ij}^* \to 0}U_{ij}^{E,DPD}(\textbf{r}_{ij}^*) &=& \frac{1}{4\pi\epsilon_0} \frac{q_iq_j}{R_c}\beta \,,\label{eq:59}\\
\lim\limits_{\textbf{r}_{ij}^*\to 0}F_{ij}^{E,DPD}(\textbf{r}_{ij}^*) &=& 0 \label{eq:60}\,.
\end{eqnarray}
From Eq.~\ref{eq:59} and Eq.~\ref{eq:60}, we can specify that the adoption of Slater-type charge density distributions in DPD simulations removes the divergence of electrostatic interactions at $\textbf{r}_{ij}^* =0$, indicating that both electrostatic energies and forces between Slater-type charge density distributions are finite quantities.

\par
By matching electrostatic interactions between Slater-type charge density distributions at $\textbf{r}_{ij}^*=0$ with previous work~\cite{groot2003electrostatic}, it gives $\beta=1.125$.
From the relation of $\beta=R_c/\lambda_e$, we can get $\lambda_e = 6.954~\textrm{\AA}$, which is consistent with the electrostatic smearing radii used in Gonz\'alez-Melchor's computational model~\cite{gonzalez2006electrostatic}.
It should be noted that as charge density distributions in simulation systems are affected by hydrodynamic flow~\cite{pagonabarraga2010recent}, these proposed methods provide a natural coupling between electrostatics and fluid motion.

\begin{figure}[t]
\centering\includegraphics[width=0.75\textwidth]{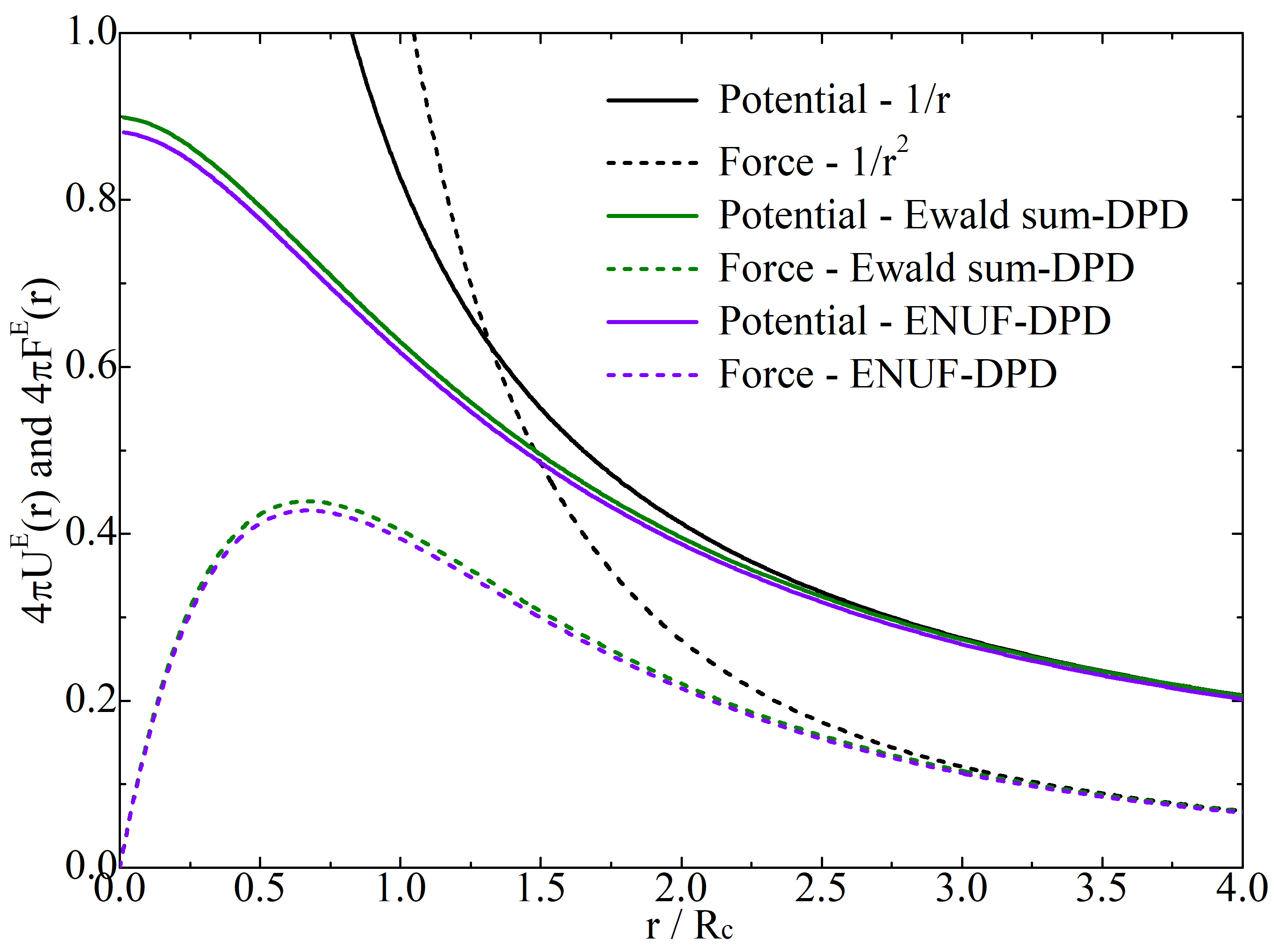}
\caption{Electrostatic potential and force between Slater-type charge density distributions calculated from the ENUF and the Ewald summation methods. The Coulombic potential and force, both of which diverge at $r=0$, are included for a comparative purpose. Both electrostatic potential and force expressions are plotted for two equal sign charge density distributions.}\label{fig:enuf_dpd_energy_force}
\end{figure}

\par
Fig.~\ref{fig:enuf_dpd_energy_force} presents the reduced electrostatic potential and force between Slater-type charge density distributions.
It is clearly demonstrated that at short distance $r < 3.0R_c$ the electrostatic energy and force calculated using the ENUF-DPD method are comparable with those obtained from the Ewald summation method.
In addition, both the ENUF-DPD and the Ewald summation methods give indistinguishable electrostatic energy and force at $r\ge 3.0R_c$.
Therefore, the ENUF-DPD method can capture essential characteristics of electrostatic interactions as the Ewald summation method does in DPD simulations in describing phase behaviors of charged soft matter systems at extended spatiotemporal scales~\cite{groot2003electrostatic, gonzalez2006electrostatic, ibergay2009electrostatic, yan2009dissipative, warren2013screening, mao2015modeling, gavrilov2016dissipative, terron2016electrostatics, vaiwala2017electrostatic, eslami2019gaussian}.

\subsubsection{Determination of physical parameters for the ENUF-DPD method}

\par
The implementation of the ENUF-DPD method uses similar parameters as the ENUF method does, and the correlations between these physical parameters are described by Eq.~\ref{eq:31} and Eq.~\ref{eq:32} with pre-determined oversampling factor $\sigma_s$ and parameter $m$ controlling the number of terms in spatial domain approximation~\cite{dutt1993fast, hedman2006ewald, weeber2019accelerating}.
However, due to the fact that $n_c$ should be an integer and $r_c$, which is the short range cutoff for conservative interactions between dissipative particles and is related to the volume of dissipative particles determined from specific coarse-graining schemes, should be a suitable value for the link-cell list update scheme in DPD simulations, we adopted another procedure to determine these parameters for the ENUF-DPD method.

\par
First, due to the soft repulsive feature of conservative force $\mathbf{F}^C$ in the DPD method, we adopted $\delta = 1.0\times10^{-4}$ for the computational accuracy parameter, which is enough to keep acceptable accuracy in describing electrostatic interactions between charge density distributions in DPD simulations~\cite{hedman2006ewald}.

\par
Second, we determined a suitable value for the short range cutoff $r_c$.
Gonz\'alez-Melchor~\emph{et al.}~\cite{gonzalez2006electrostatic} adopted $1.08 R_c$ and $3.0R_c$, respectively, as electrostatic smearing radii and real space cutoff for electrostatic interactions between Slater-type charge density distributions calculated using the Ewald summation method.
In the ENUF-DPD method, as specified in Eq.~\ref{eq:55} and Eq.~\ref{eq:56}, electrostatic energy $U_{ij}^{E,DPD}(\textbf{r}_{ij}^*)$ and force $F_{ij}^{E,DPD}(\textbf{r}_{ij}^*)$ are scaled with correction factors, $B_U$ and $B_F$, respectively, both of which are $\textbf{r}$-dependent.
This indicates that reciprocal space summations of electrostatic energy $U_{ij}^{E,K,DPD}(\textbf{r}_{ij}^*)$ and force $F_{ij}^{E,K,DPD}(\textbf{r}_{ij}^*)$ are also scaled with the corresponding correction factors.
It is noteworthy that the electrostatic energy and force we obtained from conjugated and transposed FFT algorithms are the total influence of the other charged dissipative particles on particle $i$.
It is difficult to differentiate their individual contributions since the corresponding correction factors for the other charged dissipative particles are related to their relative distance to particle $i$.
But if we choose suitable $r_c$, beyond which two correction factors $B_U$ and $B_F$ approximate to $1.0$, the total reciprocal space summations of electrostatic energy $B_UU_{ij}^{E,K,DPD}(\textbf{r}_{ij}^*)$ and force $B_FF_{ij}^{E,K,DPD}(\textbf{r}_{ij}^*)$ can be approximately expressed as $U_{ij}^{E,K,DPD}(\textbf{r}_{ij}^*)$ and $F_{ij}^{E,K,DPD}(\textbf{r}_{ij}^*)$, respectively.
Such an approximation enables us to directly use conjugated and transposed FFT results as reciprocal space summations.
It was found that both $B_U$ and $B_F$ converge to unity when $r \ge 3.0 R_c$, which is consistent with those for electrostatic energy and force between Slater-type charge density distributions shown in Fig.~\ref{fig:enuf_dpd_energy_force}.
Therefore in the ENUF-DPD scheme, $r_c = 3.0 R_c$ is taken as the cutoff for real space summations of electrostatic interactions.
Such an adoption indicates that both $B_U$ and $B_F$ are only applied on real space summations of electrostatic interactions within a cutoff of $r_c =3.0 R_c$.

\par
Since the Fourier-based Ewald summation methods utilize FFT to evaluate reciprocal space summations, it is more appropriate to choose a suitable value for the Ewald convergence parameter $\alpha$, with which we can minimize the total computational time in the calculation of electrostatic interactions from real and reciprocal space summations.
The choice of $\alpha$ is system-dependent and is related to trade-offs between accuracy and computational speed.
Based on Eq.~\ref{eq:31} and Eq.~\ref{eq:32}, as well as the determined $r_c = 3.0R_c$, one can deduce that $\alpha\geq 0.12~\textrm{\AA}^{-1}$.
Although the electrostatic energy is invariant to $\alpha$, the value of $\alpha$ indeed affects the total time in calculating electrostatic interactions.
In order to find a suitable value for $\alpha$, we carried out a set of trial simulations to evaluate the Madelung constant of a face-centered cubic (FCC) crystal lattice consisting of $4000$ charged particles, half of which are positively charged with partial charge of $+1.0$ and the other half have negative partial charge of $-1.0$, respectively.
It was revealed that for a wide range of $\alpha$ values the calculated Madelung values coincide with the theoretical value~\cite{lu2003computer}.
The lowest acceptable value, $\alpha=0.20~\textrm{\AA}^{-1}$, is then adopted in the subsequent DPD simulations to minimize the total computational effort.

\par
In the last step, the parameter $n_c$ was determined together with additional two parameters ($\sigma_s$ and $m$) controlling approximation errors in NFFT.
It has been shown in MD simulations that $\sigma_s=2$ is adequate to provide reliable computational accuracy in the calculation of electrostatic interactions between charged particles~\cite{hedman2006ewald}.
Therefore, this value is used in DPD simulations to keep a comparable accuracy in handling electrostatic interactions between charge density distributions at mesoscopic level.

\par
\begin{figure}[t]
\centering\includegraphics[width=0.95\textwidth]{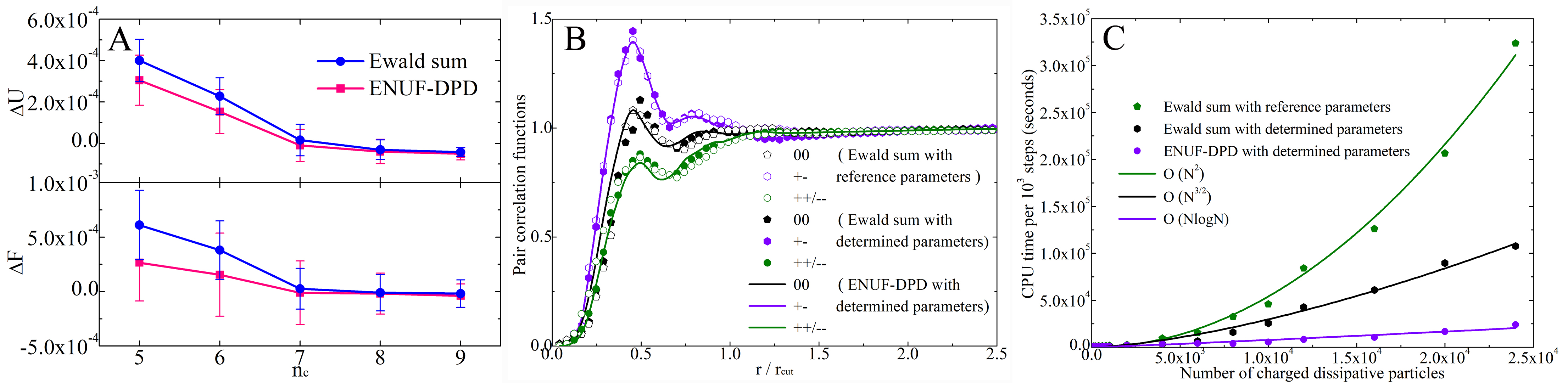}
\caption{(A) The errors in electrostatic energy ($\Delta U=\frac{(U^E-U^E_{ref})}{U^E_{ref}}$) and force ($\Delta F=\frac{\bar{F}^E-\bar{F}^{E,ref}}{\bar{F}^{E,ref}}$) calculated using the ENUF-DPD and the Ewald summation methods with $m=2$ and various $n_c$ are compared with those calculated from the standard Ewald summation method with reference parameters.
$U^E_{ref}$ ($\bar{F}^{E,ref}$) is the total (averaged) electrostatic energy (force) on charged dissipative particles calculated from the Ewald summation method with reference parameters, and $U^E$ ($\bar{F}^E$) is the total (averaged) electrostatic energy (force) on charged dissipative particles calculated via either the ENUF-DPD or the Ewald summation methods with determined parameters.
(B) The pair correlation functions between different types of dissipative particles calculated from the ENUF-DPD and the Ewald summation methods with determined parameters, as well as the standard Ewald summation method with reference parameters.
(C) The computational complexity of the ENUF-DPD method with pre-determined physical parameters.}\label{fig:enuf_dpd_uf_rdf_scaling}
\end{figure}

Additional trial simulations were performed on bulk electrolyte systems to determine parameters $n_c$ and $m$.
It is shown in Fig.~\ref{fig:enuf_dpd_uf_rdf_scaling}A that the relative errors of electrostatic energies and forces calculated from the ENUF-DPD and the Ewald summation methods with $m=2$ and $n_c\geq7$ fluctuate within pre-determined accuracy values, indicating that these two methods behave in a same manner in describing electrostatic interactions between charge density distributions with adopted physical parameters.
With larger $m$ and $n_c$ values, we can further increase the accuracy of electrostatic interactions in DPD simulations, which, however, leads to increased computational time in handling electrostatic interactions.
By compromising the accuracy and computational efficiency of the ENUF-DPD method, $m=2$ and $n_c=7$ were used in following DPD simulations.

\par
A third set of trial simulations were performed to study microstructural properties of dilute aqueous electrolyte solution with a salt concentration of $0.6$ mol using the standard Ewald summation method with reference parameters, and the ENUF-DPD and the Ewald summation methods with above determined parameters.
It is shown in Fig.~\ref{fig:enuf_dpd_uf_rdf_scaling}B that pair correlation functions for the same pair particles calculated from three methods exhibit similar tendencies.
A peculiar feature is that there is no ion cluster formation at distance close to $r=0$.
In addition, pair correlation functions between Slater-type charge density distributions satisfy $g_{+-}(r)g_{++/--}(r)=g^2_{00}(r)$, where $+$, $-$, and $0$ correspond to positive, negative, and neutral dissipative particles in simulation systems, respectively.
This indicates that microstructures between charged particles are related to the effective electrostatic potentials between different particle pairs~\cite{groot2003electrostatic}.

\par
By a systemic variation of the number of charged dissipative particles in simulation systems, we estimated the scaling behavior of the ENUF-DPD method with pre-determined parameters.
Fig.~\ref{fig:enuf_dpd_uf_rdf_scaling}C presents the averaged time per $10^3$ steps in DPD simulations as a function of the number of charged dissipative particles.
It is shown that the standard Ewald summation method with reference parameters scales as $\mathcal{O}(N^2)$, and its computational complexity is reduced to $\mathcal{O}(N^{3/2})$ with pre-determined optimal parameters, which is consistent with that observed in previous studies~\cite{gonzalez2006electrostatic, ibergay2009electrostatic}.
The ENUF-DPD method with the above determined parameters exhibits an excellent computational efficiency in describing electrostatic interactions between charge density distributions at extended spatiotemporal levels, and scales as $\mathcal{O}(N\log N)$, which is in line with the scaling behavior of FFT in treating electrostatic interactions.
Theoretically, the computational complexity of NFFT is $\mathcal{O}(M_{\Pi}\log M_{\Pi} + \log(N/\delta))$, where $M_{\Pi}$ is the total number of data-points in the index set, and $\delta$ is the desired computational accuracy and also a function of $m$ for fixed over-sampling factor $\sigma_s$~\cite{dutt1993fast}.
Combining the definition of $M_{\Pi}$ and the relationship in Eq.~\ref{eq:31} and Eq.~\ref{eq:32}, we can get $M_{\Pi}\propto n_c^3\propto N$.
Therefore the theoretical complexity of the ENUF-DPD method is $\mathcal{O}(N\log N+\log(N/\delta))$, which is well reproduced from trial DPD simulations.

\clearpage
\section{Parallelization of the ENUF Method in CPU and GPU Frameworks}

\subsection{Development of the CU-NFFT}

\par
It is shown in the previous section that the ENUF and the ENUF-DPD methods are mainly responsible for the computation of reciprocal space summations of electrostatic energies and forces via conjugated and transposed NFFTs.
More specifically, the estimation of electrostatic energies in the ENUF and the ENUF-DPD methods needs a forward NFFT, and the evaluation of electrostatic forces on charged particles or charge density distributions needs three inverse NFFTs.
These two processes overwhelmingly dominate the computational efficiency of NFFT.
Therefore, the speedup of NFFT calculations is another procedure to improve the efficiency of the ENUF and the ENUF-DPD methods in handling electrostatic interactions at extended spatiotemporal scales.

\par
\begin{figure}[t]
\centering\includegraphics[width=0.95\textwidth]{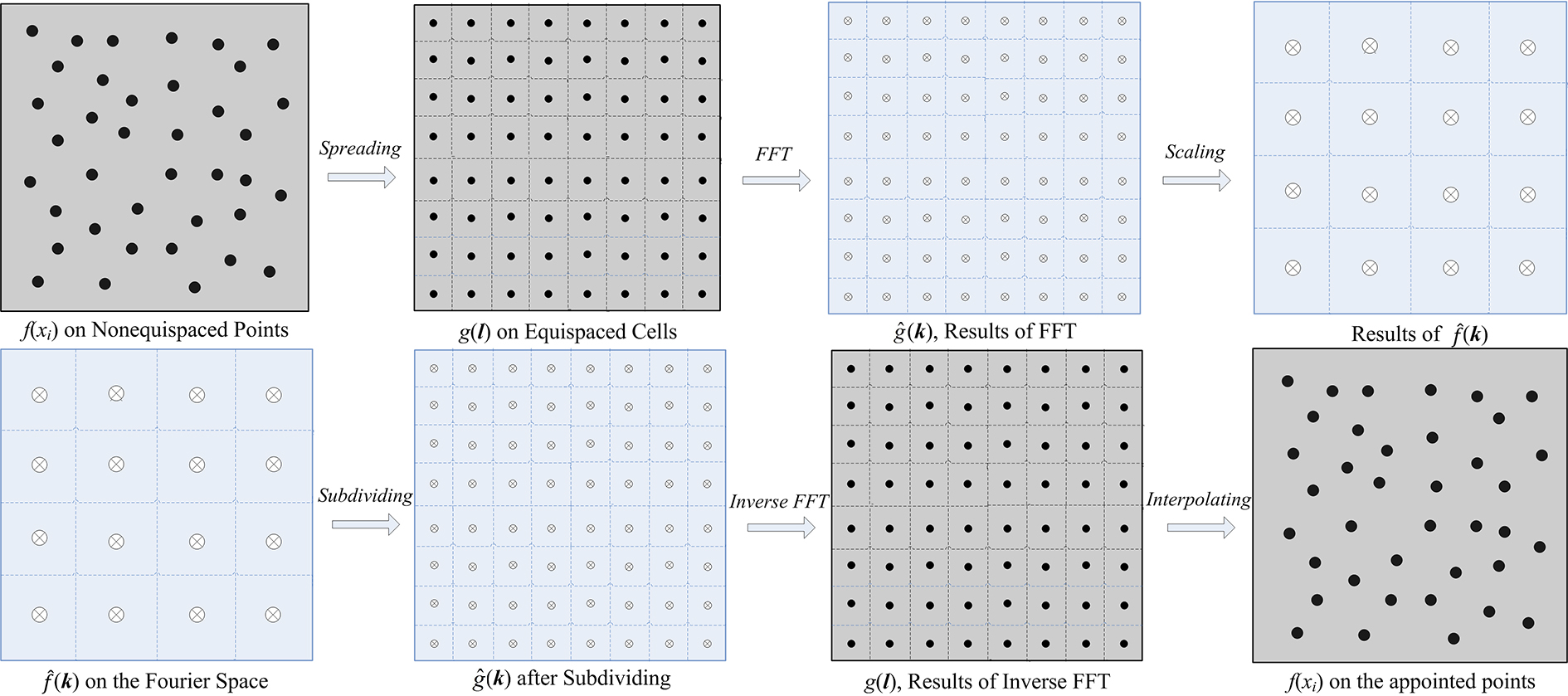}
\caption{The forward (upper panel) and inverse (lower panel) NFFTs are decomposed into spreading, FFT, and scaling steps, and into subdividing, inverse FFT, and interpolating steps, respectively, in the proposed $gridding$ algorithm.}\label{fig:gridding_algorithm}
\end{figure}

Following the work of Greengard and Lee~\cite{greengard2004accelerating}, we proposed a $gridding$ algorithm to accelerate NFFT~\cite{yang2018new}.
The main idea of the $gridding$ algorithm is to transform non-equispaced data-points in 3D space into equivalent equispaced ones via an approximation scheme, and thereafter to accelerate NFFT calculations using standard FFT algorithms~\cite{yang2018new}.
The approximation scheme mapping data from non-equispaced to equispaced matrices is performed on the basis of a window function $\phi$ that is well localized both in spatial and in frequency domains, respectively.
The forward NFFT for the estimation of electrostatic energies between charged particles or charge density distributions is decomposed into spreading, FFT, and scaling steps (upper panel in Fig.~\ref{fig:gridding_algorithm}), which first spreads the values of $f(x_i)$ from non-equispaced data-points $x_i$ to equispaced and over-sampled cells using a periodic window function $\phi$, then performs forward NFFT on equispaced cells using standard FFT algorithms, and finally scales the computational FFT results to obtain $\hat{f}(\textbf{k})$ in a given frequency domain.

\par
Correspondingly, the inverse NFFT for the estimation of electrostatic forces on charged particles or charge density distributions is divided into subdividing, inverse FFT, and interpolating steps, respectively (lower panel in Fig.~\ref{fig:gridding_algorithm}).
In the inverse NFFT calculations, the values of $\hat{f}(\textbf{k})$ are first subdivided into $\hat{g}(\textbf{k})$ in an over-sampled cells, which are further converted to $g_l$ in real space using an inverse FFT algorithm.
Then a interpolation process is executed on the given data-points $x_i$ to obtain the values of $f(x_i)$ using the periodic window function $\phi$.
In the implementations of spreading and interpolating steps, a Gaussian window function is adopted due to the fact that its periodic version has a uniformly convergent Fourier series and is also well localized both in spatial and frequency domains~\cite{yang2018new}.

\par
\begin{figure}[t]
\centering\includegraphics[width=0.95\textwidth]{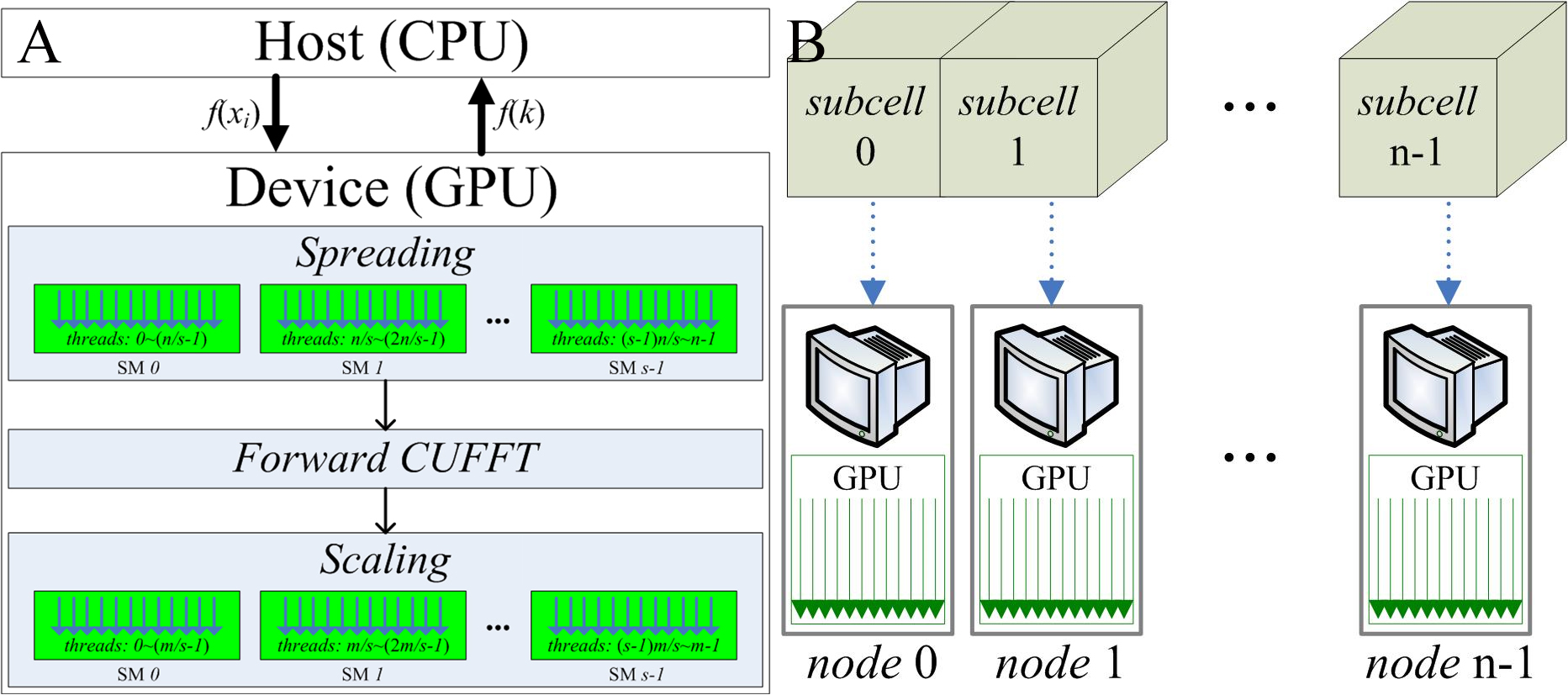}
\caption{Schematic diagrams for (A) the forward CU-NFFT algorithm and (B) the hybrid parallel scheme for the HP-ENUF method.}\label{fig:cunfft_hpenuf}
\end{figure}

The NFFT and the proposed $gridding$ algorithm have been implemented and paralleled using NVIDIA GPU via CUDA-C language and is named as CU-NFFT~\cite{yang2018new}.
The parallel codes of CUDA threads called kernels are designed for data-parallel processing to speed up computations using GPU.
As shown in a representative diagram in Fig.~\ref{fig:cunfft_hpenuf}A, the kernels map data elements to parallel processing threads, which are executed $s$ times in parallel using $s$ CUDA threads with high arithmetic intensity.
In CU-NFFT, all CUDA threads are performed on physically separated devices (GPU) that operate as co-processors of the host (CPU) to run CUDA-C program.
The CU-NFFT algorithm is composed of three main steps: allocating global memory in device (GPU) and loading data $f(x_i)$ from host (CPU) to device (GPU), executing kernel functions using CU-FFT, and retrieving computational results $\hat{f}_k$ from device (GPU) to host (CPU) before releasing global memory.
More specifically, the spreading kernels divide the task into $n$ threads that are evenly partitioned into $s$ steaming multiprocessors (SMs) for  concurrent computations (Fig.~\ref{fig:cunfft_hpenuf}A). 
Once all SMs finalize their spreading processes, the forward CU-NFFT in CUDA library starts to conduct a standard Fourier transformation.
Subsequently, all scaling kernels execute their tasks with $m$ threads in a similar manner.
Correspondingly, the inverse CU-NFFT is performed in a similar manner for subdividing, inverse CU-FFT, and interpolating kernels.
It is noteworthy that an advantage of the proposed CU-NFFT algorithm is that both the forward and the inverse procedures are executed concurrently in GPU without interruption of transferring data between CPU and GPU so as to improve the computational efficiency of CU-NFFT.

\subsection{Implementation of the CU-ENUF method}

\par
Based on CU-NFFT algorithm, we have implemented the ENUF method in GPU, which is termed as CU-ENUF because in this method both real and reciprocal space summations are handled using GPU and CUDA technology~\cite{yang2016accelerating}.
In another word, the CU-ENUF method is an essential ENUF method paralleled with GPU and CUDA threads.
In real space summations, a $NearDistance$ algorithm was developed to effectively reduce neighbor list size so as minimize computational time in searching interacting particle pairs within a desired cutoff distance, whereas in reciprocal space calculations, the CU-NFFT replaces traditional NFFT for an efficient evaluation of electrostatic energies and forces among charged particles.
Both real and reciprocal space summations are accelerated by using GPU and CUDA technology.
An additional procedure was proposed to determine optimal simulation parameters so that the CU-ENUF method can achieve a good efficiency in handling long range electrostatic interactions among charged particles at extended spatiotemporal scales.

\par
In real space summations of electrostatic energies and forces among charged particles (Eq.~\ref{eq:17} and Eq.~\ref{eq:29}), we need to sum up all contributions of interacting particles within a cutoff sphere with a radius of $r_c$.
In the linked-cell list scheme~\cite{eastwood1981computer, hockney1988particle}, the cubic simulation box is decomposed into a regular lattice of small cells, and the side length of these cells is slightly larger than the cutoff $r_c$.
Herein, we adopted a simple neighbor-list scheme to further refine the size of these cells, with which one can further decrease the computational time in searching interacting particles nearby.
A set of trial MD and DPD simulations indicated that in dense granular systems the adoption of an average distance between interacting particles as the side length of the decomposed cells will be a optimal balance of cell partitioning and particle counting~\cite{yang2016accelerating}.
The proposed neighbor-list scheme embedded in linked-cell list scheme was implemented in real space summations of electrostatic interactions using the CU-ENUF method.

\par
For the calculations of reciprocal space summations of electrostatic energies and forces between charged particles using the CU-ENUF method, the computations of structure factor $S(n)$ and $\mathbf{F}^{E,K}$, corresponding to a forward FFT and three inverse FFTs, are calculated using NFFT (for CPU) and accelerated using CU-NFFT (for GPU), respectively.
In the CU-ENUF method, reciprocal space summations of electrostatic interactions are transformed into parallel structure capable of running GPU acceleration based on CUDA technology.
The parallel structure accepts particle charges and positions as input from CPU, and produces total electrostatic energies and forces on charged particles, the latter of which are thereafter combined with those determined from real space summations as output to CPU.

\par
Additional art of programming were included in the CU-ENUF method to further optimize its efficiency in handling electrostatic interactions.
For examples, some shared data, $i.e.$, mainly the values of window functions in CU-NFFT, are computed in local memory instead of being calculated and stored in global memory so as to avoid scarifying efficiency when accessing to global memory of GPU.
Many pre-processing and post-processing computations in CPU are computed using GPU. 
In addition, the calculations of electrostatic forces in 3D space are packed into a single calculation entity in the inverse CU-NFFT procedure, which can reduce a large amount of execution time for initialization and redundant loops for three independent inverse NFFT calls in the ENUF method.
These improvements, together with pre-determined optimal simulation parameters, render the CU-ENUF method a pure CUDA-based program and have a comparable or a better performance than the particle-mesh Ewald summation method and the ENUF method using CPU for computations.

\par
The CU-ENUF method has been implemented as a computational module in the GALAMOST package for the computation of electrostatic interactions among charged particles in model simulation systems~\cite{zhu2013galamost, wang2018electrostatic, zhu2018employing}.
Both real and reciprocal space summations are accelerated using one GPU card without any participation of CPUs.
With the GALAMOST package, the CU-ENUF method can be adopted to perform CG MD and DPD simulations of charged soft matter systems at microcanonical, canonical, and isothermal-isobaric ensembles under periodic boundary conditions.
In each step calculation, the GALAMOST package maximizes the amount of computations using GPU card and minimizes communications between GPU and CPUs, except compulsory I/O performance.
Benchmarks on representative IL systems demonstrated that the performance of the PPPM Ewald summation method appears better than the CU-ENUF method by roughly 50\% for small and intermediate simulation systems with the number of ion pairs less than 0.2 million, whereas these two methods exhibit comparable computational efficiencies in handling electrostatic interactions in large simulation systems with number of ion pairs exceeding 0.5 million~\cite{wang2018electrostatic}.
In addition, the performance of the ENUF method in the GALAMOST package is better than the PME method in the GROMACS package using one GPU and upto 28 CPU processors  for the computations of electrostatic interactions in small simulation systems.
However, at current stage, it is difficult to accurately quantify the computational performances of the GALAMOST and GROMACS packages for charged soft matter systems since these two packages support different features with different computational demands.
Additional parallelization strategy will be explored to further improve the computational efficiency of the ENUF related methods, and the compatibility of the GALAMOST and the GROMACS packages such that we can perform consistent multiscale modelling of charged soft matter systems at micro- and mesoscopic levels to explore their striking phase behaviors at extended spatiotemporal scales using GPU and CUDA technology

\subsection{Architecture of the HP-ENUF method}

\par
Although the CU-ENUF method achieves a qualitative leap compared with other particle-mesh based methods in handling electrostatic interactions between charged particles, its computational efficiency is limited to the throughput capacity of GPU for simulation systems at extreme spatiotemporal scales.
Therefore, we proposed a hybrid parallel scheme combining multiple CPU and GPU devices to upgrade the CU-ENUF method, which is described as HP-ENUF method~\cite{yang2017hybrid, yang2020hybrid}.
Similarly to the CU-ENUF method~\cite{yang2016accelerating}, a GPU-optimized particle-data structure is employed in the HP-ENUF method so that all computations are mainly performed using multiple GPU devices.
The HP-ENUF method enables direct communications between GPU devices within different computer nodes via NVIDIA GPUDirect technology supported by CUDA-aware MPI (Message Passing Interface) library, which eliminates unnecessary data transfer between CPU and GPU devices.

\par
The hybrid parallel scheme in the HP-ENUF method consists of multiple MPI ranks, each of which includes a CPU node and a GPU node responsible for calculations of electrostatic energies and forces between charged particles in specific domains (subcells) in simulation systems.
A schematic diagram  of the hybrid parallel scheme for the HP-ENUF method is illustrated in Fig.~\ref{fig:cunfft_hpenuf}B.
Both real and reciprocal space summations of electrostatic interactions are first paralleled via a domain decomposition scheme, which is implemented using MPI libraries on multiple CPU nodes, and thereafter paralleled via GPU threads in each CPU node.
Using the domain decomposition scheme, each simulation task (large simulation system) is uniformly decomposed into several subtasks (small simulation subcells) that are partitioned to different CPU nodes for CPU-parallel computation.
Each CPU node is responsible for calculations of electrostatic interactions between charged particles in its own subcell and its communication with other CPU nodes is relatively low.
The MPI library is used to implement CPU parallel strategy (process level), and each MPI rank delivers subtask (the data of its own subcell) to the corresponding GPU card, and this subtask is executed in GPU-parallel computation (thread level) using CUDA technology.

\par
For each subcell, the real space summations of electrostatic interactions between charged particles in the HP-ENUF method is the same as that we used in the CU-ENUF method.
However, unlike independent calculations of real space summations, the computations of reciprocal space summations of electrostatic interactions in each subcell depend on interactions of charged particles with those in the other subcells, indicating that CU-NFFT must be performed for charged particles in all subcells of the simulation system.
It should be mentioned that an additive property of NFFT for the calculation of structure factor $S(\textbf{n})$ is that it can be decomposed into several sub-NFFTs with arbitrary particle distributions, which makes the parallel computation of CU-NFFT feasible in the HP-ENUF method.
Assuming that $N$ charged particles are uniformly distributed in $n$ subcells, there are $C = N/n$ charged particles in each subcell.
The total structural factor for the whole simulation system is mathematically composed of $n$ partial structural factors determined from $n$ subcells as
\begin{eqnarray}
S(\textbf{n})
&=&\sum_{i=1}^{N}q_ie^{-\frac{2\pi\imath}{L}\textbf{n}\cdot\textbf{r}_i}\nonumber\\
&=&\sum_{i=1}^{C}q_i e^{-\frac{2\pi\imath}{L}\textbf{n}\cdot\textbf{r}_i}+ \sum_{i=C+1}^{2C}q_i e^{-\frac{2\pi\imath}{L}\textbf{n}\cdot\textbf{r}_i}+ \cdots + \sum_{(n-1)C+1}^{nC}q_i e^{-\frac{2\pi\imath}{L}\textbf{n}\cdot\textbf{r}_i}\\
&=&S_1(\textbf{n})+S_2(\textbf{n})+\cdots+S_n(\textbf{n})\,.
\end{eqnarray}
Once $S_i(\textbf{n})$ is calculated from node $i$, all partial structural factors will be collected and summarized in a specific computer node, and thereafter the total $S(\textbf{n})$ will be broadcasted to all computer nodes to calculate electrostatic interactions between charged particles in each subcell.

\par
All in all, as a significant extension of the CU-ENUF method, the HP-ENUF scheme successfully removes the throughput capacity of a single GPU, and is capable of conducting efficient simulations of charged soft matter systems at extended spatiotemporal scales.
In addition, the HP-ENUF method is constructed with concise software architectures using C and CUDA C language, which makes it pretty transferable to other popular CUDA-accelerated packages, such as HOOMD and LAMMPS.

\clearpage
\section{Applications of the ENUF Related Methods in Modelling Charged Soft Matter Systems}

\par
The ENUF method and its derivatives can capture essential characteristics of electrostatic interactions between charged particles and charge density distributions at extended spatiotemporal scales, and have been adopted to explore representative properties of charged soft matter systems, such as the effect of charge fractions of polyelectrolytes, ion concentration and counterion valency of added salts on conformational properties of polyelectrolytes~\cite{wang2013implementation}, the binding structures of dendrimers on bilayer membranes and the corresponding permeation mechanisms~\cite{wang2012specific}, and the heterogeneous structures and dynamics in ILs matrices and how electrostatic interactions between charged particles affect these properties at extended spatiotemporal scales~\cite{wang2018electrostatic}.

\subsection{Polyelectrolyte conformational properties}

\par
\begin{figure}[t]
\centering\includegraphics[width=0.95\textwidth]{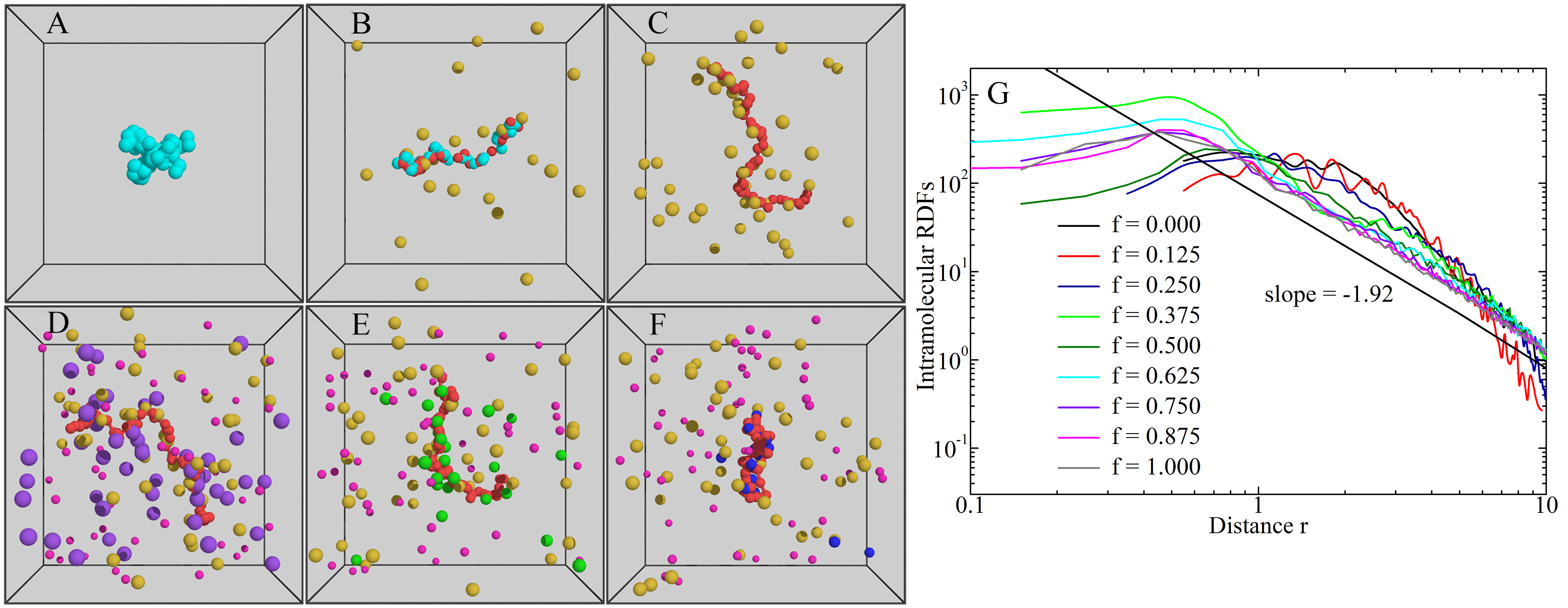}
\caption{Typical conformations of polyelectrolytes with different charge fraction $f$ (upper panels), and with varied charge valency of added salts (lower panels) at a condition where the total charge of salt counterions is equal to that of polyelectrolyte. (A) $f=0.0$, (B) $f=0.5$, and (C) $f=1.0$. The red and cyan spheres indicate charged and neutral particles on polyelectrolytes, respectively. The counterions of polyelectrolytes are represented by yellow spheres.
(D) (1:1) salt, (E) (2:1) salt, and (F) (3:1) salt. Monovalent, divalent and trivalent ions of added salts are represented by purple, green and blue spheres, respectively. All counterions of added salts are presented by magenta spheres. (G) Intramolecular pair correlation functions between charge particles on polyelectrolytes with varied charge fraction $f$.}\label{fig:pe_conf_rdf}
\end{figure}

Electrostatic interactions between charged particles on polyelectrolytes lead to rich conformational properties of polyelectrolytes~\cite{wang2013implementation}, which are qualitatively different from those of neutral polymers~\cite{dobrynin2005theory, jusufi2009colloquium}.
CG MD simulations demonstrated that the size of polyelectrolyte increases with increasing the degree of ionization of polyelectrolyte, exhibiting a structural transition of polyelectrolyte from collapse (Fig.~\ref{fig:pe_conf_rdf}A) to fully extended conformation (Fig.~\ref{fig:pe_conf_rdf}C).
These computational results are qualitatively consistent with experimental observations~\cite{roiter2005afm} and theoretical predictions~\cite{liao2006counterion} for weakly charged polyelectrolytes.

\par
The intramolecular pair correlation functions between charged particles on polyelectrolytes reveal distinct tendencies with an increase of charge fraction of polyelectrolytes.
The intramolecular correlations in initial zone ($r<1$) are dominated by soft conservative repulsions.
In the regime of $r>1$, two striking tendencies are observed in simulations and shown in Fig.~\ref{fig:pe_conf_rdf}G.
For polyelectrolytes with a small charge fraction, a small scaling-like domain is observed and then followed by a terminal correlation range.
In contrast, polyelectrolytes with a large charge fraction exhibit a scaling behavior over the entire range.
These simulation results are consistent with the theoretical description of weakly charged polyelectrolytes deduced from scaling theory~\cite{rubinstein2003polymer}.

\par
In the absence of inorganic salts, polyelectrolytes adopt extended conformations, owing to strong electrostatic repulsions between charged particles on polyelectrolytes.
However, these electrostatic interactions are partially screened upon addition of salts into solution~\cite{sanders2005structure}.
Both ion concentration and valency of salt ions can significantly affect conformational properties of polyelectrolytes due to strong electrostatic correlations between multivalent ions and charged particles on polyelectrolytes.
The condensation ability of trivalent ions on polyelectrolytes (Fig.~\ref{fig:pe_conf_rdf}F) is much stronger than that of monovalent ions (Fig.~\ref{fig:pe_conf_rdf}D), leading to a decrease of osmotic pressure and a conformational collapse of polyelectrolytes in solution~\cite{mei2006collapse, roiter2010single}.
It is noted that a gradual increase in multivalent counterion concentration after a threshold leads to a structural transition of polyelectrolytes from fully collapsed to semi-swelled conformations, which is akin to the redissolution behavior of multichain aggregates and is attributed to a competitive feature of counterions with varied charge valencies in condensating polyelectrolytes~\cite{sanders2005structure, mei2006collapse, roiter2010single}.

\par
The effect of ion concentration and charge valency of counterions on polyelectrolyte conformations can be 
specified by the Debye screening length in polyelectrolyte solution.
The addition of inorganic salts with multivalent counterions leads to a short Debye screening length~\cite{yan2009dissipative}, demonstrating that electrostatic interactions between charge particles beyond a certain distance separated are screened and hence are no longer long-ranged interactions.
Therefore, it is very likely that a finite cutoff for electrostatic interactions, or a screened interaction potential (like the Yukawa potential) between charge particles, can be used in describing electrostatic interactions in ion-concentrated simulation systems.

\subsection{Dendrimer-lipid membrane complexes}

\par
The ENUF-DPD method were adopted to investigate specific binding structures of dendrimers on amphiphilic bilayer membranes~\cite{wang2012specific}.
Polyamidoamine (PAMAM) dendrimers have hollow core and dense shell structures, and are promising nano-vehicles to protect small drug molecules during delivery process~\cite{percec2002self, ma2013theoretical}.
Moreover, PAMAM dendrimers undergo conformational transitions from dense shell to dense core under external stimuli, facilitating the release of drug molecules to specific targets~\cite{liu2009pamam}.
When PAMAM dendrimers are used as nano-devices for nonviral gene delivery and antitumor therapeutics, the central issue is how they interact with cell membranes and how to control structures of dendrimer-membrane complexes during drug delivery~\cite{ma2013theoretical}.

\par
\begin{figure}[t]
\centering\includegraphics[width=0.95\textwidth]{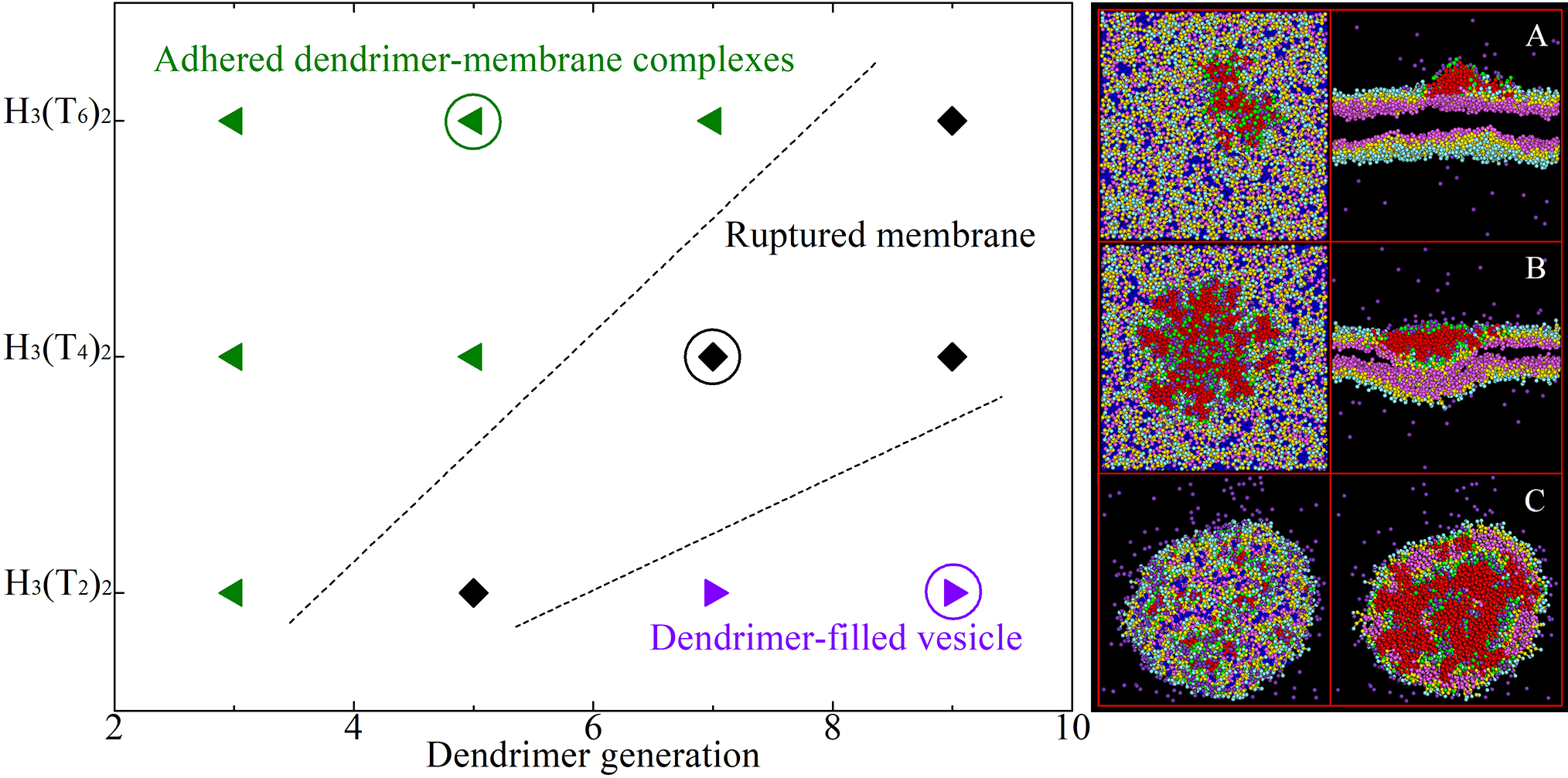}
\caption{Phase diagram of binding structures of charged dendrimers on bilayer membranes. In CG lipid model, H represents charged head group particles and T represents neutral tail group particles, respectively. The H$_3$(T$_4$)$_2$ amphiphile corresponds to the lipid molecule consisting of one charged head with three beads and two neutral tails, each of which consists of four beads. Snapshots (left images for top view and right images for cross-sectional view) of representative conformations of dendrimers on membranes. (A) G5 on H$_3$(T$_6$)$_2$, (B) G7 on H$_3$(T$_4$)$_2$, and (C) G9 on H$_3$(T$_2$)$_2$.}\label{fig:den_lipid_diagram_conf}
\end{figure}

Mutually consistent CG models of PAMAM dendrimers and dimyristoylphosphatidylcholine (DMPC) lipid molecules were constructed based on volume criteria and chemical identities.
These CG models could qualitatively describe conformational properties of charged dendrimers and surface tension of atomistic DMPC lipid membranes, respectively~\cite{maiti2004structure, lee2011effects}.
DPD simulation results revealed that the permeability of dendrimers across tensionless DMPC bilayer membranes is enhanced upon increasing dendrimer size.
The $3^{rd}$ generation (G3) PAMAM dendrimer rests on DMPC bilayer membranes without any impact on membranes, whereas G5 dendrimer spreads on bilayer membranes and attracts some lipid head groups into the vicinity of dendrimer cations.
For larger generation dendrimers, significant bending of bilayer membranes is observed due to a strong coordination of dendrimers with membranes (Fig.~\ref{fig:den_lipid_diagram_conf}B).
The dendrimer-membrane contact region is characterized with high surface tension, facilitating the permeation of dendrimers across membranes.

\par
In addition, the conformational properties of dendrimers and binding structures of dendrimers on membranes are influenced by the length of hydrophobic moieties of lipid amphiphiles.
For dendrimers on membranes consisting of amphiphiles with shorter hydrophobic tails than DMPC molecule, G3 dendrimer can easily penetrate into membranes due to a small energy barrier formed by the corresponding amphiphiles.
Significant binding structures of G5 dendrimer onto membranes are observed, which destroys the integrity of membranes and induce a big hole on membranes.
By increasing dendrimer sizes, the dendrimer-filled vesicles (Fig.~\ref{fig:den_lipid_diagram_conf}C) are formed, which are suggested to be a possible mechanism for charged dendrimers removing amphiphiles from membranes in atomic force microscope experiments~\cite{mecke2005lipid}.
While for dendrimers on membranes consisting of amphiphiles with longer hydrophobic moieties than DMPC molecule, small dendrimers exhibit planar conformations on membranes (Fig.~\ref{fig:den_lipid_diagram_conf}A), and large dendrimers induce small cavities on membranes but can neither rupture membranes nor initiate isolated pores on membranes due to a large energy barrier formed by hydrophobic moieties of corresponding amphiphiles.
An illustrative phase diagram and representative binding structures of charged dendrimers on bilayer membranes are shown in Fig.~\ref{fig:den_lipid_diagram_conf}.

\subsection{Heterogeneous structures and dynamics in ionic liquids}

\par
Additional CG MD simulations were performed using the CU-ENUF method to address heterogeneous structures and dynamics in ILs, and how electrostatic interactions among charged particles affect these properties in IL matrices~\cite{wang2018electrostatic}.
Room temperature ILs are fascinating molten salts solely composed of ion species with distinct molecular symmetry and charge delocalization, having their melting points below $100^{\circ}\mathrm{C}$~\cite{armand2009ionic, castner2011ionic, wang2020chemrev}.
Recent years have seen a great enthusiasm for ILs regarding their utilities as facilitating functional materials in diverse applications including material synthesis and catalysis, micro-lubrication and nanotribology, gas adsorption and separation, and electrochemical devices for energy storage and harvesting~\cite{armand2009ionic, castner2011ionic, hayes2015structure}.
Compared with traditional molten salts, like sodium chloride, one fascinating feature of ILs is that they exhibit distinct heterogeneous microstructures and dynamics spanning multiple length and time scales in bulk region and in confined environments~\cite{armand2009ionic, wang2014heterogeneous, hayes2015structure, wang2017interfacial, wang2018competitive, wang2019multigranular, wang2020chemrev}.
Both experimental and computational characterizations revealed that mesoscopic liquid organization of ILs is characterized by either sponge-like interpenetrating polar and apolar networks or segregated polar (apolar) domains within apolar (polar) framework depending on the relative ratios of polar groups over apolar moieties in ion species~\cite{wang2005unique, hu2006heterogeneity, jin2010heterogeneous, ji2013effect, kim2016heterogeneous, wang2017interfacial,wang2018competitive, wang2020chemrev}.
Recent atomistic simulations of ILs demonstrated that both simulation size and simulation time do matter to get reliable collective structural and dynamical quantities of IL ions in IL matrices~\cite{gabl2012computational}.

\par
As simulations of ILs should be performed over long time scales due to sluggish dynamics of ion species in heterogeneous IL matrices, it imposes severe fundamental challenges for atomistic simulations to accurately predict dynamics and transport properties of ILs~\cite{hu2006heterogeneity, jin2010heterogeneous, gabl2012computational, wang2019multigranular}.
In this regard, we preformed extensive CG MD simulations with a modest computational cost to explore the effects of neutral chain length in cations, molecular sizes of anions, and temperatures on microstructural and dynamical quantities of ion groups in 1-alkyl-3-methylimidazolium tetrafluoroborate ([C$_n$MIM][BF$_4$]) ILs at extended spatiotemporal scales~\cite{wang2013multiscale, wang2018electrostatic}.
It was found that lengthening cation alkyl chains leads to an aggregation of neutral beads, promotes the formation of spatially heterogeneous apolar domains dispersed in ionic channels, and thereafter results in a remarkable transition of mesoscopic liquid morphologies in model IL systems from dispersed neutral (apolar) beads in a 3D framework of ion channels to that characterized by bi-continuous interpenetrating polar and apolar networks in liquid matrices.
Such a microstructural evolution in model IL matrices can be rationalized by a competition of short-range collective interactions between neutral beads and long-range Coulombic interactions among charged particles in constituent ions.

\par
For dynamical quantities, translational diffusion of ion groups presents a gradual decrease upon lengthening cation alkyl chains and enlarging molecular sizes of anions.
The temperature dependence of diffusion coefficients of all representative groups in ion models is described by a classical Arrhenius feature.
The rotational dynamics of cations in varied IL matrices are characterized by a bi-exponential structural relaxation behavior.
The correlation times for these two rotational modes are significantly temperature dependent, and are strongly related to cation structures, indicating a rotational heterogeneity of charged and neutral moieties of cations in heterogeneous ionic environments.

\par
\begin{figure}[t]
\centering\includegraphics[width=0.95\textwidth]{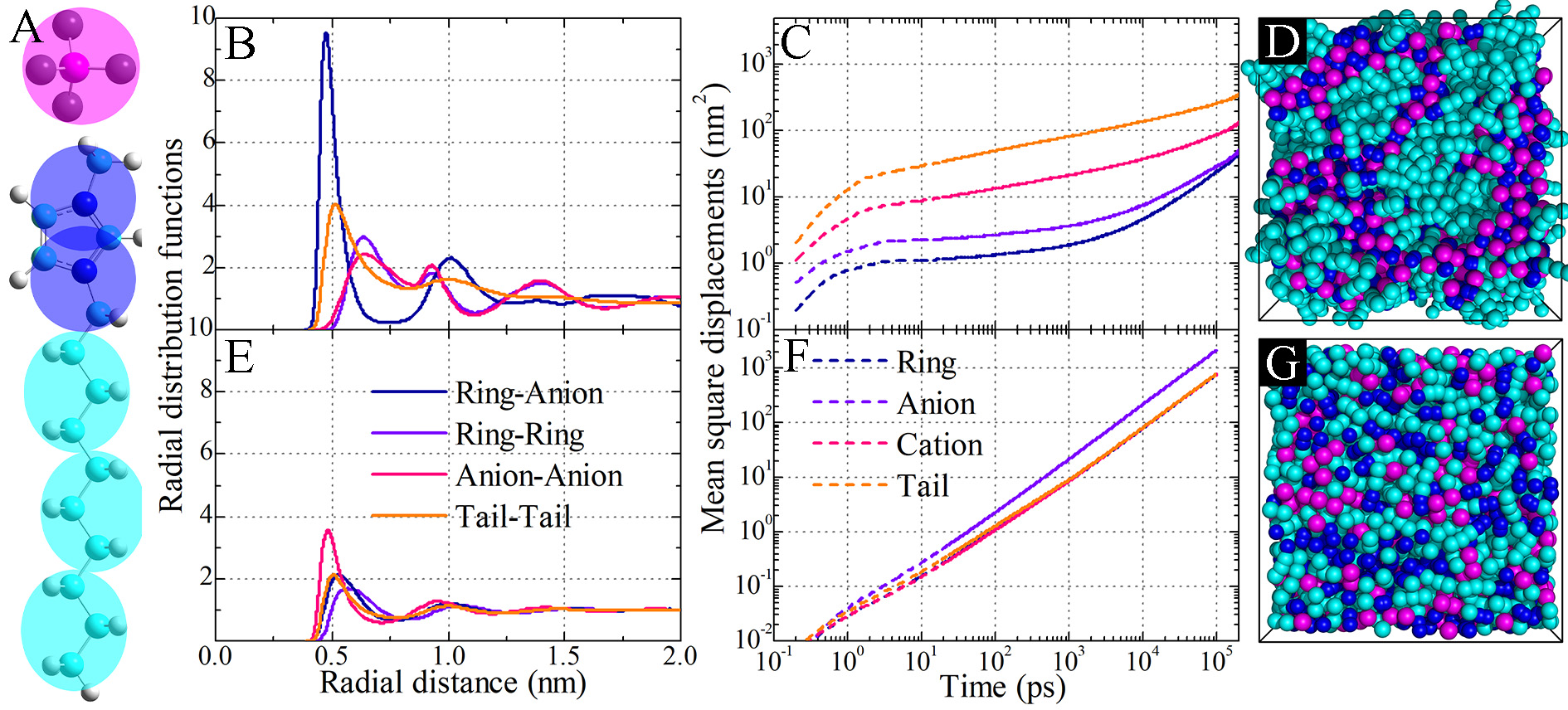}
\caption{(A) A schematic representation of CG model for [C$_{10}$MIM][BF$_4$] IL. The blue and cyan beads are labelled as “ring” and “alkane”, respectively, representing charged (imidazolium ring and two closest methyl groups) and neutral (alkyl units) moieties in cations. Comparison of (B, E) radial distribution functions and (C, F) mean square displacements of representative groups in [C$_{10}$MIM][BF$_4$] IL (upper panels) and its neutral counterpart (lower panels) at 400 K, and (D, G) representative snapshots of these two model systems.}\label{fig:il_rdfmsd_conf}
\end{figure}

Additional CG MD simulations were carried out on neutral analogues of model [C$_{10}$MIM][BF$_4$] IL (Fig.~\ref{fig:il_rdfmsd_conf}A) in which all electrostatic interactions between charged particles were switched off.
The removal of electrostatic interactions between ion models tends to loosen liquid structures leading to an almost homogeneous and interleaved distribution of neutral~\lq\lq ion\rq\rq~models resembling a distorted ion lattice in simulation systems~\cite{roy2010dynamics, kim2016heterogeneous, wang2018electrostatic}.
This visual microstructural change is manifested in detailed characterizations, such as liquid densities, radial distribution functions, and translational dynamics of representative beads in neutral and ion models, as clearly shown in Fig.~\ref{fig:il_rdfmsd_conf}.
These observations highlight a critical role of electrostatic interactions in describing collective structural and dynamical quantities of model IL systems.

\clearpage
\section{Concluding Remarks and Outlook}

\par
Computer simulations provide a unique insight into molecular interactions and structural and dynamical quantities responsible for many peculiar properties of materials and biological systems at multiple spatiotemporal scales.
The quality of computer simulations is essentially determined by an accurate description of intra- and intermolecular interactions in these molecular systems, among which the electrostatic interactions between charged particles deserve special attention because of these long-ranged interactions play a prominent role in determining structures and states of simulation systems.
The efficient calculations of electrostatic interactions in model systems subjected to partial or full periodic boundary conditions have been a daunting task.
A wide variety of theoretical approaches, ranging from quantum mechanical $ab~initio$ methods, classic Maxwell theory of electromagnetism, generalized Born algorithms, to phenomenological modifications of Coulomb’s law, and computational methods, including particle-mesh based methods, have been developed for electrostatic analysis at different resolution levels~\cite{allen1989computer, frenkel1996understanding}.

\par
The standard Ewald summation method does a remarkable job in splitting the very slowly converged Coulomb potential into two parts converging fast exponentially, which make the calculations of electrostatic interactions in computer simulations feasible, and therefore this method has been widely used to handle long range electrostatic interactions in modelling charged simulation systems.
However, the standard Ewald summation method suffers from its computational demanding feature due to the long range nature of electrostatic interactions.
Although plenty of methods have been proposed with different computational schemes to tackle long range interactions in simulation systems with partial or full periodic boundary conditions, only a certain subset of them has entered into widely used molecular dynamics codes for scientific computing with varied computational efficiencies.
This fact is related to the long standing and continuously improved implementations of selected methods and also the large effort needed to propose and implement new approaches.

\par
As an attractive alternative approach to particle-mesh based schemes which show a linear scaling feature, we sketched an ENUF method, which has been implemented in particle-based simulation packages to speedup calculations of electrostatic energies and forces between charged particles at micro- and mesoscopic levels without resorting to a nonphysical truncation of the Coulomb potential.
The ENUF method and its derivatives conserve both energy and momentum to floating point accuracy, capture essential characteristics of electrostatic interactions between charged particles, and exhibit an outstanding computational complexity of $\mathcal{O}(N\log N)$ with optimal physical parameters.
The ENUF method are further adopted to investigate the dependence of conformational properties of polyelectrolytes on charge fraction of polyelectrolyte and counterion valency and concentration of added salts, and the effect of dendrimer sizes and variation of hydrophobic tails of lipid molecules on the specific binding structures of dendrimers on amphiphilic membranes.

\par
Taking advantages of GPU and CUDA technology, we have upgraded the GPU version of NFFT and the ENUF method, which are termed as CU-NFFT and CU-ENUF, respectively, and are specialized to improve computational efficiencies in computations of electrostatic energies and forces among charged particles using the GALAMOST package.
In addition, several (hybrid) parallelization strategies based on $gridding$ and $NearDistance$ algorithms were developed to effectively partition hardware memories and balance computational loads between CPU and (multi) GPU nodes.
In addition, the CU-ENUF and HP-ENUF methods are developed using C and CUDA C language and are constructed with concise software architecture, which render them having significant transferabilities to other popular CUDA-enabled packages, such as GROMACS, LAMMPS, and HOOMD.
It is expected that the ENUF related methods will be used in varied computational communities and updated by researchers to expand their visibilities and applications in handling electrostatic interactions between charged particles at extended spatiotemporal scales.

\section*{Acknowledgment}
We thank Prof. Zhong-Yuan Lu (Jilin University, China) for insightful discussion.
S.-C. Yang acknowledges financial supported from Science and Technology Research Project Fund from Education Department of Jilin Province, China (JJKH20190696KJ), and from State Key Laboratory of Supramolecular Structure and Materials of Jilin University (SKLSSM202031).
Y.-L. Zhu acknowledges financial supported from the National Natural Science Foundation of China (21774129).
A. Laaksonen acknowledges Swedish Science Council for financial support (2019-03865), and partial support from a grant from Ministry of Research and Innovation of Romania (CNCS - UEFISCDI, project number PN-III-P4-ID-PCCF-2016-0050, within PNCDI III). 
Y.-L. Wang gratefully acknowledges financial support from Knut and Alice Wallenberg Foundation (KAW 2018.0380).

\bibliography{enuf_review}

\begin{thebibliography}{111}
\expandafter\ifx\csname natexlab\endcsname\relax\def\natexlab#1{#1}\fi
\expandafter\ifx\csname bibnamefont\endcsname\relax
  \def\bibnamefont#1{#1}\fi
\expandafter\ifx\csname bibfnamefont\endcsname\relax
  \def\bibfnamefont#1{#1}\fi
\expandafter\ifx\csname citenamefont\endcsname\relax
  \def\citenamefont#1{#1}\fi
\expandafter\ifx\csname url\endcsname\relax
  \def\url#1{\texttt{#1}}\fi
\expandafter\ifx\csname urlprefix\endcsname\relax\def\urlprefix{URL }\fi
\providecommand{\bibinfo}[2]{#2}
\providecommand{\eprint}[2][]{\url{#2}}

\bibitem[{\citenamefont{Allen and Tildesley}(1989)}]{allen1989computer}
\bibinfo{author}{\bibfnamefont{M.~P.} \bibnamefont{Allen}} \bibnamefont{and}
  \bibinfo{author}{\bibfnamefont{D.~J.} \bibnamefont{Tildesley}},
  \emph{\bibinfo{title}{Computer Simulation of Liquids}}
  (\bibinfo{publisher}{Oxford University Press, New York},
  \bibinfo{year}{1989}).

\bibitem[{\citenamefont{Frenkel and Smit}(1996)}]{frenkel1996understanding}
\bibinfo{author}{\bibfnamefont{D.}~\bibnamefont{Frenkel}} \bibnamefont{and}
  \bibinfo{author}{\bibfnamefont{B.}~\bibnamefont{Smit}},
  \emph{\bibinfo{title}{Understanding molecular simulation: From algorithms to
  applications}} (\bibinfo{publisher}{Academic Press New York},
  \bibinfo{year}{1996}).

\bibitem[{\citenamefont{Sharp and Honig}(1990)}]{sharp1990electrostatic}
\bibinfo{author}{\bibfnamefont{K.}~\bibnamefont{Sharp}} \bibnamefont{and}
  \bibinfo{author}{\bibfnamefont{B.}~\bibnamefont{Honig}},
  \bibinfo{journal}{Annu. Rev. Biophys. Chem.} \textbf{\bibinfo{volume}{19}},
  \bibinfo{pages}{301} (\bibinfo{year}{1990}).

\bibitem[{\citenamefont{Holm et~al.}(2001)\citenamefont{Holm, K{\'e}kicheff,
  and Podgornik}}]{holm2001electrostatic}
\bibinfo{author}{\bibfnamefont{C.}~\bibnamefont{Holm}},
  \bibinfo{author}{\bibfnamefont{P.}~\bibnamefont{K{\'e}kicheff}},
  \bibnamefont{and}
  \bibinfo{author}{\bibfnamefont{R.}~\bibnamefont{Podgornik}},
  \emph{\bibinfo{title}{Electrostatic effects in soft matter and biophysics}}
  (\bibinfo{publisher}{Springer}, \bibinfo{year}{2001}).

\bibitem[{\citenamefont{Naji et~al.}(2005)\citenamefont{Naji, Jungblut,
  Moreira, and Netz}}]{naji2005electrostatic}
\bibinfo{author}{\bibfnamefont{A.}~\bibnamefont{Naji}},
  \bibinfo{author}{\bibfnamefont{S.}~\bibnamefont{Jungblut}},
  \bibinfo{author}{\bibfnamefont{A.~G.} \bibnamefont{Moreira}},
  \bibnamefont{and} \bibinfo{author}{\bibfnamefont{R.~R.} \bibnamefont{Netz}},
  \bibinfo{journal}{Physica A} \textbf{\bibinfo{volume}{352}},
  \bibinfo{pages}{131} (\bibinfo{year}{2005}).

\bibitem[{\citenamefont{Sainis et~al.}(2008)\citenamefont{Sainis, Merrill, and
  Dufresne}}]{sainis2008electrostatic}
\bibinfo{author}{\bibfnamefont{S.~K.} \bibnamefont{Sainis}},
  \bibinfo{author}{\bibfnamefont{J.~W.} \bibnamefont{Merrill}},
  \bibnamefont{and} \bibinfo{author}{\bibfnamefont{E.~R.}
  \bibnamefont{Dufresne}}, \bibinfo{journal}{Langmuir}
  \textbf{\bibinfo{volume}{24}}, \bibinfo{pages}{13334} (\bibinfo{year}{2008}).

\bibitem[{\citenamefont{Kobrak and Li}(2010)}]{kobrak2010electrostatic}
\bibinfo{author}{\bibfnamefont{M.~N.} \bibnamefont{Kobrak}} \bibnamefont{and}
  \bibinfo{author}{\bibfnamefont{H.}~\bibnamefont{Li}}, \bibinfo{journal}{Phys.
  Chem. Chem. Phys.} \textbf{\bibinfo{volume}{12}}, \bibinfo{pages}{1922}
  (\bibinfo{year}{2010}).

\bibitem[{\citenamefont{Wang et~al.}(2018)\citenamefont{Wang, Zhu, Lu, and
  Laaksonen}}]{wang2018electrostatic}
\bibinfo{author}{\bibfnamefont{Y.-L.} \bibnamefont{Wang}},
  \bibinfo{author}{\bibfnamefont{Y.-L.} \bibnamefont{Zhu}},
  \bibinfo{author}{\bibfnamefont{Z.-Y.} \bibnamefont{Lu}}, \bibnamefont{and}
  \bibinfo{author}{\bibfnamefont{A.}~\bibnamefont{Laaksonen}},
  \bibinfo{journal}{Soft Matter} \textbf{\bibinfo{volume}{14}},
  \bibinfo{pages}{4252} (\bibinfo{year}{2018}).

\bibitem[{\citenamefont{Ewald}(1921)}]{ewald1921berechnung}
\bibinfo{author}{\bibfnamefont{P.}~\bibnamefont{Ewald}},
  \bibinfo{journal}{Annalen der Physik} \textbf{\bibinfo{volume}{369}},
  \bibinfo{pages}{253} (\bibinfo{year}{1921}).

\bibitem[{\citenamefont{Eastwood and Hockney}(1981)}]{eastwood1981computer}
\bibinfo{author}{\bibfnamefont{J.~W.} \bibnamefont{Eastwood}} \bibnamefont{and}
  \bibinfo{author}{\bibfnamefont{R.~W.} \bibnamefont{Hockney}},
  \bibinfo{journal}{New York: Mc GrawHill}  (\bibinfo{year}{1981}).

\bibitem[{\citenamefont{Hockney and Eastwood}(1988)}]{hockney1988particle}
\bibinfo{author}{\bibfnamefont{R.~W.} \bibnamefont{Hockney}} \bibnamefont{and}
  \bibinfo{author}{\bibfnamefont{J.~W.} \bibnamefont{Eastwood}},
  \bibinfo{journal}{Computer simulation using particles}
  (\bibinfo{year}{1988}).

\bibitem[{\citenamefont{Deserno and Holm}(1998)}]{deserno1998mesh2}
\bibinfo{author}{\bibfnamefont{M.}~\bibnamefont{Deserno}} \bibnamefont{and}
  \bibinfo{author}{\bibfnamefont{C.}~\bibnamefont{Holm}}, \bibinfo{journal}{J.
  Chem. Phys.} \textbf{\bibinfo{volume}{109}}, \bibinfo{pages}{7694}
  (\bibinfo{year}{1998}).

\bibitem[{\citenamefont{Brown et~al.}(2012)\citenamefont{Brown, Kohlmeyer,
  Plimpton, and Tharrington}}]{brown2012implementing}
\bibinfo{author}{\bibfnamefont{W.~M.} \bibnamefont{Brown}},
  \bibinfo{author}{\bibfnamefont{A.}~\bibnamefont{Kohlmeyer}},
  \bibinfo{author}{\bibfnamefont{S.~J.} \bibnamefont{Plimpton}},
  \bibnamefont{and} \bibinfo{author}{\bibfnamefont{A.~N.}
  \bibnamefont{Tharrington}}, \bibinfo{journal}{Comput. Phys. Commun.}
  \textbf{\bibinfo{volume}{183}}, \bibinfo{pages}{449} (\bibinfo{year}{2012}).

\bibitem[{\citenamefont{Darden et~al.}(1993)\citenamefont{Darden, York, and
  Pedersen}}]{darden1993particle}
\bibinfo{author}{\bibfnamefont{T.}~\bibnamefont{Darden}},
  \bibinfo{author}{\bibfnamefont{D.}~\bibnamefont{York}}, \bibnamefont{and}
  \bibinfo{author}{\bibfnamefont{L.}~\bibnamefont{Pedersen}},
  \bibinfo{journal}{J. Chem. Phys.} \textbf{\bibinfo{volume}{98}},
  \bibinfo{pages}{10089} (\bibinfo{year}{1993}).

\bibitem[{\citenamefont{Batcho et~al.}(2001)\citenamefont{Batcho, Case, and
  Schlick}}]{batcho2001optimized}
\bibinfo{author}{\bibfnamefont{P.~F.} \bibnamefont{Batcho}},
  \bibinfo{author}{\bibfnamefont{D.~A.} \bibnamefont{Case}}, \bibnamefont{and}
  \bibinfo{author}{\bibfnamefont{T.}~\bibnamefont{Schlick}},
  \bibinfo{journal}{J. Chem. Phys.} \textbf{\bibinfo{volume}{115}},
  \bibinfo{pages}{4003} (\bibinfo{year}{2001}).

\bibitem[{\citenamefont{Essmann et~al.}(1995)\citenamefont{Essmann, Perera,
  Berkowitz, Darden, Lee, and Pedersen}}]{essmann1995smooth}
\bibinfo{author}{\bibfnamefont{U.}~\bibnamefont{Essmann}},
  \bibinfo{author}{\bibfnamefont{L.}~\bibnamefont{Perera}},
  \bibinfo{author}{\bibfnamefont{M.~L.} \bibnamefont{Berkowitz}},
  \bibinfo{author}{\bibfnamefont{T.}~\bibnamefont{Darden}},
  \bibinfo{author}{\bibfnamefont{H.}~\bibnamefont{Lee}}, \bibnamefont{and}
  \bibinfo{author}{\bibfnamefont{L.~G.} \bibnamefont{Pedersen}},
  \bibinfo{journal}{J. Chem. Phys.} \textbf{\bibinfo{volume}{103}},
  \bibinfo{pages}{8577} (\bibinfo{year}{1995}).

\bibitem[{\citenamefont{Duan and Krasny}(2000)}]{duan2000ewald}
\bibinfo{author}{\bibfnamefont{Z.-H.} \bibnamefont{Duan}} \bibnamefont{and}
  \bibinfo{author}{\bibfnamefont{R.}~\bibnamefont{Krasny}},
  \bibinfo{journal}{J. Chem. Phys.} \textbf{\bibinfo{volume}{113}},
  \bibinfo{pages}{3492} (\bibinfo{year}{2000}).

\bibitem[{\citenamefont{Shan et~al.}(2005)\citenamefont{Shan, Klepeis,
  Eastwood, Dror, and Shaw}}]{shan2005gaussian}
\bibinfo{author}{\bibfnamefont{Y.}~\bibnamefont{Shan}},
  \bibinfo{author}{\bibfnamefont{J.~L.} \bibnamefont{Klepeis}},
  \bibinfo{author}{\bibfnamefont{M.~P.} \bibnamefont{Eastwood}},
  \bibinfo{author}{\bibfnamefont{R.~O.} \bibnamefont{Dror}}, \bibnamefont{and}
  \bibinfo{author}{\bibfnamefont{D.~E.} \bibnamefont{Shaw}},
  \bibinfo{journal}{J. Chem. Phys.} \textbf{\bibinfo{volume}{122}},
  \bibinfo{pages}{054101} (\bibinfo{year}{2005}).

\bibitem[{\citenamefont{Harvey and
  De~Fabritiis}(2009)}]{harvey2009implementation}
\bibinfo{author}{\bibfnamefont{M.}~\bibnamefont{Harvey}} \bibnamefont{and}
  \bibinfo{author}{\bibfnamefont{G.}~\bibnamefont{De~Fabritiis}},
  \bibinfo{journal}{J. Chem. Theory Comput.} \textbf{\bibinfo{volume}{5}},
  \bibinfo{pages}{2371} (\bibinfo{year}{2009}).

\bibitem[{\citenamefont{Wang et~al.}(2010)\citenamefont{Wang, Dommert, and
  Holm}}]{wang2010optimizing}
\bibinfo{author}{\bibfnamefont{H.}~\bibnamefont{Wang}},
  \bibinfo{author}{\bibfnamefont{F.}~\bibnamefont{Dommert}}, \bibnamefont{and}
  \bibinfo{author}{\bibfnamefont{C.}~\bibnamefont{Holm}}, \bibinfo{journal}{J.
  Chem. Phys.} \textbf{\bibinfo{volume}{133}}, \bibinfo{pages}{034117}
  (\bibinfo{year}{2010}).

\bibitem[{\citenamefont{Hedman and Laaksonen}(2006)}]{hedman2006ewald}
\bibinfo{author}{\bibfnamefont{F.}~\bibnamefont{Hedman}} \bibnamefont{and}
  \bibinfo{author}{\bibfnamefont{A.}~\bibnamefont{Laaksonen}},
  \bibinfo{journal}{Chem. Phys. Lett.} \textbf{\bibinfo{volume}{425}},
  \bibinfo{pages}{142} (\bibinfo{year}{2006}).

\bibitem[{\citenamefont{Dutt and Rokhlin}(1993)}]{dutt1993fast}
\bibinfo{author}{\bibfnamefont{A.}~\bibnamefont{Dutt}} \bibnamefont{and}
  \bibinfo{author}{\bibfnamefont{V.}~\bibnamefont{Rokhlin}},
  \bibinfo{journal}{SIAM J. Sci. Stat. Comput.} \textbf{\bibinfo{volume}{14}},
  \bibinfo{pages}{1368} (\bibinfo{year}{1993}).

\bibitem[{\citenamefont{Dutt and Rokhlin}(1995)}]{dutt1995fast}
\bibinfo{author}{\bibfnamefont{A.}~\bibnamefont{Dutt}} \bibnamefont{and}
  \bibinfo{author}{\bibfnamefont{V.}~\bibnamefont{Rokhlin}},
  \bibinfo{journal}{Appl. Comput. Harmon. Analysis}
  \textbf{\bibinfo{volume}{2}}, \bibinfo{pages}{85} (\bibinfo{year}{1995}).

\bibitem[{\citenamefont{Weeber et~al.}(2019)\citenamefont{Weeber, Nestler,
  Weik, Pippig, Potts, and Holm}}]{weeber2019accelerating}
\bibinfo{author}{\bibfnamefont{R.}~\bibnamefont{Weeber}},
  \bibinfo{author}{\bibfnamefont{F.}~\bibnamefont{Nestler}},
  \bibinfo{author}{\bibfnamefont{F.}~\bibnamefont{Weik}},
  \bibinfo{author}{\bibfnamefont{M.}~\bibnamefont{Pippig}},
  \bibinfo{author}{\bibfnamefont{D.}~\bibnamefont{Potts}}, \bibnamefont{and}
  \bibinfo{author}{\bibfnamefont{C.}~\bibnamefont{Holm}}, \bibinfo{journal}{J.
  Comput. Phys.} \textbf{\bibinfo{volume}{391}}, \bibinfo{pages}{243}
  (\bibinfo{year}{2019}).

\bibitem[{\citenamefont{Wang et~al.}(2014{\natexlab{a}})\citenamefont{Wang,
  Hedman, Porcu, Mocci, and Laaksonen}}]{wang2014non}
\bibinfo{author}{\bibfnamefont{Y.-L.} \bibnamefont{Wang}},
  \bibinfo{author}{\bibfnamefont{F.}~\bibnamefont{Hedman}},
  \bibinfo{author}{\bibfnamefont{M.}~\bibnamefont{Porcu}},
  \bibinfo{author}{\bibfnamefont{F.}~\bibnamefont{Mocci}}, \bibnamefont{and}
  \bibinfo{author}{\bibfnamefont{A.}~\bibnamefont{Laaksonen}},
  \bibinfo{journal}{Appl. Math.} \textbf{\bibinfo{volume}{5}},
  \bibinfo{pages}{520} (\bibinfo{year}{2014}{\natexlab{a}}).

\bibitem[{\citenamefont{Wang}(2013)}]{wang2013electrostatic}
\bibinfo{author}{\bibfnamefont{Y.-L.} \bibnamefont{Wang}}, Ph.D. thesis,
  \bibinfo{school}{Department of Materials and Environmental Chemistry,
  Stockholm University} (\bibinfo{year}{2013}).

\bibitem[{\citenamefont{Wang et~al.}(2013{\natexlab{a}})\citenamefont{Wang,
  Laaksonen, and Lu}}]{wang2013implementation}
\bibinfo{author}{\bibfnamefont{Y.-L.} \bibnamefont{Wang}},
  \bibinfo{author}{\bibfnamefont{A.}~\bibnamefont{Laaksonen}},
  \bibnamefont{and} \bibinfo{author}{\bibfnamefont{Z.-Y.} \bibnamefont{Lu}},
  \bibinfo{journal}{J. Comput. Phys.} \textbf{\bibinfo{volume}{235}},
  \bibinfo{pages}{666} (\bibinfo{year}{2013}{\natexlab{a}}).

\bibitem[{\citenamefont{Wang et~al.}(2012)\citenamefont{Wang, Lu, and
  Laaksonen}}]{wang2012specific}
\bibinfo{author}{\bibfnamefont{Y.-L.} \bibnamefont{Wang}},
  \bibinfo{author}{\bibfnamefont{Z.-Y.} \bibnamefont{Lu}}, \bibnamefont{and}
  \bibinfo{author}{\bibfnamefont{A.}~\bibnamefont{Laaksonen}},
  \bibinfo{journal}{Phys. Chem. Chem. Phys.} \textbf{\bibinfo{volume}{14}},
  \bibinfo{pages}{8348} (\bibinfo{year}{2012}).

\bibitem[{\citenamefont{Shaw et~al.}(2007)\citenamefont{Shaw, Deneroff, Dror,
  Kuskin, Larson, Salmon, Young, Batson, Bowers, and Chao}}]{shaw2007anton}
\bibinfo{author}{\bibfnamefont{D.~E.} \bibnamefont{Shaw}},
  \bibinfo{author}{\bibfnamefont{M.~M.} \bibnamefont{Deneroff}},
  \bibinfo{author}{\bibfnamefont{R.~O.} \bibnamefont{Dror}},
  \bibinfo{author}{\bibfnamefont{J.~S.} \bibnamefont{Kuskin}},
  \bibinfo{author}{\bibfnamefont{R.~H.} \bibnamefont{Larson}},
  \bibinfo{author}{\bibfnamefont{J.~K.} \bibnamefont{Salmon}},
  \bibinfo{author}{\bibfnamefont{C.}~\bibnamefont{Young}},
  \bibinfo{author}{\bibfnamefont{B.}~\bibnamefont{Batson}},
  \bibinfo{author}{\bibfnamefont{K.~J.} \bibnamefont{Bowers}},
  \bibnamefont{and} \bibinfo{author}{\bibfnamefont{J.~C.} \bibnamefont{Chao}},
  \bibinfo{journal}{ACM SIGARCH Computer Architecture News}
  \textbf{\bibinfo{volume}{35}}, \bibinfo{pages}{1} (\bibinfo{year}{2007}).

\bibitem[{\citenamefont{Kirk}(2007)}]{kirk2007nvidia}
\bibinfo{author}{\bibfnamefont{D.}~\bibnamefont{Kirk}}, \bibinfo{journal}{ISMM}
  \textbf{\bibinfo{volume}{7}}, \bibinfo{pages}{103} (\bibinfo{year}{2007}).

\bibitem[{\citenamefont{Liu et~al.}(2007)\citenamefont{Liu, Schmidt, Voss, and
  M{\"u}ller-Wittig}}]{liu2007molecular}
\bibinfo{author}{\bibfnamefont{W.}~\bibnamefont{Liu}},
  \bibinfo{author}{\bibfnamefont{B.}~\bibnamefont{Schmidt}},
  \bibinfo{author}{\bibfnamefont{G.}~\bibnamefont{Voss}}, \bibnamefont{and}
  \bibinfo{author}{\bibfnamefont{W.}~\bibnamefont{M{\"u}ller-Wittig}}, in
  \emph{\bibinfo{booktitle}{International Conference on High-Performance
  Computing}} (\bibinfo{organization}{Springer}, \bibinfo{year}{2007}), pp.
  \bibinfo{pages}{185--196}.

\bibitem[{\citenamefont{Vetter et~al.}(2011)\citenamefont{Vetter, Glassbrook,
  Dongarra, Schwan, Loftis, McNally, Meredith, Rogers, Roth, Spafford
  et~al.}}]{vetter2011keeneland}
\bibinfo{author}{\bibfnamefont{J.~S.} \bibnamefont{Vetter}},
  \bibinfo{author}{\bibfnamefont{R.}~\bibnamefont{Glassbrook}},
  \bibinfo{author}{\bibfnamefont{J.}~\bibnamefont{Dongarra}},
  \bibinfo{author}{\bibfnamefont{K.}~\bibnamefont{Schwan}},
  \bibinfo{author}{\bibfnamefont{B.}~\bibnamefont{Loftis}},
  \bibinfo{author}{\bibfnamefont{S.}~\bibnamefont{McNally}},
  \bibinfo{author}{\bibfnamefont{J.}~\bibnamefont{Meredith}},
  \bibinfo{author}{\bibfnamefont{J.}~\bibnamefont{Rogers}},
  \bibinfo{author}{\bibfnamefont{P.}~\bibnamefont{Roth}},
  \bibinfo{author}{\bibfnamefont{K.}~\bibnamefont{Spafford}},
  \bibnamefont{et~al.}, \bibinfo{journal}{Computing in Science \& Engineering}
  \textbf{\bibinfo{volume}{13}}, \bibinfo{pages}{90} (\bibinfo{year}{2011}).

\bibitem[{\citenamefont{Blumers et~al.}(2017)\citenamefont{Blumers, Tang, Li,
  Li, and Karniadakis}}]{blumers2017gpu}
\bibinfo{author}{\bibfnamefont{A.~L.} \bibnamefont{Blumers}},
  \bibinfo{author}{\bibfnamefont{Y.-H.} \bibnamefont{Tang}},
  \bibinfo{author}{\bibfnamefont{Z.}~\bibnamefont{Li}},
  \bibinfo{author}{\bibfnamefont{X.}~\bibnamefont{Li}}, \bibnamefont{and}
  \bibinfo{author}{\bibfnamefont{G.~E.} \bibnamefont{Karniadakis}},
  \bibinfo{journal}{Comput. Phys. Commun.} \textbf{\bibinfo{volume}{217}},
  \bibinfo{pages}{171} (\bibinfo{year}{2017}).

\bibitem[{\citenamefont{Anderson et~al.}(2008)\citenamefont{Anderson, Lorenz,
  and Travesset}}]{anderson2008general}
\bibinfo{author}{\bibfnamefont{J.~A.} \bibnamefont{Anderson}},
  \bibinfo{author}{\bibfnamefont{C.~D.} \bibnamefont{Lorenz}},
  \bibnamefont{and}
  \bibinfo{author}{\bibfnamefont{A.}~\bibnamefont{Travesset}},
  \bibinfo{journal}{J. Comput. Phys.} \textbf{\bibinfo{volume}{227}},
  \bibinfo{pages}{5342} (\bibinfo{year}{2008}).

\bibitem[{\citenamefont{Friedrichs et~al.}(2009)\citenamefont{Friedrichs,
  Eastman, Vaidyanathan, Houston, Legrand, Beberg, Ensign, Bruns, and
  Pande}}]{friedrichs2009accelerating}
\bibinfo{author}{\bibfnamefont{M.~S.} \bibnamefont{Friedrichs}},
  \bibinfo{author}{\bibfnamefont{P.}~\bibnamefont{Eastman}},
  \bibinfo{author}{\bibfnamefont{V.}~\bibnamefont{Vaidyanathan}},
  \bibinfo{author}{\bibfnamefont{M.}~\bibnamefont{Houston}},
  \bibinfo{author}{\bibfnamefont{S.}~\bibnamefont{Legrand}},
  \bibinfo{author}{\bibfnamefont{A.~L.} \bibnamefont{Beberg}},
  \bibinfo{author}{\bibfnamefont{D.~L.} \bibnamefont{Ensign}},
  \bibinfo{author}{\bibfnamefont{C.~M.} \bibnamefont{Bruns}}, \bibnamefont{and}
  \bibinfo{author}{\bibfnamefont{V.~S.} \bibnamefont{Pande}},
  \bibinfo{journal}{J. Comput. Chem.} \textbf{\bibinfo{volume}{30}},
  \bibinfo{pages}{864} (\bibinfo{year}{2009}).

\bibitem[{\citenamefont{Brown et~al.}(2010)\citenamefont{Brown, Hampton,
  Agarwal, Wang, Crozier, and Plimpton}}]{brown2010porting}
\bibinfo{author}{\bibfnamefont{W.~M.} \bibnamefont{Brown}},
  \bibinfo{author}{\bibfnamefont{S.}~\bibnamefont{Hampton}},
  \bibinfo{author}{\bibfnamefont{P.}~\bibnamefont{Agarwal}},
  \bibinfo{author}{\bibfnamefont{P.}~\bibnamefont{Wang}},
  \bibinfo{author}{\bibfnamefont{P.}~\bibnamefont{Crozier}}, \bibnamefont{and}
  \bibinfo{author}{\bibfnamefont{S.}~\bibnamefont{Plimpton}},
  \bibinfo{journal}{Sandia National Laboratories, Tech. Rep}
  (\bibinfo{year}{2010}).

\bibitem[{\citenamefont{Stone et~al.}(2010)\citenamefont{Stone, Hardy,
  Ufimtsev, and Schulten}}]{stone2010gpu}
\bibinfo{author}{\bibfnamefont{J.~E.} \bibnamefont{Stone}},
  \bibinfo{author}{\bibfnamefont{D.~J.} \bibnamefont{Hardy}},
  \bibinfo{author}{\bibfnamefont{I.~S.} \bibnamefont{Ufimtsev}},
  \bibnamefont{and} \bibinfo{author}{\bibfnamefont{K.}~\bibnamefont{Schulten}},
  \bibinfo{journal}{J. Mol. Graph. Model} \textbf{\bibinfo{volume}{29}},
  \bibinfo{pages}{116} (\bibinfo{year}{2010}).

\bibitem[{\citenamefont{Gotz et~al.}(2012)\citenamefont{Gotz, Williamson, Xu,
  Poole, Le~Grand, and Walker}}]{gotz2012routine}
\bibinfo{author}{\bibfnamefont{A.~W.} \bibnamefont{Gotz}},
  \bibinfo{author}{\bibfnamefont{M.~J.} \bibnamefont{Williamson}},
  \bibinfo{author}{\bibfnamefont{D.}~\bibnamefont{Xu}},
  \bibinfo{author}{\bibfnamefont{D.}~\bibnamefont{Poole}},
  \bibinfo{author}{\bibfnamefont{S.}~\bibnamefont{Le~Grand}}, \bibnamefont{and}
  \bibinfo{author}{\bibfnamefont{R.~C.} \bibnamefont{Walker}},
  \bibinfo{journal}{J. Chem. Theory Comput.} \textbf{\bibinfo{volume}{8}},
  \bibinfo{pages}{1542} (\bibinfo{year}{2012}).

\bibitem[{\citenamefont{Salomon-Ferrer
  et~al.}(2013)\citenamefont{Salomon-Ferrer, Gotz, Poole, Le~Grand, and
  Walker}}]{salomon2013routine}
\bibinfo{author}{\bibfnamefont{R.}~\bibnamefont{Salomon-Ferrer}},
  \bibinfo{author}{\bibfnamefont{A.~W.} \bibnamefont{Gotz}},
  \bibinfo{author}{\bibfnamefont{D.}~\bibnamefont{Poole}},
  \bibinfo{author}{\bibfnamefont{S.}~\bibnamefont{Le~Grand}}, \bibnamefont{and}
  \bibinfo{author}{\bibfnamefont{R.~C.} \bibnamefont{Walker}},
  \bibinfo{journal}{J. Chem. Theory Comput.} \textbf{\bibinfo{volume}{9}},
  \bibinfo{pages}{3878} (\bibinfo{year}{2013}).

\bibitem[{\citenamefont{Abraham et~al.}(2015)\citenamefont{Abraham, Murtola,
  Schulz, P{\'a}ll, Smith, Hess, and Lindahl}}]{abraham2015gromacs}
\bibinfo{author}{\bibfnamefont{M.~J.} \bibnamefont{Abraham}},
  \bibinfo{author}{\bibfnamefont{T.}~\bibnamefont{Murtola}},
  \bibinfo{author}{\bibfnamefont{R.}~\bibnamefont{Schulz}},
  \bibinfo{author}{\bibfnamefont{S.}~\bibnamefont{P{\'a}ll}},
  \bibinfo{author}{\bibfnamefont{J.~C.} \bibnamefont{Smith}},
  \bibinfo{author}{\bibfnamefont{B.}~\bibnamefont{Hess}}, \bibnamefont{and}
  \bibinfo{author}{\bibfnamefont{E.}~\bibnamefont{Lindahl}},
  \bibinfo{journal}{SoftwareX} \textbf{\bibinfo{volume}{1}},
  \bibinfo{pages}{19} (\bibinfo{year}{2015}).

\bibitem[{\citenamefont{Zhu et~al.}(2013)\citenamefont{Zhu, Liu, Li, Qian,
  Milano, and Lu}}]{zhu2013galamost}
\bibinfo{author}{\bibfnamefont{Y.-L.} \bibnamefont{Zhu}},
  \bibinfo{author}{\bibfnamefont{H.}~\bibnamefont{Liu}},
  \bibinfo{author}{\bibfnamefont{Z.-W.} \bibnamefont{Li}},
  \bibinfo{author}{\bibfnamefont{H.-J.} \bibnamefont{Qian}},
  \bibinfo{author}{\bibfnamefont{G.}~\bibnamefont{Milano}}, \bibnamefont{and}
  \bibinfo{author}{\bibfnamefont{Z.-Y.} \bibnamefont{Lu}}, \bibinfo{journal}{J.
  Comput. Chem.} \textbf{\bibinfo{volume}{34}}, \bibinfo{pages}{2197}
  (\bibinfo{year}{2013}).

\bibitem[{\citenamefont{Zhu et~al.}(2018)\citenamefont{Zhu, Pan, Li, Liu, Qian,
  Zhao, Lu, and Sun}}]{zhu2018employing}
\bibinfo{author}{\bibfnamefont{Y.-L.} \bibnamefont{Zhu}},
  \bibinfo{author}{\bibfnamefont{D.}~\bibnamefont{Pan}},
  \bibinfo{author}{\bibfnamefont{Z.-W.} \bibnamefont{Li}},
  \bibinfo{author}{\bibfnamefont{H.}~\bibnamefont{Liu}},
  \bibinfo{author}{\bibfnamefont{H.-J.} \bibnamefont{Qian}},
  \bibinfo{author}{\bibfnamefont{Y.}~\bibnamefont{Zhao}},
  \bibinfo{author}{\bibfnamefont{Z.-Y.} \bibnamefont{Lu}}, \bibnamefont{and}
  \bibinfo{author}{\bibfnamefont{Z.-Y.} \bibnamefont{Sun}},
  \bibinfo{journal}{Mol. Phys.} \textbf{\bibinfo{volume}{116}},
  \bibinfo{pages}{1065} (\bibinfo{year}{2018}).

\bibitem[{\citenamefont{Yang et~al.}(2018)\citenamefont{Yang, Qian, and
  Lu}}]{yang2018new}
\bibinfo{author}{\bibfnamefont{S.-C.} \bibnamefont{Yang}},
  \bibinfo{author}{\bibfnamefont{H.-J.} \bibnamefont{Qian}}, \bibnamefont{and}
  \bibinfo{author}{\bibfnamefont{Z.-Y.} \bibnamefont{Lu}},
  \bibinfo{journal}{Appl. Comput. Harmon. Anal.} \textbf{\bibinfo{volume}{44}},
  \bibinfo{pages}{273} (\bibinfo{year}{2018}).

\bibitem[{\citenamefont{Yang et~al.}(2016)\citenamefont{Yang, Wang, Jiao, Qian,
  and Lu}}]{yang2016accelerating}
\bibinfo{author}{\bibfnamefont{S.-C.} \bibnamefont{Yang}},
  \bibinfo{author}{\bibfnamefont{Y.-L.} \bibnamefont{Wang}},
  \bibinfo{author}{\bibfnamefont{G.-S.} \bibnamefont{Jiao}},
  \bibinfo{author}{\bibfnamefont{H.-J.} \bibnamefont{Qian}}, \bibnamefont{and}
  \bibinfo{author}{\bibfnamefont{Z.-Y.} \bibnamefont{Lu}}, \bibinfo{journal}{J.
  Comput. Chem.} \textbf{\bibinfo{volume}{37}}, \bibinfo{pages}{378}
  (\bibinfo{year}{2016}).

\bibitem[{\citenamefont{Yang et~al.}(2017)\citenamefont{Yang, Lu, Qian, Wang,
  and Han}}]{yang2017hybrid}
\bibinfo{author}{\bibfnamefont{S.-C.} \bibnamefont{Yang}},
  \bibinfo{author}{\bibfnamefont{Z.-Y.} \bibnamefont{Lu}},
  \bibinfo{author}{\bibfnamefont{H.-J.} \bibnamefont{Qian}},
  \bibinfo{author}{\bibfnamefont{Y.-L.} \bibnamefont{Wang}}, \bibnamefont{and}
  \bibinfo{author}{\bibfnamefont{J.-P.} \bibnamefont{Han}},
  \bibinfo{journal}{Comput. Phys. Commun.} \textbf{\bibinfo{volume}{220}},
  \bibinfo{pages}{376} (\bibinfo{year}{2017}).

\bibitem[{\citenamefont{Yang and Wang}(2020)}]{yang2020hybrid}
\bibinfo{author}{\bibfnamefont{S.-C.} \bibnamefont{Yang}} \bibnamefont{and}
  \bibinfo{author}{\bibfnamefont{Y.-L.} \bibnamefont{Wang}},
  \bibinfo{journal}{ArXiv:2001.01583}  (\bibinfo{year}{2020}).

\bibitem[{Note1()}]{Note1}
Note1, \bibinfo{note}{the functional form can be chosen arbitrarily as long as
  this function leads to two fast decaying terms. Herein we choose the Gaussian
  charge density distribution function as an example to extract electrostatic
  energies and forces between charged particles.}

\bibitem[{\citenamefont{Cooley and Tukey}(1965)}]{cooley1965algorithm}
\bibinfo{author}{\bibfnamefont{J.}~\bibnamefont{Cooley}} \bibnamefont{and}
  \bibinfo{author}{\bibfnamefont{J.}~\bibnamefont{Tukey}},
  \bibinfo{journal}{Math. Comput.} \textbf{\bibinfo{volume}{19}},
  \bibinfo{pages}{297} (\bibinfo{year}{1965}).

\bibitem[{\citenamefont{Benedetto and Ferreira}(2001)}]{benedetto2001modern}
\bibinfo{author}{\bibfnamefont{J.~J.} \bibnamefont{Benedetto}}
  \bibnamefont{and} \bibinfo{author}{\bibfnamefont{P.~J.}
  \bibnamefont{Ferreira}}, \emph{\bibinfo{title}{Modern Sampling Theory:
  Mathematics and Applications}} (\bibinfo{publisher}{Springer Science \&
  Business Media}, \bibinfo{year}{2001}).

\bibitem[{\citenamefont{Nestler}(2016)}]{nestler2016parameter}
\bibinfo{author}{\bibfnamefont{F.}~\bibnamefont{Nestler}},
  \bibinfo{journal}{Front. Phys.} \textbf{\bibinfo{volume}{4}},
  \bibinfo{pages}{28} (\bibinfo{year}{2016}).

\bibitem[{\citenamefont{Hofmann et~al.}(2017)\citenamefont{Hofmann, Nestler,
  and Pippig}}]{hofmann2017nfft}
\bibinfo{author}{\bibfnamefont{M.}~\bibnamefont{Hofmann}},
  \bibinfo{author}{\bibfnamefont{F.}~\bibnamefont{Nestler}}, \bibnamefont{and}
  \bibinfo{author}{\bibfnamefont{M.}~\bibnamefont{Pippig}},
  \bibinfo{journal}{Appl. Numer. Math.} \textbf{\bibinfo{volume}{122}},
  \bibinfo{pages}{39} (\bibinfo{year}{2017}).

\bibitem[{\citenamefont{Hoogerbrugge and
  Koelman}(1992)}]{hoogerbrugge1992simulating}
\bibinfo{author}{\bibfnamefont{P.}~\bibnamefont{Hoogerbrugge}}
  \bibnamefont{and} \bibinfo{author}{\bibfnamefont{J.}~\bibnamefont{Koelman}},
  \bibinfo{journal}{Europhys. Lett.} \textbf{\bibinfo{volume}{19}},
  \bibinfo{pages}{155} (\bibinfo{year}{1992}).

\bibitem[{\citenamefont{Koelman and Hoogerbrugge}(1993)}]{koelman1993dynamic}
\bibinfo{author}{\bibfnamefont{J.}~\bibnamefont{Koelman}} \bibnamefont{and}
  \bibinfo{author}{\bibfnamefont{P.}~\bibnamefont{Hoogerbrugge}},
  \bibinfo{journal}{Europhys. Lett.} \textbf{\bibinfo{volume}{21}},
  \bibinfo{pages}{363} (\bibinfo{year}{1993}).

\bibitem[{\citenamefont{Espanol and Warren}(1995)}]{espanol1995statistical}
\bibinfo{author}{\bibfnamefont{P.}~\bibnamefont{Espanol}} \bibnamefont{and}
  \bibinfo{author}{\bibfnamefont{P.}~\bibnamefont{Warren}},
  \bibinfo{journal}{Europhys. Lett.} \textbf{\bibinfo{volume}{30}},
  \bibinfo{pages}{191} (\bibinfo{year}{1995}).

\bibitem[{\citenamefont{Groot and Warren}(1997)}]{groot1997dissipative}
\bibinfo{author}{\bibfnamefont{R.}~\bibnamefont{Groot}} \bibnamefont{and}
  \bibinfo{author}{\bibfnamefont{P.}~\bibnamefont{Warren}},
  \bibinfo{journal}{J. Chem. Phys.} \textbf{\bibinfo{volume}{107}},
  \bibinfo{pages}{4423} (\bibinfo{year}{1997}).

\bibitem[{\citenamefont{Pagonabarraga and
  Frenkel}(2001)}]{pagonabarraga2001dissipative}
\bibinfo{author}{\bibfnamefont{I.}~\bibnamefont{Pagonabarraga}}
  \bibnamefont{and} \bibinfo{author}{\bibfnamefont{D.}~\bibnamefont{Frenkel}},
  \bibinfo{journal}{J. Chem. Phys.} \textbf{\bibinfo{volume}{115}},
  \bibinfo{pages}{5015} (\bibinfo{year}{2001}).

\bibitem[{\citenamefont{Lu and Wang}(2013)}]{lu2013introduction}
\bibinfo{author}{\bibfnamefont{Z.-Y.} \bibnamefont{Lu}} \bibnamefont{and}
  \bibinfo{author}{\bibfnamefont{Y.-L.} \bibnamefont{Wang}}, in
  \emph{\bibinfo{booktitle}{Biomolecular Simulations}}
  (\bibinfo{publisher}{Springer}, \bibinfo{year}{2013}), pp.
  \bibinfo{pages}{617--633}.

\bibitem[{\citenamefont{Espanol and Warren}(2017)}]{espanol2017perspective}
\bibinfo{author}{\bibfnamefont{P.}~\bibnamefont{Espanol}} \bibnamefont{and}
  \bibinfo{author}{\bibfnamefont{P.~B.} \bibnamefont{Warren}},
  \bibinfo{journal}{J. Chem. Phys.} \textbf{\bibinfo{volume}{146}},
  \bibinfo{pages}{150901} (\bibinfo{year}{2017}).

\bibitem[{\citenamefont{Groot and Madden}(1998)}]{groot1998dynamic}
\bibinfo{author}{\bibfnamefont{R.}~\bibnamefont{Groot}} \bibnamefont{and}
  \bibinfo{author}{\bibfnamefont{T.}~\bibnamefont{Madden}},
  \bibinfo{journal}{J. Chem. Phys.} \textbf{\bibinfo{volume}{108}},
  \bibinfo{pages}{8713} (\bibinfo{year}{1998}).

\bibitem[{\citenamefont{Qian et~al.}(2005)\citenamefont{Qian, Lu, Chen, Li, and
  Sun}}]{qian2005computer}
\bibinfo{author}{\bibfnamefont{H.}~\bibnamefont{Qian}},
  \bibinfo{author}{\bibfnamefont{Z.}~\bibnamefont{Lu}},
  \bibinfo{author}{\bibfnamefont{L.}~\bibnamefont{Chen}},
  \bibinfo{author}{\bibfnamefont{Z.}~\bibnamefont{Li}}, \bibnamefont{and}
  \bibinfo{author}{\bibfnamefont{C.}~\bibnamefont{Sun}},
  \bibinfo{journal}{Macromolecules} \textbf{\bibinfo{volume}{38}},
  \bibinfo{pages}{1395} (\bibinfo{year}{2005}).

\bibitem[{\citenamefont{Rekvig et~al.}(2004)\citenamefont{Rekvig, Hafskjold,
  and Smit}}]{rekvig2004chain}
\bibinfo{author}{\bibfnamefont{L.}~\bibnamefont{Rekvig}},
  \bibinfo{author}{\bibfnamefont{B.}~\bibnamefont{Hafskjold}},
  \bibnamefont{and} \bibinfo{author}{\bibfnamefont{B.}~\bibnamefont{Smit}},
  \bibinfo{journal}{Phys. Rev. Lett.} \textbf{\bibinfo{volume}{92}},
  \bibinfo{pages}{116101} (\bibinfo{year}{2004}).

\bibitem[{\citenamefont{Whittle and Travis}(2010)}]{whittle2010dynamic}
\bibinfo{author}{\bibfnamefont{M.}~\bibnamefont{Whittle}} \bibnamefont{and}
  \bibinfo{author}{\bibfnamefont{K.}~\bibnamefont{Travis}},
  \bibinfo{journal}{J. Chem. Phys.} \textbf{\bibinfo{volume}{132}},
  \bibinfo{pages}{124906} (\bibinfo{year}{2010}).

\bibitem[{\citenamefont{Mai-Duy et~al.}(2015)\citenamefont{Mai-Duy, Phan-Thien,
  and Khoo}}]{mai2015investigation}
\bibinfo{author}{\bibfnamefont{N.}~\bibnamefont{Mai-Duy}},
  \bibinfo{author}{\bibfnamefont{N.}~\bibnamefont{Phan-Thien}},
  \bibnamefont{and} \bibinfo{author}{\bibfnamefont{B.~C.} \bibnamefont{Khoo}},
  \bibinfo{journal}{Comput. Phys. Commun.} \textbf{\bibinfo{volume}{189}},
  \bibinfo{pages}{37} (\bibinfo{year}{2015}).

\bibitem[{\citenamefont{Kranenburg and Smit}(2005)}]{kranenburg2005phase}
\bibinfo{author}{\bibfnamefont{M.}~\bibnamefont{Kranenburg}} \bibnamefont{and}
  \bibinfo{author}{\bibfnamefont{B.}~\bibnamefont{Smit}}, \bibinfo{journal}{J.
  Phys. Chem. B} \textbf{\bibinfo{volume}{109}}, \bibinfo{pages}{6553}
  (\bibinfo{year}{2005}).

\bibitem[{\citenamefont{Shillcock and Lipowsky}(2005)}]{shillcock2005tension}
\bibinfo{author}{\bibfnamefont{J.}~\bibnamefont{Shillcock}} \bibnamefont{and}
  \bibinfo{author}{\bibfnamefont{R.}~\bibnamefont{Lipowsky}},
  \bibinfo{journal}{Nat. Mater.} \textbf{\bibinfo{volume}{4}},
  \bibinfo{pages}{225} (\bibinfo{year}{2005}).

\bibitem[{\citenamefont{de~Meyer and Smit}(2009)}]{de2009effect}
\bibinfo{author}{\bibfnamefont{F.}~\bibnamefont{de~Meyer}} \bibnamefont{and}
  \bibinfo{author}{\bibfnamefont{B.}~\bibnamefont{Smit}},
  \bibinfo{journal}{Proc. Natl. Acad. Sci. U. S. A.}
  \textbf{\bibinfo{volume}{106}}, \bibinfo{pages}{3654} (\bibinfo{year}{2009}).

\bibitem[{\citenamefont{Li et~al.}(2012)\citenamefont{Li, Popel, and
  Karniadakis}}]{li2012blood}
\bibinfo{author}{\bibfnamefont{X.}~\bibnamefont{Li}},
  \bibinfo{author}{\bibfnamefont{A.~S.} \bibnamefont{Popel}}, \bibnamefont{and}
  \bibinfo{author}{\bibfnamefont{G.~E.} \bibnamefont{Karniadakis}},
  \bibinfo{journal}{Phys. Biology} \textbf{\bibinfo{volume}{9}},
  \bibinfo{pages}{026010} (\bibinfo{year}{2012}).

\bibitem[{\citenamefont{Li et~al.}(2017)\citenamefont{Li, Li, Chang,
  Lykotrafitis, and Em~Karniadakis}}]{li2017computational}
\bibinfo{author}{\bibfnamefont{X.}~\bibnamefont{Li}},
  \bibinfo{author}{\bibfnamefont{H.}~\bibnamefont{Li}},
  \bibinfo{author}{\bibfnamefont{H.-Y.} \bibnamefont{Chang}},
  \bibinfo{author}{\bibfnamefont{G.}~\bibnamefont{Lykotrafitis}},
  \bibnamefont{and}
  \bibinfo{author}{\bibfnamefont{G.}~\bibnamefont{Em~Karniadakis}},
  \bibinfo{journal}{J. Biomechanical Eng.} \textbf{\bibinfo{volume}{139}}
  (\bibinfo{year}{2017}).

\bibitem[{\citenamefont{Groot}(2003)}]{groot2003electrostatic}
\bibinfo{author}{\bibfnamefont{R.}~\bibnamefont{Groot}}, \bibinfo{journal}{J.
  Chem. Phys.} \textbf{\bibinfo{volume}{118}}, \bibinfo{pages}{11265}
  (\bibinfo{year}{2003}).

\bibitem[{\citenamefont{Gonz{\'a}lez-Melchor
  et~al.}(2006)\citenamefont{Gonz{\'a}lez-Melchor, Mayoral, Vel{\'a}zquez, and
  Alejandre}}]{gonzalez2006electrostatic}
\bibinfo{author}{\bibfnamefont{M.}~\bibnamefont{Gonz{\'a}lez-Melchor}},
  \bibinfo{author}{\bibfnamefont{E.}~\bibnamefont{Mayoral}},
  \bibinfo{author}{\bibfnamefont{M.}~\bibnamefont{Vel{\'a}zquez}},
  \bibnamefont{and}
  \bibinfo{author}{\bibfnamefont{J.}~\bibnamefont{Alejandre}},
  \bibinfo{journal}{J. Chem. Phys.} \textbf{\bibinfo{volume}{125}},
  \bibinfo{pages}{224107} (\bibinfo{year}{2006}).

\bibitem[{\citenamefont{Ibergay et~al.}(2009)\citenamefont{Ibergay, Malfreyt,
  and Tildesley}}]{ibergay2009electrostatic}
\bibinfo{author}{\bibfnamefont{C.}~\bibnamefont{Ibergay}},
  \bibinfo{author}{\bibfnamefont{P.}~\bibnamefont{Malfreyt}}, \bibnamefont{and}
  \bibinfo{author}{\bibfnamefont{D.~J.} \bibnamefont{Tildesley}},
  \bibinfo{journal}{J. Chem. Theory Comput.} \textbf{\bibinfo{volume}{5}},
  \bibinfo{pages}{3245} (\bibinfo{year}{2009}).

\bibitem[{\citenamefont{Warren et~al.}(2013)\citenamefont{Warren, Vlasov,
  Anton, and Masters}}]{warren2013screening}
\bibinfo{author}{\bibfnamefont{P.~B.} \bibnamefont{Warren}},
  \bibinfo{author}{\bibfnamefont{A.}~\bibnamefont{Vlasov}},
  \bibinfo{author}{\bibfnamefont{L.}~\bibnamefont{Anton}}, \bibnamefont{and}
  \bibinfo{author}{\bibfnamefont{A.~J.} \bibnamefont{Masters}},
  \bibinfo{journal}{J. Chem. Phys.} \textbf{\bibinfo{volume}{138}},
  \bibinfo{pages}{204907} (\bibinfo{year}{2013}).

\bibitem[{\citenamefont{Terr{\'o}n-Mej{\'\i}a
  et~al.}(2016)\citenamefont{Terr{\'o}n-Mej{\'\i}a, L{\'o}pez-Rend{\'o}n, and
  Goicochea}}]{terron2016electrostatics}
\bibinfo{author}{\bibfnamefont{K.~A.} \bibnamefont{Terr{\'o}n-Mej{\'\i}a}},
  \bibinfo{author}{\bibfnamefont{R.}~\bibnamefont{L{\'o}pez-Rend{\'o}n}},
  \bibnamefont{and} \bibinfo{author}{\bibfnamefont{A.~G.}
  \bibnamefont{Goicochea}}, \bibinfo{journal}{J. Phys.: Condens. Matter}
  \textbf{\bibinfo{volume}{28}}, \bibinfo{pages}{425101}
  (\bibinfo{year}{2016}).

\bibitem[{\citenamefont{Vaiwala et~al.}(2017)\citenamefont{Vaiwala, Jadhav, and
  Thaokar}}]{vaiwala2017electrostatic}
\bibinfo{author}{\bibfnamefont{R.}~\bibnamefont{Vaiwala}},
  \bibinfo{author}{\bibfnamefont{S.}~\bibnamefont{Jadhav}}, \bibnamefont{and}
  \bibinfo{author}{\bibfnamefont{R.}~\bibnamefont{Thaokar}},
  \bibinfo{journal}{J. Chem. Phys.} \textbf{\bibinfo{volume}{146}},
  \bibinfo{pages}{124904} (\bibinfo{year}{2017}).

\bibitem[{\citenamefont{Eslami et~al.}(2019)\citenamefont{Eslami, Khani, and
  Muller-Plathe}}]{eslami2019gaussian}
\bibinfo{author}{\bibfnamefont{H.}~\bibnamefont{Eslami}},
  \bibinfo{author}{\bibfnamefont{M.}~\bibnamefont{Khani}}, \bibnamefont{and}
  \bibinfo{author}{\bibfnamefont{F.}~\bibnamefont{Muller-Plathe}},
  \bibinfo{journal}{J. Chem. Theory Comput.} \textbf{\bibinfo{volume}{15}},
  \bibinfo{pages}{4197} (\bibinfo{year}{2019}).

\bibitem[{\citenamefont{Pagonabarraga et~al.}(2010)\citenamefont{Pagonabarraga,
  Rotenberg, and Frenkel}}]{pagonabarraga2010recent}
\bibinfo{author}{\bibfnamefont{I.}~\bibnamefont{Pagonabarraga}},
  \bibinfo{author}{\bibfnamefont{B.}~\bibnamefont{Rotenberg}},
  \bibnamefont{and} \bibinfo{author}{\bibfnamefont{D.}~\bibnamefont{Frenkel}},
  \bibinfo{journal}{Phys. Chem. Chem. Phys.} \textbf{\bibinfo{volume}{12}},
  \bibinfo{pages}{9566} (\bibinfo{year}{2010}).

\bibitem[{\citenamefont{Yan and Zhang}(2009)}]{yan2009dissipative}
\bibinfo{author}{\bibfnamefont{L.-T.} \bibnamefont{Yan}} \bibnamefont{and}
  \bibinfo{author}{\bibfnamefont{X.}~\bibnamefont{Zhang}},
  \bibinfo{journal}{Soft Matter} \textbf{\bibinfo{volume}{5}},
  \bibinfo{pages}{2101} (\bibinfo{year}{2009}).

\bibitem[{\citenamefont{Mao et~al.}(2015)\citenamefont{Mao, Lee, Vishnyakov,
  and Neimark}}]{mao2015modeling}
\bibinfo{author}{\bibfnamefont{R.}~\bibnamefont{Mao}},
  \bibinfo{author}{\bibfnamefont{M.-T.} \bibnamefont{Lee}},
  \bibinfo{author}{\bibfnamefont{A.}~\bibnamefont{Vishnyakov}},
  \bibnamefont{and} \bibinfo{author}{\bibfnamefont{A.~V.}
  \bibnamefont{Neimark}}, \bibinfo{journal}{J. Phys. Chem. B}
  \textbf{\bibinfo{volume}{119}}, \bibinfo{pages}{11673}
  (\bibinfo{year}{2015}).

\bibitem[{\citenamefont{Gavrilov et~al.}(2016)\citenamefont{Gavrilov,
  Chertovich, and Kramarenko}}]{gavrilov2016dissipative}
\bibinfo{author}{\bibfnamefont{A.}~\bibnamefont{Gavrilov}},
  \bibinfo{author}{\bibfnamefont{A.}~\bibnamefont{Chertovich}},
  \bibnamefont{and} \bibinfo{author}{\bibfnamefont{E.~Y.}
  \bibnamefont{Kramarenko}}, \bibinfo{journal}{J. Chem. Phys.}
  \textbf{\bibinfo{volume}{145}}, \bibinfo{pages}{174101}
  (\bibinfo{year}{2016}).

\bibitem[{\citenamefont{Lu and Hentschke}(2003)}]{lu2003computer}
\bibinfo{author}{\bibfnamefont{Z.~Y.} \bibnamefont{Lu}} \bibnamefont{and}
  \bibinfo{author}{\bibfnamefont{R.}~\bibnamefont{Hentschke}},
  \bibinfo{journal}{Phys. Rev. E} \textbf{\bibinfo{volume}{67}},
  \bibinfo{pages}{061807} (\bibinfo{year}{2003}).

\bibitem[{\citenamefont{Greengard and Lee}(2004)}]{greengard2004accelerating}
\bibinfo{author}{\bibfnamefont{L.}~\bibnamefont{Greengard}} \bibnamefont{and}
  \bibinfo{author}{\bibfnamefont{J.-Y.} \bibnamefont{Lee}},
  \bibinfo{journal}{SIAM Rev.} \textbf{\bibinfo{volume}{46}},
  \bibinfo{pages}{443} (\bibinfo{year}{2004}).

\bibitem[{\citenamefont{Dobrynin and Rubinstein}(2005)}]{dobrynin2005theory}
\bibinfo{author}{\bibfnamefont{A.}~\bibnamefont{Dobrynin}} \bibnamefont{and}
  \bibinfo{author}{\bibfnamefont{M.}~\bibnamefont{Rubinstein}},
  \bibinfo{journal}{Prog. Polym. Sci.} \textbf{\bibinfo{volume}{30}},
  \bibinfo{pages}{1049} (\bibinfo{year}{2005}).

\bibitem[{\citenamefont{Jusufi and Likos}(2009)}]{jusufi2009colloquium}
\bibinfo{author}{\bibfnamefont{A.}~\bibnamefont{Jusufi}} \bibnamefont{and}
  \bibinfo{author}{\bibfnamefont{C.}~\bibnamefont{Likos}},
  \bibinfo{journal}{Rev. Mod. Phys.} \textbf{\bibinfo{volume}{81}},
  \bibinfo{pages}{1753} (\bibinfo{year}{2009}).

\bibitem[{\citenamefont{Roiter and Minko}(2005)}]{roiter2005afm}
\bibinfo{author}{\bibfnamefont{Y.}~\bibnamefont{Roiter}} \bibnamefont{and}
  \bibinfo{author}{\bibfnamefont{S.}~\bibnamefont{Minko}}, \bibinfo{journal}{J.
  Am. Chem. Soc.} \textbf{\bibinfo{volume}{127}}, \bibinfo{pages}{15688}
  (\bibinfo{year}{2005}).

\bibitem[{\citenamefont{Liao et~al.}(2006)\citenamefont{Liao, Dobrynin, and
  Rubinstein}}]{liao2006counterion}
\bibinfo{author}{\bibfnamefont{Q.}~\bibnamefont{Liao}},
  \bibinfo{author}{\bibfnamefont{A.}~\bibnamefont{Dobrynin}}, \bibnamefont{and}
  \bibinfo{author}{\bibfnamefont{M.}~\bibnamefont{Rubinstein}},
  \bibinfo{journal}{Macromolecules} \textbf{\bibinfo{volume}{39}},
  \bibinfo{pages}{1920} (\bibinfo{year}{2006}).

\bibitem[{\citenamefont{Rubinstein and Colby}(2003)}]{rubinstein2003polymer}
\bibinfo{author}{\bibfnamefont{M.}~\bibnamefont{Rubinstein}} \bibnamefont{and}
  \bibinfo{author}{\bibfnamefont{R.~H.} \bibnamefont{Colby}},
  \emph{\bibinfo{title}{Polymer physics}} (\bibinfo{publisher}{Oxford
  University Press, New York}, \bibinfo{year}{2003}).

\bibitem[{\citenamefont{Sanders et~al.}(2005)\citenamefont{Sanders,
  Gu{\'a}queta, Angelini, Lee, Slimmer, Luijten, and
  Wong}}]{sanders2005structure}
\bibinfo{author}{\bibfnamefont{L.~K.} \bibnamefont{Sanders}},
  \bibinfo{author}{\bibfnamefont{C.}~\bibnamefont{Gu{\'a}queta}},
  \bibinfo{author}{\bibfnamefont{T.~E.} \bibnamefont{Angelini}},
  \bibinfo{author}{\bibfnamefont{J.-W.} \bibnamefont{Lee}},
  \bibinfo{author}{\bibfnamefont{S.~C.} \bibnamefont{Slimmer}},
  \bibinfo{author}{\bibfnamefont{E.}~\bibnamefont{Luijten}}, \bibnamefont{and}
  \bibinfo{author}{\bibfnamefont{G.~C.} \bibnamefont{Wong}},
  \bibinfo{journal}{Phys. Rev. Lett.} \textbf{\bibinfo{volume}{95}},
  \bibinfo{pages}{108302} (\bibinfo{year}{2005}).

\bibitem[{\citenamefont{Mei et~al.}(2006)\citenamefont{Mei, Lauterbach,
  Hoffmann, Borisov, Ballauff, and Jusufi}}]{mei2006collapse}
\bibinfo{author}{\bibfnamefont{Y.}~\bibnamefont{Mei}},
  \bibinfo{author}{\bibfnamefont{K.}~\bibnamefont{Lauterbach}},
  \bibinfo{author}{\bibfnamefont{M.}~\bibnamefont{Hoffmann}},
  \bibinfo{author}{\bibfnamefont{O.~V.} \bibnamefont{Borisov}},
  \bibinfo{author}{\bibfnamefont{M.}~\bibnamefont{Ballauff}}, \bibnamefont{and}
  \bibinfo{author}{\bibfnamefont{A.}~\bibnamefont{Jusufi}},
  \bibinfo{journal}{Phys. Rev. Lett.} \textbf{\bibinfo{volume}{97}},
  \bibinfo{pages}{158301} (\bibinfo{year}{2006}).

\bibitem[{\citenamefont{Roiter et~al.}(2010)\citenamefont{Roiter, Trotsenko,
  Tokarev, and Minko}}]{roiter2010single}
\bibinfo{author}{\bibfnamefont{Y.}~\bibnamefont{Roiter}},
  \bibinfo{author}{\bibfnamefont{O.}~\bibnamefont{Trotsenko}},
  \bibinfo{author}{\bibfnamefont{V.}~\bibnamefont{Tokarev}}, \bibnamefont{and}
  \bibinfo{author}{\bibfnamefont{S.}~\bibnamefont{Minko}}, \bibinfo{journal}{J.
  Am. Chem. Soc.} \textbf{\bibinfo{volume}{132}}, \bibinfo{pages}{13660}
  (\bibinfo{year}{2010}).

\bibitem[{\citenamefont{Percec et~al.}(2002)\citenamefont{Percec, Glodde, Bera,
  Miura, Shiyanovskaya, Singer, Balagurusamy, Heiney, Schnell, Rapp
  et~al.}}]{percec2002self}
\bibinfo{author}{\bibfnamefont{V.}~\bibnamefont{Percec}},
  \bibinfo{author}{\bibfnamefont{M.}~\bibnamefont{Glodde}},
  \bibinfo{author}{\bibfnamefont{T.}~\bibnamefont{Bera}},
  \bibinfo{author}{\bibfnamefont{Y.}~\bibnamefont{Miura}},
  \bibinfo{author}{\bibfnamefont{I.}~\bibnamefont{Shiyanovskaya}},
  \bibinfo{author}{\bibfnamefont{K.}~\bibnamefont{Singer}},
  \bibinfo{author}{\bibfnamefont{V.}~\bibnamefont{Balagurusamy}},
  \bibinfo{author}{\bibfnamefont{P.}~\bibnamefont{Heiney}},
  \bibinfo{author}{\bibfnamefont{I.}~\bibnamefont{Schnell}},
  \bibinfo{author}{\bibfnamefont{A.}~\bibnamefont{Rapp}}, \bibnamefont{et~al.},
  \bibinfo{journal}{Nature} \textbf{\bibinfo{volume}{419}},
  \bibinfo{pages}{384} (\bibinfo{year}{2002}).

\bibitem[{\citenamefont{Tian and Ma}(2013)}]{ma2013theoretical}
\bibinfo{author}{\bibfnamefont{W.-D.} \bibnamefont{Tian}} \bibnamefont{and}
  \bibinfo{author}{\bibfnamefont{Y.-Q.} \bibnamefont{Ma}},
  \bibinfo{journal}{Chem. Soc. Rev.} \textbf{\bibinfo{volume}{42}},
  \bibinfo{pages}{705} (\bibinfo{year}{2013}).

\bibitem[{\citenamefont{Liu et~al.}(2009)\citenamefont{Liu, Bryantsev, Diallo,
  and Goddard~III}}]{liu2009pamam}
\bibinfo{author}{\bibfnamefont{Y.}~\bibnamefont{Liu}},
  \bibinfo{author}{\bibfnamefont{V.}~\bibnamefont{Bryantsev}},
  \bibinfo{author}{\bibfnamefont{M.}~\bibnamefont{Diallo}}, \bibnamefont{and}
  \bibinfo{author}{\bibfnamefont{W.}~\bibnamefont{Goddard~III}},
  \bibinfo{journal}{J. Am. Chem. Soc.} \textbf{\bibinfo{volume}{131}},
  \bibinfo{pages}{2798} (\bibinfo{year}{2009}).

\bibitem[{\citenamefont{Maiti et~al.}(2004)\citenamefont{Maiti,
  {\c{C}}aǧ{\i}n, Wang, and Goddard}}]{maiti2004structure}
\bibinfo{author}{\bibfnamefont{P.~K.} \bibnamefont{Maiti}},
  \bibinfo{author}{\bibfnamefont{T.}~\bibnamefont{{\c{C}}aǧ{\i}n}},
  \bibinfo{author}{\bibfnamefont{G.}~\bibnamefont{Wang}}, \bibnamefont{and}
  \bibinfo{author}{\bibfnamefont{W.~A.} \bibnamefont{Goddard}},
  \bibinfo{journal}{Macromolecules} \textbf{\bibinfo{volume}{37}},
  \bibinfo{pages}{6236} (\bibinfo{year}{2004}).

\bibitem[{\citenamefont{Lee and Larson}(2011)}]{lee2011effects}
\bibinfo{author}{\bibfnamefont{H.}~\bibnamefont{Lee}} \bibnamefont{and}
  \bibinfo{author}{\bibfnamefont{R.~G.} \bibnamefont{Larson}},
  \bibinfo{journal}{Macromolecules} \textbf{\bibinfo{volume}{44}},
  \bibinfo{pages}{2291} (\bibinfo{year}{2011}).

\bibitem[{\citenamefont{Mecke et~al.}(2005)\citenamefont{Mecke, Majoros, Patri,
  Baker~Jr, Holl, and Orr}}]{mecke2005lipid}
\bibinfo{author}{\bibfnamefont{A.}~\bibnamefont{Mecke}},
  \bibinfo{author}{\bibfnamefont{I.}~\bibnamefont{Majoros}},
  \bibinfo{author}{\bibfnamefont{A.}~\bibnamefont{Patri}},
  \bibinfo{author}{\bibfnamefont{J.}~\bibnamefont{Baker~Jr}},
  \bibinfo{author}{\bibfnamefont{M.}~\bibnamefont{Holl}}, \bibnamefont{and}
  \bibinfo{author}{\bibfnamefont{B.}~\bibnamefont{Orr}},
  \bibinfo{journal}{Langmuir} \textbf{\bibinfo{volume}{21}},
  \bibinfo{pages}{10348} (\bibinfo{year}{2005}).

\bibitem[{\citenamefont{Armand et~al.}(2009)\citenamefont{Armand, Endres,
  MacFarlane, Ohno, and Scrosati}}]{armand2009ionic}
\bibinfo{author}{\bibfnamefont{M.}~\bibnamefont{Armand}},
  \bibinfo{author}{\bibfnamefont{F.}~\bibnamefont{Endres}},
  \bibinfo{author}{\bibfnamefont{D.}~\bibnamefont{MacFarlane}},
  \bibinfo{author}{\bibfnamefont{H.}~\bibnamefont{Ohno}}, \bibnamefont{and}
  \bibinfo{author}{\bibfnamefont{B.}~\bibnamefont{Scrosati}},
  \bibinfo{journal}{Nat. Mater.} \textbf{\bibinfo{volume}{8}},
  \bibinfo{pages}{621} (\bibinfo{year}{2009}).

\bibitem[{\citenamefont{Castner~Jr et~al.}(2011)\citenamefont{Castner~Jr,
  Margulis, Maroncelli, and Wishart}}]{castner2011ionic}
\bibinfo{author}{\bibfnamefont{E.~W.} \bibnamefont{Castner~Jr}},
  \bibinfo{author}{\bibfnamefont{C.~J.} \bibnamefont{Margulis}},
  \bibinfo{author}{\bibfnamefont{M.}~\bibnamefont{Maroncelli}},
  \bibnamefont{and} \bibinfo{author}{\bibfnamefont{J.~F.}
  \bibnamefont{Wishart}}, \bibinfo{journal}{Annu. Rev. Phys. Chem.}
  \textbf{\bibinfo{volume}{62}}, \bibinfo{pages}{85} (\bibinfo{year}{2011}).

\bibitem[{\citenamefont{Wang et~al.}(2020)\citenamefont{Wang, Li, Sarman,
  Mocci, Lu, Yuan, Laaksonen, and Fayer}}]{wang2020chemrev}
\bibinfo{author}{\bibfnamefont{Y.-L.} \bibnamefont{Wang}},
  \bibinfo{author}{\bibfnamefont{B.}~\bibnamefont{Li}},
  \bibinfo{author}{\bibfnamefont{S.}~\bibnamefont{Sarman}},
  \bibinfo{author}{\bibfnamefont{F.}~\bibnamefont{Mocci}},
  \bibinfo{author}{\bibfnamefont{Z.-Y.} \bibnamefont{Lu}},
  \bibinfo{author}{\bibfnamefont{J.}~\bibnamefont{Yuan}},
  \bibinfo{author}{\bibfnamefont{A.}~\bibnamefont{Laaksonen}},
  \bibnamefont{and} \bibinfo{author}{\bibfnamefont{M.~D.} \bibnamefont{Fayer}},
  \bibinfo{journal}{Chem. Rev.} \textbf{\bibinfo{volume}{120}}
  (\bibinfo{year}{2020}).

\bibitem[{\citenamefont{Hayes et~al.}(2015)\citenamefont{Hayes, Warr, and
  Atkin}}]{hayes2015structure}
\bibinfo{author}{\bibfnamefont{R.}~\bibnamefont{Hayes}},
  \bibinfo{author}{\bibfnamefont{G.~G.} \bibnamefont{Warr}}, \bibnamefont{and}
  \bibinfo{author}{\bibfnamefont{R.}~\bibnamefont{Atkin}},
  \bibinfo{journal}{Chem. Rev.} \textbf{\bibinfo{volume}{115}},
  \bibinfo{pages}{6357} (\bibinfo{year}{2015}).

\bibitem[{\citenamefont{Wang et~al.}(2014{\natexlab{b}})\citenamefont{Wang, Lu,
  and Laaksonen}}]{wang2014heterogeneous}
\bibinfo{author}{\bibfnamefont{Y.-L.} \bibnamefont{Wang}},
  \bibinfo{author}{\bibfnamefont{Z.-Y.} \bibnamefont{Lu}}, \bibnamefont{and}
  \bibinfo{author}{\bibfnamefont{A.}~\bibnamefont{Laaksonen}},
  \bibinfo{journal}{Phys. Chem. Chem. Phys.} \textbf{\bibinfo{volume}{16}},
  \bibinfo{pages}{20731} (\bibinfo{year}{2014}{\natexlab{b}}).

\bibitem[{\citenamefont{Wang et~al.}(2017)\citenamefont{Wang, Golets, Li,
  Sarman, and Laaksonen}}]{wang2017interfacial}
\bibinfo{author}{\bibfnamefont{Y.-L.} \bibnamefont{Wang}},
  \bibinfo{author}{\bibfnamefont{M.}~\bibnamefont{Golets}},
  \bibinfo{author}{\bibfnamefont{B.}~\bibnamefont{Li}},
  \bibinfo{author}{\bibfnamefont{S.}~\bibnamefont{Sarman}}, \bibnamefont{and}
  \bibinfo{author}{\bibfnamefont{A.}~\bibnamefont{Laaksonen}},
  \bibinfo{journal}{ACS Appl. Mater. Interfaces} \textbf{\bibinfo{volume}{9}},
  \bibinfo{pages}{4976} (\bibinfo{year}{2017}).

\bibitem[{\citenamefont{Wang}(2018)}]{wang2018competitive}
\bibinfo{author}{\bibfnamefont{Y.-L.} \bibnamefont{Wang}}, \bibinfo{journal}{J.
  Phys. Chem. B} \textbf{\bibinfo{volume}{122}}, \bibinfo{pages}{6570}
  (\bibinfo{year}{2018}).

\bibitem[{\citenamefont{Wang et~al.}(2019)\citenamefont{Wang, Sarman, Golets,
  Mocci, Lu, and Laaksonen}}]{wang2019multigranular}
\bibinfo{author}{\bibfnamefont{Y.-L.} \bibnamefont{Wang}},
  \bibinfo{author}{\bibfnamefont{S.}~\bibnamefont{Sarman}},
  \bibinfo{author}{\bibfnamefont{M.}~\bibnamefont{Golets}},
  \bibinfo{author}{\bibfnamefont{F.}~\bibnamefont{Mocci}},
  \bibinfo{author}{\bibfnamefont{Z.-Y.} \bibnamefont{Lu}}, \bibnamefont{and}
  \bibinfo{author}{\bibfnamefont{A.}~\bibnamefont{Laaksonen}},
  \bibinfo{journal}{Ionic Liquids: Synthesis, Properties, Technologies and
  Applications} p.~\bibinfo{pages}{55} (\bibinfo{year}{2019}).

\bibitem[{\citenamefont{Wang and Voth}(2005)}]{wang2005unique}
\bibinfo{author}{\bibfnamefont{Y.}~\bibnamefont{Wang}} \bibnamefont{and}
  \bibinfo{author}{\bibfnamefont{G.~A.} \bibnamefont{Voth}},
  \bibinfo{journal}{J. Am. Chem. Soc.} \textbf{\bibinfo{volume}{127}},
  \bibinfo{pages}{12192} (\bibinfo{year}{2005}).

\bibitem[{\citenamefont{Hu and Margulis}(2006)}]{hu2006heterogeneity}
\bibinfo{author}{\bibfnamefont{Z.}~\bibnamefont{Hu}} \bibnamefont{and}
  \bibinfo{author}{\bibfnamefont{C.~J.} \bibnamefont{Margulis}},
  \bibinfo{journal}{Proc. Natl. Acad. Sci. U. S. A.}
  \textbf{\bibinfo{volume}{103}}, \bibinfo{pages}{831} (\bibinfo{year}{2006}).

\bibitem[{\citenamefont{Jin et~al.}(2010)\citenamefont{Jin, Li, and
  Maroncelli}}]{jin2010heterogeneous}
\bibinfo{author}{\bibfnamefont{H.}~\bibnamefont{Jin}},
  \bibinfo{author}{\bibfnamefont{X.}~\bibnamefont{Li}}, \bibnamefont{and}
  \bibinfo{author}{\bibfnamefont{M.}~\bibnamefont{Maroncelli}},
  \bibinfo{journal}{J. Phys. Chem. B} \textbf{\bibinfo{volume}{114}},
  \bibinfo{pages}{11370} (\bibinfo{year}{2010}).

\bibitem[{\citenamefont{Ji et~al.}(2013)\citenamefont{Ji, Shi, Wang, and
  Saielli}}]{ji2013effect}
\bibinfo{author}{\bibfnamefont{Y.}~\bibnamefont{Ji}},
  \bibinfo{author}{\bibfnamefont{R.}~\bibnamefont{Shi}},
  \bibinfo{author}{\bibfnamefont{Y.}~\bibnamefont{Wang}}, \bibnamefont{and}
  \bibinfo{author}{\bibfnamefont{G.}~\bibnamefont{Saielli}},
  \bibinfo{journal}{J. Phys. Chem. B} \textbf{\bibinfo{volume}{117}},
  \bibinfo{pages}{1104} (\bibinfo{year}{2013}).

\bibitem[{\citenamefont{Kim et~al.}(2016)\citenamefont{Kim, Park, and
  Jung}}]{kim2016heterogeneous}
\bibinfo{author}{\bibfnamefont{S.}~\bibnamefont{Kim}},
  \bibinfo{author}{\bibfnamefont{S.-W.} \bibnamefont{Park}}, \bibnamefont{and}
  \bibinfo{author}{\bibfnamefont{Y.}~\bibnamefont{Jung}},
  \bibinfo{journal}{Phys. Chem. Chem. Phys.} \textbf{\bibinfo{volume}{18}},
  \bibinfo{pages}{6486} (\bibinfo{year}{2016}).

\bibitem[{\citenamefont{Gabl et~al.}(2012)\citenamefont{Gabl, Schr{\"o}der, and
  Steinhauser}}]{gabl2012computational}
\bibinfo{author}{\bibfnamefont{S.}~\bibnamefont{Gabl}},
  \bibinfo{author}{\bibfnamefont{C.}~\bibnamefont{Schr{\"o}der}},
  \bibnamefont{and}
  \bibinfo{author}{\bibfnamefont{O.}~\bibnamefont{Steinhauser}},
  \bibinfo{journal}{J. Chem. Phys.} \textbf{\bibinfo{volume}{137}},
  \bibinfo{pages}{094501} (\bibinfo{year}{2012}).

\bibitem[{\citenamefont{Wang et~al.}(2013{\natexlab{b}})\citenamefont{Wang,
  Lyubartsev, Lu, and Laaksonen}}]{wang2013multiscale}
\bibinfo{author}{\bibfnamefont{Y.-L.} \bibnamefont{Wang}},
  \bibinfo{author}{\bibfnamefont{A.}~\bibnamefont{Lyubartsev}},
  \bibinfo{author}{\bibfnamefont{Z.-Y.} \bibnamefont{Lu}}, \bibnamefont{and}
  \bibinfo{author}{\bibfnamefont{A.}~\bibnamefont{Laaksonen}},
  \bibinfo{journal}{Phys. Chem. Chem. Phys.} \textbf{\bibinfo{volume}{15}},
  \bibinfo{pages}{7701} (\bibinfo{year}{2013}{\natexlab{b}}).

\bibitem[{\citenamefont{Roy et~al.}(2010)\citenamefont{Roy, Patel, Conte, and
  Maroncelli}}]{roy2010dynamics}
\bibinfo{author}{\bibfnamefont{D.}~\bibnamefont{Roy}},
  \bibinfo{author}{\bibfnamefont{N.}~\bibnamefont{Patel}},
  \bibinfo{author}{\bibfnamefont{S.}~\bibnamefont{Conte}}, \bibnamefont{and}
  \bibinfo{author}{\bibfnamefont{M.}~\bibnamefont{Maroncelli}},
  \bibinfo{journal}{J. Phys. Chem. B} \textbf{\bibinfo{volume}{114}},
  \bibinfo{pages}{8410} (\bibinfo{year}{2010}).

\end{thebibliography}

\end{document}